\pdfoutput=1
\documentclass[11pt]{article}
\usepackage[utf8]{inputenc}
\usepackage{cite}
\usepackage{hyperref}
\hypersetup{colorlinks=true,linkcolor=bluecyan,citecolor=red3,urlcolor=green3,pdfencoding=auto,linktocpage}
\usepackage{amsmath,amssymb,amsbsy,amstext,amsthm,simplewick,amsfonts}
\usepackage{mathrsfs}
\usepackage{graphicx}
\usepackage{wrapfig}
\usepackage{upgreek}
\usepackage{bm} 
\usepackage{framed}
\usepackage{bbm}
\usepackage{textcomp}
\usepackage{adjustbox}
\usepackage{makecell}
\usepackage{tcolorbox}
\usepackage{empheq}
\usepackage[normalem]{ulem}
\usepackage{cancel}
\usepackage{enumitem}
\usepackage{braket} 
\usepackage{array}
\usepackage{dsfont}
\usepackage{physics}
\usepackage{ulem}
\usepackage{xcolor}
\usepackage{caption}
\usepackage{subcaption}
\usepackage{tocloft}
\usepackage{tikz}
\usepackage{ulem}
\usepackage[labelfont=bf]{caption}
\usetikzlibrary{decorations.markings}
\usetikzlibrary{decorations.pathmorphing}

\usepackage[framemethod=default]{mdframed}

\newmdenv[backgroundcolor=gray!10,%
skipabove=5pt,%
skipbelow=5pt,%
leftmargin=2pt,%
rightmargin=2pt,%
innertopmargin=-6pt,%
innerbottommargin=5pt,%
innerleftmargin=5pt,%
innerrightmargin=5pt,%
splittopskip=0pt,%
splitbottomskip=0pt,%
linewidth=0pt,%
nobreak=true]%
{keyeqn}

\graphicspath{{Fig/}}
\definecolor{lightgreen}{cmyk}{0.2, 0, 0.2, 0.2}
\definecolor{lightgray}{cmyk}{0.1,0.2,0,0.1}
\definecolor{lightgray2}{cmyk}{0.1,0.1,0,0.1}
\definecolor{greyish2}{rgb}{.96,.96,.96}
\definecolor{bluecyan}{RGB}{0, 100, 200}
\definecolor{blue3}{RGB}{31,119,180}
\definecolor{red3}{RGB}{214,39,40}
\definecolor{orange3}{RGB}{255,127,14}
\definecolor{green3}{RGB}{44,160,44}
\definecolor{red2}{RGB}{255,0,0}
\definecolor{green2}{RGB}{0,170,0}
\definecolor{blue2}{RGB}{0,128,255}
\definecolor{magenta2}{RGB}{191,64,191}
\definecolor{purple2}{RGB}{112,48,160}
\definecolor{orange2}{RGB}{255,192,0}
\definecolor{blue2}{RGB}{117,223,230}
\definecolor{red4}{RGB}{186,60,71}

\newcommand{\email}[1]{\footnote{\href{mailto:#1}{\nolinkurl{#1}}}}

\renewcommand{\eqref}[1]{(\ref{#1})}

\def\bfk{\mathbf{k}}
\def\bfx{\mathbf{x}}

\def \a{\alpha}
\def \b{\beta}

\def \sa{\mathsf{a}}
\def \sb{\mathsf{b}}
\def \sc{\mathsf{c}}

\def \dd{\mathrm{d}}
\def\bgm{\begin{matrix}}
\def\edm{\end{matrix}}

\newcommand{\pFqcomma}{\mskip\pFqmuskip}
\newcommand*\pregFq[6][8]{%
	\begingroup 
	\pFqmuskip=#1mu\relax
	\mathcode`\,=\string"8000
	\begingroup\lccode`\~=`\,
	\lowercase{\endgroup\let~}\pFqcomma
	{}_{#2}\mathbf{F}_{#3}{\left[\genfrac..{0pt}{}{#4}{#5};#6\right]}%
	\endgroup
}

\newmuskip\pFqmuskip
\newcommand*\pFq[6][8]{%
	\begingroup 
	\pFqmuskip=#1mu\relax
	\mathcode`\,=\string"8000
	\begingroup\lccode`\~=`\,
	\lowercase{\endgroup\let~}\pFqcomma
	{}_{#2}F_{#3}{\left[\genfrac..{0pt}{}{#4}{#5};#6\right]}%
	\endgroup
}

\newcommand*\pdressFq[6][8]{%
	\begingroup 
	\pFqmuskip=#1mu\relax
	\mathcode`\,=\string"8000
	\begingroup\lccode`\~=`\,
	\lowercase{\endgroup\let~}\pFqcomma
	{}_{#2}\mathcal{F}_{#3}{\left[\genfrac..{0pt}{}{#4}{#5};#6\right]}%
	\endgroup
}
\numberwithin{equation}{section}

\allowdisplaybreaks[1]
\begin{document}

	\begin{titlepage}
\setcounter{page}{1} \baselineskip=15.5pt 
\thispagestyle{empty}
		\renewcommand*{\thefootnote}{\fnsymbol{footnote}}

    \begin{center}
        {\fontsize{15.5}{15.5} {\bf{Fermionic Bubble Loop in Cosmological Collider Revisited:}}\\
        [11pt]
       \fontsize{14}{14}{\textit{Exact signals from spectral and Mellin-Barnes methods}}}
    \end{center}
    \vskip 18pt
    \begin{center}
    \noindent
    {\fontsize{12}{18} \selectfont 
    Shuntaro Aoki 
   \footnote[1]{\href{mailto:shuntaro.aoki@riken.jp}{shuntaro.aoki@riken.jp}}$^{,a}$, 
    Zhehan Qin
   \footnote[2]{\href{mailto:qzh21@mails.tsinghua.edu.cn}{qzh21@mails.tsinghua.edu.cn}}$^{,b}$,
    Masahide Yamaguchi 
   \footnote[3]{\href{mailto:gucci@ibs.re.kr}{gucci@ibs.re.kr}}$^{,c,d,e}$
    and Yuhang Zhu 
    \footnote[4]{\href{mailto:yhzhu@ibs.re.kr}{yhzhu@ibs.re.kr}}$^{,c}$
    }
\end{center}
		
    \begin{center}
    \vskip 8pt
		$a$ \textit{RIKEN Center for Interdisciplinary Theoretical and Mathematical
Sciences (iTHEMS),
 Wako, Saitama 351-0198, Japan} \\[5pt]
		$b$ \textit{Department of Physics, Tsinghua University, Beijing 100084, China}  \\[5pt]
		$c$  \textit{Cosmology, Gravity and Astroparticle Physics Group,\\
            Center for Theoretical Physics of the Universe,\\
            Institute for Basic Science, Daejeon 34126, Korea}        \\[5pt]
        $d$ \textit{Department of Physics, Institute of Science Tokyo,
            2-12-1 Ookayama, Meguro-ku, \\Tokyo 152-8551, Japan} \\ [5pt]
        $e$ \textit{ Department of Physics and IPAP, Yonsei University, 50 Yonsei-ro, Seodaemun-gu,\\ Seoul 03722, Korea}
    \end{center}
		
\noindent\rule{\textwidth}{0.4pt}
     \noindent \textbf{Abstract}~~~
     Fermionic degrees of freedom are essential ingredients in cosmological collider physics and are well motivated by many phenomenological models beyond the Standard Model, but their signals remain largely unexplored due to the difficulty of computing loop diagrams. In this work, we ask how fermionic bubble loops contribute to cosmological collider signals and provide an exact answer for arbitrary couplings. 
     We develop two parallel analytical methods whose agreement provides a non-trivial check of the result. The first method is similar in spirit to spectral decomposition and is built directly from an identity for the product of propagators, which turns the bubble signal into an infinite sum of tree-level exchange signals. The second method is based on the Mellin-Barnes representation, where the result is reconstructed from the residues of distinct families of poles. We also show that the fermionic bubble  can be generated from the scalar bubble by the action of appropriate differential operators. As a phenomenologically important application, we consider Yukawa interactions between fermions and the inflaton, finding that the resulting bispectrum signal vanishes identically. Through the spectral decomposition, this vanishing can be traced to a field redefinition of the associated tree-level counterparts.

		
	\end{titlepage} 
	

	\setcounter{tocdepth}{2}
	{
		\tableofcontents
	}

    \renewcommand*{\thefootnote}{\arabic{footnote}}
\setcounter{footnote}{0} 

	\newpage

	
\section{Introduction}\label{sec: intro}
We have long been seeking to understand the fundamental laws of nature. Over the past century, one of the principal paths has been through collider experiments, exploring new physics at ever shorter distances and higher energies. However, there are questions that no terrestrial experiment may ever be able to answer, for which the universe itself may have already performed the experiment in its earliest epochs, encoding the results for us in the sky. To this end, we look into the statistics of fluctuations in the Cosmic Microwave Background (CMB) and Large Scale Structures (LSS), tracing them back to their initial values set by inflation. These are characterised  by the so-called primordial \textit{cosmological correlators}, the $n$-point correlation functions of curvature perturbations $\zeta$ and gravitons $\gamma_{ij}$, evaluated at the end of inflation. Although we can only access these boundary data, they encode detailed information about the rich physical processes that took place during inflation.

The intrinsic energy scale of inflation, characterised by the Hubble parameter $H$, can reach as high as $10^{13}$ GeV. In such a rapidly expanding and high energy environment, heavy degrees of freedom can be excited on shell via spontaneous particle production. Though quickly diluted and short-lived, their presence is felt through their interactions with the inflaton or graviton fields, leaving distinctive imprints on the higher point correlators with $n\ge3$, or equivalently, the primordial non-Gaussianities. By analysing the kinematic dependence of these primordial fossils, one can in principle extract the mass, spin, parity, and other properties of new heavy particles. This programme, known as \textit{cosmological collider (CC) physics}\cite{Chen:2009zp,Baumann:2011nk,Noumi:2012vr,Arkani-Hamed:2015bza}, has attracted considerable attention over the past decade, with a wide range of new physics models and signatures having been explored \cite{Gong:2013sma, Chen:2015lza,Chen:2016uwp,Lee:2016vti,An:2017hlx,Arkani-Hamed:2018kmz,Lu:2019tjj,Hook:2019zxa,Liu:2019fag, Kumar:2019ebj,Wang:2019gbi,Wang:2020ioa,Aoki:2020zbj,Pinol:2021aun,Cui:2021iie,Reece:2022soh,Chen:2022vzh,Qin:2022lva,Aoki:2023tjm,Jazayeri:2023xcj,Jazayeri:2023kji,Pinol:2023oux,Ema:2023dxm,Tong:2023krn,Chakraborty:2023eoq,Aoki:2024jha,Wu:2024wti,Craig:2024qgy,McCulloch:2024hiz,Pajer:2024ckd,Wang:2025qww,Aoki:2025uff,Kumar:2025anx,Jazayeri:2025vlv,Xianyu:2025lbk,Colas:2025ind,Bodas:2025vpb,Cheung:2025dmc,Ferreira:2026tyj}, and recent works have started to search for cosmological collider signals in observational data \cite{Cabass:2024wob,Goldstein:2024bky,Sohn:2024xzd,Goldstein:2025eyj,Anbajagane:2025uro,Suman:2025tpv,Green:2026yev,Kumar:2026ogn,Philcox:2026bfa,Kumar:2026dih,Ferreira:2026rsl,You:2026xoq}.

To extract useful information buried inside correlators and confront it with cosmological data, a precise theoretical understanding of the signatures generated by different processes is essential. Yet the computation of correlators involving massive spinning fields is notoriously challenging within traditional methods such as the Schwinger-Keldysh formalism\cite{Chen:2017ryl}, which requires evaluating complicated nested time integrals over products of mode functions for these fields. These are typically Hankel functions, and in more general settings may become even more complicated Whittaker functions\cite{Qin:2022fbv,Aoki:2023wdc,Stefanyszyn:2023qov}. In cases where some symmetries are absent, modified dispersion relations arise and analytical solutions may not even exist. The situation has, however, improved dramatically in recent years, with all kinds of new theoretical techniques being developed. These include, for example, solving differential equations from a boundary perspective \cite{Arkani-Hamed:2018kmz,Baumann:2019oyu,Baumann:2020dch,Baumann:2022jpr,Pimentel:2022fsc,Jazayeri:2022kjy,Qin:2022fbv,Wang:2022eop,Qin:2023ejc,Aoki:2024uyi,Liu:2024str,
Qin:2025xct,Xianyu:2025lbk} or integro-differential equations when non-trivial time dependence is introduced \cite{Jazayeri:2025vlv}, the reformulation of correlators in Mellin space \cite{Sleight:2019mgd,Sleight:2019hfp,Sleight:2020obc,Sleight:2021plv,Premkumar:2021mlz}, partial Mellin-Barnes (MB) representations to simplify bulk time integrals \cite{Qin:2022lva,Qin:2022fbv,Qin:2024gtr}, dispersive methods \cite{Liu:2024xyi,Chowdhury:2026upp,Das:2026vfv}, spectral decomposition \cite{Marolf:2010zp,Xianyu:2022jwk,Loparco:2023rug,Werth:2024mjg,
Qin:2025xct,Zhang:2025nzd} and de Sitter momentum space\cite{Belrhali:2026ktb,Belrhali:2026rkn}. See also \cite{Pajer:2020wnj,Pajer:2020wxk,Cabass:2021fnw,Xianyu:2023ytd,Wang:2025qfh,Cespedes:2025ple,Baumann:2026atn,Chen:2023iix,Grimm:2024tbg,Chen:2024glu,Chen:2026dqp} for an incomplete list of references.  However, most of these developments have been mainly focused on massive fields with \textit{integer} spins, whereas fermionic fields remain largely unexplored.

Fermions are, after all, the fundamental constituents of matter, and their diverse phenomenology on the cosmological correlators has indeed been explored in many directions. To name a few, the Standard Model background from fermions on cosmological colliders has been systematically discussed around ten years ago\cite{Chen:2016hrz,Chen:2016uwp}, and subsequently extended to inflaton induced Standard Model fermion dynamics and non-trivial Higgs vacua\cite{Hook:2019vcn,Hook:2019zxa}. Neutrinos, as a representative fermionic example, have also been extensively investigated in \cite{Chen:2018xck}, with later work further exploring their connections to leptogenesis\cite{Cui:2021iie} and seesaw motivated scenarios\cite{You:2024hit,Han:2024qbw,You:2026xoq}. Other directions include studies related to supersymmetry\cite{Alexander:2019vtb}, CP violation in modular inflation\cite{Aoki:2026olh}, and non-perturbative fermionic dynamics such as BCS-like condensation \cite{Tong:2023krn,Fujikura:2025xgl}. 

However, since fermionic fields always appear in pairs, their leading contribution to cosmological correlators is necessarily at loop level. Previous studies of fermionic signatures have thus been forced to rely on approximations, such as expanding the mode functions in the late time limit, whose precision remains unclear. On the other hand, as discussed above, significant theoretical progress has been made in recent years, and  cosmological correlators at the loop-level have themselves attracted growing attention \cite{Lee:2023jby,Xianyu:2022jwk,Qin:2023bjk,Cespedes:2023aal,Salcedo:2022aal,AguiSalcedo:2023nds,Qin:2024gtr,Bhowmick:2025mxh,Zhang:2025nzd,Jain:2025maa,Cespedes:2025ple,Pimentel:2026kqc,Grafe:2026qsm,Farren:2026hao,Baumann:2026atn,Chen:2026dqp,Westerdijk:2026msm}. It is therefore both timely and necessary to revisit the problem of fermionic correlators with these new tools in hand. In addition, in the presence of the so-called chemical potential \cite{Wang:2019gbi,Sou:2021juh}, loop effects on cosmological correlators can become sizeable \cite{Chen:2018xck,Hook:2019zxa,Tong:2023krn,Aoki:2026olh}, despite the fact that loop diagrams are normally suppressed by loop factors.  Although we do not pursue this possibility here and instead focus on the minimal setup without a chemical potential, our analysis provides an important first step toward incorporating such effects in future studies. 

In this work, we present the computation of the cosmological collider signals arising from fermionic bubble loops with the general couplings as shown in Figure \ref{Fig: SK_diagrams}, approached through two entirely different methods. The results no longer rely on approximations and capture the full oscillatory signal of the correlators.
\vspace{0.37cm}

\textbf{Spectral method.} The first method is rooted in the K\"{a}ll\'{e}n-Lehmann representation in de Sitter spacetime \cite{Marolf:2010zp,DiPietro:2021sjt,Sleight:2021plv,Loparco:2023rug}. Through this spectral decomposition, the bubble loop can be recast as the sum over tree-level diagrams with varying masses, each of which we are by now well equipped to handle analytically. This strategy has already been successfully applied to integer-spin fields, including scalars and massive spin one fields\cite{Xianyu:2022jwk,Zhang:2025nzd,Grafe:2026qsm}, and the fermionic spectral representation is also found recently in  \cite{Altshuler:2025qmk}.

In our work, we first observe that the fermionic bubble is related to its scalar counterpart through the action of certain time differential operators, which in principle allows the full result, including all cosmological collider signals and background contributions, to be extracted directly from previous scalar results. In the actual calculation, we take a slightly different path, building on certain mathematical identities for hypergeometric functions \cite{ProductofHyper,Bros:2011vh},  which does not require prior knowledge of the spectral density and is potentially more straightforward to generalise. More precisely, this identity relates the product of two hypergeometric functions to an infinite series in a single one. Since the position space propagators are themselves of this form, the decomposition follows immediately. The subsequent Fourier transform to momentum space requires more care, and we will present a detailed treatment of this step. We also address a subtle ordering issue between the spectral and time integrals, showing that it leaves the cosmological collider signals unaffected. The method is first illustrated in the scalar case, where full agreement with existing results is found\cite{Xianyu:2022jwk}, and then applied to the fermionic case to obtain the complete cosmological collider signals from the bubble loop.
\vspace{0.37cm}

\textbf{Mellin-Barnes transformation method.} The second method proceeds analogously to the scalar case~\cite{Qin:2024gtr}, employing a partial Mellin-Barnes transformation for the massive fermionic propagators. Following this transformation, the special functions are converted into Gamma functions, which are considerably more tractable. As a result, both the momentum and time integrals become essentially trivial, and the origins of the various contributions are rendered transparent. For the two classes of cosmological collider signals to be defined below, the non-local contributions can be directly identified from the pole structure of the contour integral, whereas the extraction of the local contributions requires an additional application of the (first) Barnes' lemma \cite{BarnesLemma1}. This approach may also be well suited to scenarios that deviate from exact de Sitter spacetime, although the trade-off is that it typically leads to multi-layered series summations.  Finally, although the two methods yield totally different expressions, we have verified in detail that they are in perfect agreement.
\vspace{0.37cm}

\textbf{Phenomenology with Yukawa couplings.} Our calculation starts from general couplings at the four-point level, and the bispectrum is obtained by taking the soft limit of one external leg. After carefully handling the cancellation of spurious divergences, we arrive at the \textit{complete} cosmological collider signals in the three-point function. For the phenomenological discussion, one of the most important and well motivated couplings is the Yukawa-type, as adopted for example in previous work on probing leptogenesis \cite{Cui:2021iie}. With the complete results now in hand, free from the approximations employed in earlier work,  we nevertheless find that:
\begin{center}
    \textit{Cosmological collider signals vanish in the bispectra generated by \\fermionic bubble loops with the Yukawa coupling.} 
\end{center}
We emphasise that this result is specific to the bubble topology with a Yukawa vertex, where the fermionic mode functions are those of pure de Sitter space, and that it applies at the level of the three-point function. We will later demonstrate this explicitly using the complete results. An intuitive understanding can also be gained from the spectral decomposition, where the fermionic bubble signal is expressed as a sum over tree-level diagrams. For Yukawa-type couplings, it maps to a series of trees with quadratic mixing. One can show that each such contribution individually vanishes via a field redefinition, and the vanishing of the full bubble then follows immediately. This suggests that the results previously obtained using approximation methods may not be reliable.
\vspace{0.37cm}

\textbf{Outline.} This paper is structured as follows: Section \ref{sec: fermion_ds} collects the necessary ingredients, covering the fermionic mode functions and propagators in de Sitter. We then discuss the relation between the fermionic and scalar bubbles, and derive the loop seed integral for generating correlators with general couplings. Section \ref{sec: signals_from_bubble} first separates the different components of the bubble loop using the cosmological collider cutting rules. The two methods, spectral decomposition and Mellin-Barnes transformation, are then applied in Sections \ref{sec: spectral} and \ref{sec: MBmethod} respectively to compute the signals from the fermionic bubble, with the results of both approaches compared in Section \ref{sec: compare}. Section \ref{sec: cancell_div} derives the three-point limit by cancelling the spurious divergences, and Section \ref{sec: yukawa} discusses the vanishing of the cosmological collider signals for Yukawa-type couplings. We conclude and outline future directions in Section \ref{sec: conclusion}. As the derivations are rather technical, we present only the key results in the main text, with the relevant mathematical formulae collected in Appendix \ref{App:MathFormulae} and the intermediate steps given in Appendix \ref{App: derivations}.

\paragraph{Notations and conventions.}
Throughout this paper, we adopt the $(-,+,+,+)$ metric sign convention  and the background spacetime is taken to be:
\begin{align}
    \mathrm{d}s^2=\frac{1}{H^2\tau^2}(-\mathrm{d}\tau^2+\mathrm{d}\mathbf{x}^2)\,,
\end{align}
where $-\infty<\tau<0$ is the conformal time and bold letters denote three-dimensional spatial vectors. For simplicity, we mostly set the Hubble constant to unity $H = 1$ unless otherwise stated.  A prime on correlation functions $\langle\cdots\rangle'$ indicates the momentum-conserving $\delta$-function and the factor $(2\pi)^3$ are omitted.
As products and ratios of Gamma functions appear frequently throughout, we adopt the compact notation:
\begin{align}
\label{gammasymbol}
\Gamma\left[\alpha_1\dots \alpha_n\right]=\Gamma(\alpha_1)\dots \Gamma(\alpha_n)\,,\qquad\qquad
\Gamma\left[\bgm \alpha_1\dots \alpha_n\\
        \beta_1\dots \beta_m\edm\right]=\dfrac{\Gamma(\alpha_1)\dots \Gamma(\alpha_n)}{\Gamma(\beta_1)\dots \Gamma(\beta_m)}\,.
\end{align}
Hypergeometric functions ${}_p F_q$ of various types appear frequently in our analytical calculations, and to avoid clutter, we introduce two additional variant forms. The regularised hypergeometric function is defined as
\begin{align}
    \pregFq{p}{q}{a_1,\cdots,a_p}{b_1,\cdots,b_q}{u}\equiv\frac{1}{
    \Gamma\left[ b_1\cdots b_q\right]
    }\times\pFq{p}{q}{a_1,\cdots,a_p}{b_1 \cdots b_q}{u}\,, \label{eq: reg_pFq}
\end{align}
and the dressed one reads:
\begin{align}
    \pdressFq{p}{q}{a_1,\cdots,a_p}{b_1,\cdots,b_q}{u}\equiv\Gamma\left[\bgm a_1\dots a_p\\
        b_1\dots b_q\edm\right]\times\pFq{p}{q}{a_1,\cdots,a_p}{b_1,\cdots,b_q}{u}\,. \label{eq: dress_pFq}
\end{align}
 We also frequently use the shorthand notations, for example $k_{1234}\equiv \sum_{i=1}^{4} k_i$ and $p_{123}\equiv p_{1}+p_2+p_3$. The dimensionless kinematic variables used in this paper are
 \begin{align}
     r_1\equiv\frac{s}{k_{12}}\,,\qquad\quad r_2\equiv\frac{s}{k_{34}}\,,
 \end{align}
 where $\mathbf s\equiv \bfk_1+\bfk_2$ denotes the $s$-channel momentum, while $k_i\equiv|\bfk_i|$ and $s\equiv |\bf s|$.

As for the fermionic conventions, the Pauli matrices are defined as $\sigma^a=(1,\vec{\sigma})$ and $\bar{\sigma}^a=(1,-\vec{\sigma})$, where $\vec{\sigma}$ denotes the standard Pauli matrices. The Dirac matrices and their algebra are given by
\begin{align}
	\gamma^a=\left(\begin{array}{cc}
		0 & \sigma^a\\
		\bar{\sigma}^a & 0
	\end{array}\right)\quad,\quad\gamma^5=\left(\begin{array}{cc}
	-1 & ~0\\
	~~0 & ~1
\end{array}\right)\,,\quad\quad \{\gamma^a,\gamma^b\}=-2\eta^{ab}\,,\quad\quad \{\gamma^5,\gamma^a\}=0~.
\end{align}
Greek indices $\mu,\nu,\rho,\ldots$ denote curved spacetime indices, while Latin indices $a,b,c,\ldots$ denote local Lorentz indices. We also adopt the van der Waerden notation for two-component spinors following the conventions of \cite{Martin:1997ns,Dreiner:2008tw}. All other variables and functions will be defined as they appear in the main text.

\section{Fermionic fields on dS background}\label{sec: fermion_ds}

In this section, we begin by reviewing basic ingredients of fermionic fields in de Sitter spacetime, starting with the quantisation of the free fermion as a warm-up. We then derive the fermionic bubble functions, which are the central objects in the following sections. We will mainly follow the notation and discussion in \cite{Tong:2023krn}. Readers may also refer to \cite{Chen:2018xck,Sou:2021juh,Ema:2023dxm,Schaub:2023scu,You:2026xoq} for further details.

\subsubsection*{Quantisation of fermions}
The free covariant Lagrangian for the Dirac fermion $\Psi$ can be expressed using vierbein (tetrad) as 
\begin{align}
	S_{\text{dS}}\left[\Psi\right]=\int \dd^4x\sqrt{-g}~\left(i\,\bar{\Psi}\gamma^a\, e^\mu_{~a}\nabla_\mu\Psi-m\bar{\Psi}\Psi\right)~,
\end{align}
where the dS vierbein is chosen as $e_{\mu}^{~a}=a(\tau)\,\delta^a_\mu$  and $e^\mu_{~a}=a^{-1}(\tau)\,\delta^\mu_a$. The spin connection is related to the vierbein via $\omega_{\mu a b}=-e^\nu_{~b}\nabla_{\mu}e_{\nu}^{~c}\,\eta_{ac}$, with $\eta_{ab}$ denoting the flat metric in the mostly positive sign convention. The covariant derivative acting on spinors is then given by $\nabla_\mu\equiv\partial_\mu+\frac{i}{4}\omega_{\mu ab}\sigma^{ab}$ with $\sigma^{a b}=\frac{i}{2}(\gamma^a\gamma^b-\gamma^b\gamma^a)$. We can rescale $\Psi\rightarrow a^{-3/2}{\widetilde{\Psi}}$, and the Lagrangian takes a simpler form as 
\begin{align}
    \mathcal{L}=i{\widetilde{\bar{\Psi}}}\gamma^a\delta^\mu_a\partial_\mu{\widetilde{\Psi}}-am\,{\widetilde{\bar{\Psi}}}{\widetilde{\Psi}}~. \label{eq:Fermion_Lagrangian}
\end{align}
For the massless fermion, the field is conformally coupled, and its equation of motion retains the same simple form as in flat spacetime after the rescaling. However, the mass term in (\ref{eq:Fermion_Lagrangian}) complicates the mode functions. For our purpose, it is often preferable to adopt the two-component Weyl spinor notation which offers technical simplicity, so we will write 
\begin{align}
	\widetilde{\Psi}=
	\begin{pmatrix}
	\xi_\alpha\\
	\chi^{\dag\dot{\alpha}}
	\end{pmatrix}~, \qquad \widetilde{\bar{\Psi}}=\left(\chi^\alpha~~\xi^{\dag}_{\dot{\alpha}}\right)\,,
\end{align}
here $\xi$ denotes the left-handed Weyl spinor, carrying an undotted index, while $\chi$ represents the right-handed Weyl spinor that has a dotted index.
Then we can rewrite the action (\ref{eq:Fermion_Lagrangian}) as
\begin{align}
S_{\text{dS}}\left[\Psi\right]=\int \dd\tau \mathrm{d}^3 \bfx\Big[\,i\xi^\dag\bar{\sigma}^a\delta^{\mu}_{a}\partial_\mu\xi+i\chi^\dag\bar{\sigma}^a\delta^{\mu}_{a}\partial_\mu\chi-a m \left(\chi\xi+\xi^\dag\chi^\dag\right)\Big]~,
\end{align}
in which $\chi\xi=\chi^\a\xi_\a=\chi^\alpha \epsilon_{\alpha\beta}\xi^\beta$, and the antisymmetric symbol $\epsilon$ follows the convention of \cite{Martin:1997ns}. Alternatively, the Dirac field can be interpreted as a combination of two fermionic fields via a suitable transformation like 
\begin{align}
	\xi=\frac{\tilde{\psi}_1+i\,\tilde{\psi}_2}{\sqrt{2}}~,\qquad \chi=\frac{\tilde{\psi}_1-i\,\tilde{\psi}_2}{\sqrt{2}}~,
\end{align}
in such a way that the Lagrangian for the free Dirac field reduces to two identical copies of the two-component spinor Lagrangian  
\begin{align}
	S_{\text{dS}}\left[\Psi\right]=S_{\psi_1}+S_{\psi_2}~,
\end{align}
 where the action of the \textit{rescaled} Weyl fermion $\tilde{\psi}$ with the Majorana mass reads 
\begin{align}
    S_{\psi}=\int \dd\tau \mathrm{d}^3 \bfx\,\Big[i\tilde\psi^\dag\bar{\sigma}^a\delta^{\mu}_{a}\partial_\mu\tilde\psi-\frac{1}{2}a m \left(\tilde\psi\tilde\psi+\tilde\psi^\dag\tilde\psi^\dag\right)\Big]~,\label{eq: Weyl_action}
\end{align}
with $\tilde{\psi}\equiv a^{3/2}\psi$ being the rescaled Weyl fermion, defined so that the kinetic term takes a form analogous to that in flat spacetime. After 
transforming to momentum space, and decomposing into the helicity basis, we obtain
\begin{align}
	\tilde{\psi}_\alpha\left(\tau,\mathbf{x}\right)=\int\frac{\dd^3\bfk}{(2\pi)^3}\sum_{\lambda=\pm}\left[h^\lambda_\alpha(\hat{\bfk})u_\lambda(k,\tau)b^\lambda_{\mathbf{k}}e^{i\mathbf{k}\cdot\mathbf{x}}+h^{\lambda\dagger}_{\dot{\gamma}}(\hat{\bfk})\bar{\sigma}^{0\dot{\gamma}\beta}\epsilon_{\alpha\beta}v^*_\lambda(k,\tau)b^\lambda_{\mathbf{k}}{}^\dagger e^{-i\mathbf{k}\cdot\mathbf{x}} \right]~,
\end{align}
here the creation and annihilation operators satisfy the anti-commutation relation
\begin{align}
\left\{b^\lambda_{\mathbf{k}},b^{\lambda'\dagger}_{\mathbf{k'}}\right\}=(2\pi)^3\delta^{\lambda\lambda'}\delta^3(\mathbf{k}-\mathbf{k}')\,,
\end{align}
and spinor-helicity eigenvectors $h^\lambda_\alpha(\hat{\bfk})$ satisfy properties
\begin{align}
	-\hat{\bfk}^i\sigma^0_{\alpha\dot{\beta}}\bar{\sigma}^{i\dot{\beta}\gamma}h^\lambda_\gamma(\hat{\bfk})=\lambda h^\lambda_\alpha(\hat{\bfk})~,\qquad h^{\lambda\dagger}_{\dot{\gamma}}(\hat{\bfk})\bar{\sigma}^{0\dot{\gamma}\beta}\sigma^i_{\beta\dot{\alpha}}\hat{\bfk}^i=\lambda h^{\lambda\dagger}_{\dot{\alpha}}(\hat{\bfk})~,
\end{align}
as well as the orthonormal conditions, 
\begin{align}
	h^{\lambda\dagger}_{\dot{\alpha}}\bar{\sigma}^{0\dot{\alpha}\beta}h^{\lambda'}_\beta=\delta^{\lambda\lambda'}~,\qquad \sum_{\lambda=\pm} h^\lambda_\alpha h^{\lambda\dagger}_{\dot{\beta}}\bar{\sigma}^{0\beta\dot{\beta}}=\delta^\beta_\alpha~,
\end{align}
the explicit forms of the helicity basis are given by
\begin{equation}
	h^+_\alpha(\mathbf{\hat k})=\left(\begin{array}{ccc}
		\cos\frac{\theta}{2}\\
		e^{i\phi}\sin\frac{\theta}{2}
	\end{array}\right)_\alpha~~,~~~h^-_\alpha(\mathbf{\hat k})=\left(\begin{array}{ccc}
		-e^{-i\phi}\sin\frac{\theta}{2}\\
		\cos\frac{\theta}{2}
	\end{array}\right)_\alpha~,
\end{equation}
with $\hat{\mathbf{ k}}$ pointing toward the $(\theta,\phi)$ direction in spherical coordinates. Another useful identity that will be frequently applied in the later discussion is 
\begin{align}
    h^{\lambda}_\alpha(\hat{\bfk})h^{\lambda\dagger}_{\dot{\beta}}(\hat{\bfk})=\frac{1}{2}\left(\sigma^0+\lambda\, \hat{\bfk}\cdot \vec{\sigma}\right)_{\alpha\dot{\beta}}\,.\label{eq: Product_Helicity}
\end{align}
Substituting the mode expansion  into the equation of motion, we can obtain a set of coupled first-order differential equations
\begin{align}
    &i u'_\lambda+\lambda k u_\lambda= am v_\lambda\,,\\
    &i v'_\lambda-\lambda k v_\lambda= am u_\lambda\,,
\end{align}
where the prime denotes the derivative with respect to $\tau$. They can also be decoupled into the pairs of the second-order differential equations
\begin{align}
    &u''_\lambda-aH u'_\lambda+\left(k^2+a^2m^2+i\lambda aH k\right)u_\lambda=0\,,\\
    &v''_{\lambda}-aHv'_{\lambda}+\left(k^2+a^2m^2-i\lambda aHk \right) v_{\lambda}=0\,.
\end{align}
By imposing the Bunch-Davies vacuum and canonical commutation relation, we can fix the mode functions of different helicities as 
\begin{keyeqn}
\begin{align}
&u_+= v_-=\frac{m}{\sqrt{-2k\tau}}\, W_{-\frac{1}{2},\,im}(2ik\tau)~,\qquad	u_-=v_+=\frac{i}{\sqrt{-2k\tau}}\, W_{\frac{1}{2},\,i m}(2ik\tau)~, \label{eq:Fermion_Modefunction}
\end{align}
\end{keyeqn}
here $W_{a,\mu}$ is the Whittaker-$W$ function and the mass $m$ is normalised by the Hubble parameter. In contrast to \cite{Tong:2023krn}, where a non-zero chemical potential is introduced via the parity-violating term $\bar{\psi}\gamma^0\gamma^5\psi$, enhancing one helicity mode while suppressing the other such that  $u_{\pm}\neq v_{\mp}$. We focus on the case without a chemical potential. The notation also differs from that of \cite{Ema:2023dxm,Schaub:2023scu},  where the mode functions are expressed in terms of the Hankel function $H^{(1)}_{im}$. We have verified those two are equivalent upon using the relations between the Whittaker functions $W_{\pm\frac{1}{2},i m}$ and Hankel functions
\begin{subequations}
\begin{align}
    W_{-\frac{1}{2},\,\mu}(-2i z)&=\frac{\sqrt{\pi} z}{2\mu}e^{(2\mu+1)\pi i/4}\,\left[ H^{(1)}_{\mu+\frac{1}{2}}(z)+i\,H^{(1)}_{\mu-\frac{1}{2}}(z)\right]~\,,\\
    W_{\frac{1}{2},\,\mu}(-2  i z)
    &=\frac{\sqrt{\pi} z}{2}e^{(2\mu-1)\pi i/4}\left[H^{(1)}_{\mu-\frac{1}{2}}(z)+i\,H^{(1)}_{\mu+\frac{1}{2}}(z)\right]~.
\end{align}\label{eq: WhittakerToHankel}
\end{subequations}
More details about the derivation of these relations can be found in Appendix~\ref{App:MathFormulae}. Expanding the fermionic mode functions in the IR limit $\tau\to 0$, one finds that they scale as $u,v\sim \tau^{\pm i m}$. This shows that spin-1/2 fields always reside in the principal series\cite{Schaub:2023scu}, with their mode functions exhibiting oscillatory behaviour regardless of how small the mass is. This stands in sharp contrast to the bosonic case, where a complex scaling dimension arises only when the mass parameter exceeds a critical threshold, such as $m>\frac{3}{2}H$ for scalars and $m>(s-\frac{1}{2}) H$ for higher spin-$s$ fields. 
\subsubsection*{Seed integral and example}
\begin{figure}[htp]
    \centering
    \begin{subfigure}{0.45\textwidth}
    \centering
\begin{tikzpicture}[baseline={([yshift=-.5ex]current bounding box.center)}, line width=1. pt, scale=3.5]
    \draw[red3, thick] (0.5,0) arc[start angle=0, end angle=180, radius=0.25];
    \draw[blue3, thick] (0.5,0) arc[start angle=360, end angle=180, radius=0.25];
    \draw[black] (0, 0) -- (-0.285, 0.305);
    \draw[black] (0, 0) -- (-0.285, -0.305);
    \node[draw, rectangle, minimum width=6.0pt, minimum height=6.0pt, inner sep=0pt, anchor=south east] at (-0.28,0.3) {};
    \node[draw, rectangle, minimum width=6.0pt, minimum height=6.0pt, inner sep=0pt, anchor=north east] at (-0.28,-0.3) {};
    \draw[black] (0.5, 0) -- (0.755, 0.305);
    \draw[black] (0.5, 0) -- (0.755, -0.305);
    \draw[draw=lightgray2, fill=lightgray2] (0, 0) circle (.035cm);
    \draw[draw=lightgray2, fill=lightgray2] (0.5, 0) circle (.035cm);
    \node[draw, rectangle, minimum width=6.0pt, minimum height=6.0pt, inner sep=0pt, anchor=south west] at (0.75,0.3) {};
    \node[draw, rectangle, minimum width=6.0pt, minimum height=6.0pt, inner sep=0pt, anchor=north west] at (0.75,-0.3) {};
    \node at (-0.12, 0.23) {$k_1$};
    \node at (-0.12, -0.23) {$k_2$};
    \node at (0.58, 0.23) {$k_3$};
    \node at (0.58, -0.23) {$k_4$};
    \node at (0.25,0.32){\textcolor{red3}{$\psi_1 , m_1$}};
    \node at (0.25,-0.32){\textcolor{blue3}{$\psi_2 , m_2$}};
    \node at (0.25,0.18){\textcolor{red3}{$\mathbf{p}_1$}};
    \node at (0.25,-0.18){\textcolor{blue3}{$\mathbf{p}_2$}};
\end{tikzpicture} 
    \end{subfigure}
        \begin{subfigure}{0.45\textwidth}
        \hspace{0.1\textwidth}
\begin{tikzpicture}[baseline={([yshift=-.5ex]current bounding box.center)}, line width=1. pt, scale=3.5]
    \draw[red3, thick] (0.5,0) arc[start angle=0, end angle=180, radius=0.25];
    \draw[blue3, thick] (0.5,0) arc[start angle=360, end angle=180, radius=0.25];
    \draw[black] (0, 0) -- (-0.285, 0.305);
    \draw[black] (0, 0) -- (-0.285, -0.305);
    \node[draw, rectangle, minimum width=6.0pt, minimum height=6.0pt, inner sep=0pt, anchor=south east] at (-0.28,0.3) {};
    \node[draw, rectangle, minimum width=6.0pt, minimum height=6.0pt, inner sep=0pt, anchor=north east] at (-0.28,-0.3) {};
    \draw[black] (0.5, 0) -- (0.85, 0);
    \draw[draw=lightgray2, fill=lightgray2] (0, 0) circle (.035cm);
    \draw[draw=lightgray2, fill=lightgray2] (0.5, 0) circle (.035cm);
    \node[draw, rectangle, minimum width=6.0pt, minimum height=6.0pt, inner sep=0pt, anchor=south west] at (0.85,-0.8pt) {};
    \node at (-0.12, 0.23) {$k_1$};
    \node at (-0.12, -0.23) {$k_2$};
    \node at (0.68, 0.08) {$k_3$};
    \node at (0.25,0.32){\textcolor{red3}{$\psi_1 , m_1$}};
    \node at (0.25,-0.32){\textcolor{blue3}{$\psi_2 , m_2$}};
    \node at (0.25,0.18){\textcolor{red3}{$\mathbf{p}_1$}};
    \node at (0.25,-0.18){\textcolor{blue3}{$\mathbf{p}_2$}};
\end{tikzpicture} 
    \end{subfigure}
    \caption{The SK diagrams for four-point and three-point correlation functions with the fermionic bubble loop. Here the vertices marked by \textcolor{lightgray2}{•} represent either time-ordered vertex (labelled by black $\bullet$) or anti-time-ordered vertex (labelled by white $\circ$). For generality, we allow two internal fermion propagators to carry different masses \textcolor{red3}{$m_1$} and \textcolor{blue3}{$m_2$}, respectively. \label{Fig: SK_diagrams}}
\end{figure}
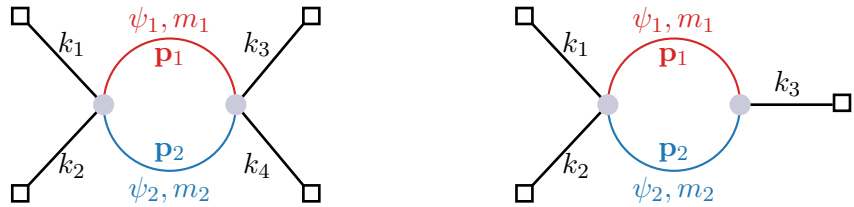
As shown in Figure \ref{Fig: SK_diagrams}, the central objects of this work are the four-point and three-point cosmological correlators involving a fermionic bubble loop. We will adopt the conventional Schwinger-Keldysh (SK) formalism \cite{Schwinger:1960qe,Feynman:1963fq,Keldysh:1964ud,Weinberg:2005vy} to carry out our calculations. A detailed discussion of the associated diagrammatic rules can be found in \cite{Chen:2017ryl}. Within the SK framework, each interaction vertex carries two possible contour labels, denoted by ``+" for the forward (time-ordered) branch and ``--" for the backward (anti-time-ordered) branch of the closed time path.

For generality, we work with the so-called seed integral, which can be defined from different perspectives. One approach is to fix the power of $\tau_i$ at each vertex, thereby specifying the coupling type, from which other cases can be obtained by acting with appropriate weight-shifting operators. In this work, we keep the fully general form, introducing parameters $q_i$ at each vertex to encode arbitrary couplings. Although this comes at the cost of heavier notation, it serves as a \textit{dictionary} that can be directly used for future phenomenological studies of fermions with different couplings. We thus define the seed integral as follows
\begin{keyeqn}
\begin{align}
\mathcal{I}_{\mathrm{\sa\sb}}^{q_1q_2}&\equiv -\sa\sb\,{s^{5+q_{12}}}\int_{-\infty}^0 \mathrm{~d} \tau_1 \mathrm{~d} \tau_2  \left(-\tau_1\right)^{q_1}\left(-\tau_2\right)^{q_2} e^{i\sa k_{12} \tau_1+i\sb k_{34} \tau_2}\times\mathcal{Q}_{\sa\sb}(s,\tau_1,\tau_2)\,.\label{eq:SeedIntegral}
\end{align}
\end{keyeqn}
Here $\sa,\sb=\pm$ denote the SK indices associated with the two vertices, and $q_1,q_2$ are constants, typically integers, that specify the couplings between the inflaton fluctuation $\varphi$ and the fermionic fields $\psi_i$. The external momenta are $k_i\equiv|\mathbf{k}_i|$, with the shorthand $k_{ij}\equiv k_i+k_j$. The internal momentum is $\mathbf{s}\equiv\mathbf{k}_1+\mathbf{k}_2$, and the prefactor $s^{5+q_{12}}$ is included to make the seed integral scale invariant. Finally, $\mathcal{Q}_{\sa\sb}$ denotes the fermionic bubble function, which encodes all fermionic dependence and will be discussed in detail later.\footnote{In this work, we focus on fermionic loops formed by pairs of the type $(\psi\psi+\psi^\dagger\psi^\dagger)$. Other cases, such as those with Pauli matrices inserted, can be treated in an analogous manner.}

With the seed integral defined in (\ref{eq:SeedIntegral}), correlators involving fermionic bubble loops can be obtained by choosing appropriate combinations of power indices $q_i$. As a concrete example, consider the interactions between the scalar inflaton $\varphi$ and fermions,
\begin{align}
\mathcal{L}_{\rm{int,3}}=\frac{a^3}{\Lambda_1}\varphi'\left(\psi_1\psi_2+\psi_2^\dagger\psi_1^\dagger\right)\,,\qquad\qquad \mathcal{L}_{\rm{int,4}}=\frac{a^{2}}{\Lambda_2^3}\varphi'{}^2\left(\psi_1\psi_2+\psi_2^\dagger\psi_1^\dagger\right)\,,\label{eq: example_interaction}
\end{align}
which are dimension-5 and dimension-7 operators, respectively. Here $\psi_1$ and $\psi_2$ are two Weyl fermions that may carry different masses, related to the rescaled field $\tilde{\psi}$ via $\psi=a^{-3/2}\tilde{\psi}$, whose action is given in (\ref{eq: Weyl_action}). The three-point  correlators with fermionic loop then can be expressed as 
\begin{align}
 \left\langle\varphi_{\mathbf{k}_1} \varphi_{\mathbf{k}_{2}} \varphi_{\mathbf{k}_{3}}\right\rangle^{\prime}&=-\frac{1}{4\Lambda_1\Lambda_2^3}\frac{1}{k_1k_2k_3^{4}}\times\lim _{k_4 \rightarrow 0}\sum_{\sa, \sb= \pm}\mathcal{I}_{\sa\sb}^{0,-2}+2 \text { perms}\,,
\end{align}
here two permutations account for the contributions from the other $t$- and $u$-channel.

The master integral $\mathcal{I}_{\sa\sb}$ to be evaluated involves both nested time integrals and loop momentum integrals, with an integrand given by a product of four Hankel functions. Because of its complexity, obtaining a fully analytical result is far from straightforward. In the next section, we aim to derive a complete analytical expression for the cosmological collider signal, capturing the full non-analytic structure of these fermionic loop cosmological correlators by employing recently developed analytical techniques.
\subsubsection*{Fermionic propagators}
To compute the fermionic bubble, one needs the set of fermionic propagators. While SK propagators for massive scalar fields are widely used and well documented in the literature, their fermionic counterparts are less commonly discussed. For the convenience of readers who may be unfamiliar with the relevant notation, we provide a brief summary below. A similar treatment of fermionic propagators can also be found in the Appendix of \cite{Chen:2018xck}.

There are different propagators depending on both the spinor indices and the SK indices. Let us start from the $\langle\psi\psi\rangle$, which is 
\begin{align}
    D_{-+,\alpha\beta}(\mathbf{k};\tau_1,\tau_2)&\equiv\int \mathrm{d}^3 \bfx\langle\psi_\alpha(\bfx,\tau_1)\psi_\beta(0,\tau_2)\rangle e^{-i \bfk \cdot \bfx}\nonumber\\
    &=\left(\tau_1\tau_2\right)^\frac{3}{2}\sum_{\lambda=\pm}u_{\lambda}(k,\tau_1)v^*_{\lambda}(k,\tau_2) h^{\lambda}_{\alpha}(\hat{\bfk})h^{\lambda\dagger}_{\dot{\gamma}}(\hat{\bfk})\bar{\sigma}^{0\dot{\gamma}\delta}\epsilon_{\beta\delta}~,\label{eq:PropagatorDmp1}
\end{align}
where in the final step the spinor indices have been made explicit, and the overall time factor arises from the rescaling $\tilde{\psi}\equiv a^{3/2}\psi$. Similarly, the other propagator is defined as follows
\begin{align}
    D_{+-,\alpha\beta}(\mathbf{k};\tau_1,\tau_2)&\equiv-\int \mathrm{d}^3 x\langle\psi_\beta(0,\tau_2)\psi_\alpha(\bfx,\tau_1)\rangle e^{-i \bfk \cdot \bfx}\nonumber\\
    &=-\left(\tau_1\tau_2\right)^{\frac{3}{2}}\sum_{\lambda=\pm}u_{\lambda}(k,\tau_2)v^*_{\lambda}(k,\tau_1) h^{\lambda}_{\beta}(-\hat{\bfk})h^{\lambda\dagger}_{\dot{\gamma}}(-\hat{\bfk})\bar{\sigma}^{0\dot{\gamma}\delta}\epsilon_{\alpha\delta}\,,
\end{align}
where the additional minus sign accounts for the anti-commuting nature of fermionic fields. These two propagators are related to each other by
\begin{align}
    D_{-+,\alpha\beta}(\bfk;\tau_1,\tau_2)=-D_{+-,\beta\alpha}(-\bfk,\tau_2,\tau_1)\,,\label{relation_-+_+-}~.
\end{align}
The time-ordered and anti-time-ordered propagators are then constructed as
\begin{align}
    &D_{++,\alpha\beta}(\bfk,\tau_1,\tau_2)=D_{-+,\alpha\beta}(\bfk,\tau_1,\tau_2)\theta(\tau_1-\tau_2)+D_{+-,\alpha\beta}(\bfk,\tau_1,\tau_2)\theta(\tau_2-\tau_1)\,,\\
    &D_{--,\alpha\beta}(\bfk,\tau_1,\tau_2)=D_{+-,\alpha\beta}(\bfk,\tau_1,\tau_2)\theta(\tau_1-\tau_2)+D_{-+,\alpha\beta}(\bfk,\tau_1,\tau_2)\theta(\tau_2-\tau_1)\,.
\end{align}
Now let us move to the other type $\langle\psi\psi^\dagger\rangle$, similarly,
\begin{align}
    D_{-+,\alpha}^{~~~~~~\dot{\beta}}(\mathbf{k},\tau_1,\tau_2)&\equiv\int \mathrm{d}^3 x\langle\psi_\alpha(\bfx,\tau_1)\psi^{\dagger\dot{\beta}}(0,\tau_2)\rangle e^{-i \bfk \cdot \bfx}\nonumber\\
    &=\left(\tau_1\tau_2\right)^{\frac{3}{2}}\sum_{s=\pm} u_{s}(k,\tau_1)u^*_{s}(k,\tau_2)h^{s}_{\alpha}(\hat{\bfk})\epsilon^{\dot{\beta}\dot{\gamma}}h^{s\dagger}_{\dot{\gamma}}(\hat{\bfk})~,\label{eq:PropagatorDmp2}\\
     D_{+-,\alpha}^{~~~~~~\dot{\beta}}(\mathbf{k},\tau_1,\tau_2)&\equiv-\int \mathrm{d}^3 x\langle\psi^{\dagger\dot{\beta}}(0,\tau_2)\psi_\alpha(\bfx,\tau_1)\rangle e^{-i \bfk \cdot \bfx}\nonumber\\
    &=-\left(\tau_1\tau_2\right)^{\frac{3}{2}}\sum_{s=\pm} v_{s}(k,\tau_2)v^*_{s}(k,\tau_1)\bar{\sigma}^{0\dot{\beta}\gamma}h^{s}_{\gamma}(-\hat{\bfk})h^{s\dagger}_{\dot{\gamma}}(-\hat{\bfk})\bar{\sigma}^{0\dot{\gamma}\delta}\epsilon_{\alpha\delta}~,
\end{align}
as for the corresponding time-ordered and anti-time-ordered propagators, they are given by
\begin{align}
    &D_{++,\alpha}^{~~~~~~\dot{\beta}}(\bfk,\tau_1,\tau_2)=D_{-+,\alpha}^{~~~~~~\dot{\beta}}(\bfk,\tau_1,\tau_2)\theta(\tau_1-\tau_2)+D_{+-,\alpha}^{~~~~~~\dot{\beta}}(\bfk,\tau_1,\tau_2)\theta(\tau_2-\tau_1)\,,\\
    &D_{--,\alpha}^{~~~~~~\dot{\beta}}(\bfk,\tau_1,\tau_2)=D_{+-,\alpha}^{~~~~~~\dot{\beta}}(\bfk,\tau_1,\tau_2)\theta(\tau_1-\tau_2)+D_{-+,\alpha}^{~~~~~~\dot{\beta}}(\bfk,\tau_1,\tau_2)\theta(\tau_2-\tau_1)\,.
\end{align}
Other types such as $\langle\psi^\dagger\psi^\dagger
\rangle$ or $\langle\psi^\dagger\psi
\rangle$ can be obtained by taking the Hermitian conjugate and properly reversing the direction of the momentum.
\subsubsection*{Fermionic bubble function}
With the fermionic propagators in hand, we are now ready to find the fermionic  bubble function, which appears in the computation of cosmological correlators involving the two-vertex fermionic  loop, as demonstrated in Figure \ref{Fig: SK_diagrams}. Explicitly, we define
\begin{align}
    \mathcal{Q}_{-+}\equiv\int \mathrm{d}^3 x\,e^{-i \mathbf{s}\cdot \mathbf{x}}\langle\mathcal{O}(x_1)\mathcal{O}(x_2)\rangle\,, \label{Q_-+}
\end{align}
here $\mathbf{x}\equiv \mathbf{x}_1-\mathbf{x}_2$, and for brevity, we also introduce the shorthand notation $x_i\equiv(\tau_i,\bfx_i)$. The scalar operator is made by a pair of fermions as $\mathcal{O}(x_i)=\psi_1(x_i) \psi_2(x_i)+\psi_2^\dagger(x_i) \psi_1^\dagger(x_i)$. Without loss of generality, we allow two Weyl fermions $\psi_1$ and $\psi_2$ to have different masses, denoted by $m_1$ and $m_2$ respectively. Then after the contraction, it reads 
\begin{align}
    &\mathcal{Q}_{-+}(s,\tau_1,\tau_2)\nonumber\\
    &=-\int \mathrm{d}^3 x\,e^{-i \mathbf{s}\cdot \mathbf{x}}\left[\bigl\langle\psi_1^\alpha(x_1)\psi_1^\beta(x_2)\bigr\rangle\bigl\langle\psi_{2\alpha}(x_1)\psi_{2\beta}(x_2)\bigr\rangle+\bigl\langle\psi_1^\alpha(x_1)\psi^\dagger_{1\dot{\beta}}(x_2)\bigr\rangle\bigl\langle\psi_{2\alpha}(x_1)\psi_{2}^{\dagger\dot{\beta}}(x_2)\bigr\rangle\right.\nonumber\\
    &\,\qquad\qquad\qquad\qquad+\big(\text{h.c.~and}~x_1\leftrightarrow x_2\big)\Big]\,,\label{eq: BubblefunctionQmp}
\end{align}
here the overall minus sign arises from exchanging the order of anti-commuting spinors, a feature already well known from fermion loops in flat spacetime. The last line is equivalent to taking the Hermitian conjugate of the first line, along with exchanging the variables $x_1$ and $x_2$. Since the spinor structure is not immediately transparent, we provide additional intermediate steps below. For example, substituting the propagator (\ref{eq:PropagatorDmp1}) into the first term of (\ref{eq: BubblefunctionQmp})
\begin{align}
    &\int \mathrm{d}^3 x\,e^{-i \mathbf{s}\cdot \mathbf{x}}\bigl\langle\psi_1^\alpha(x_1)\psi_1^\beta(x_2)\bigr\rangle\bigl\langle\psi_{2\alpha}(x_1)\psi_{2\beta}(x_2)\bigr\rangle\nonumber\\
    &=\left(\tau_1\tau_2\right)^3\sum_{\lambda,\lambda'=\pm}\int\frac{\mathrm{d}^3 \mathbf{p}_1}{(2\pi)^3}\,u_{1\lambda}(p_1,\tau_1)v^*_{1\lambda}(p_1,\tau_2)u_{2\lambda'}(p_2,\tau_1)v^*_{2\lambda'}(p_2,\tau_2)\times\mathcal{F}(\hat{p_1},\hat{p}_2,\lambda,\lambda')\,, \label{eq:Bubble_term1}
\end{align}
where $\mathbf{p}_2\equiv \mathbf{s}-\mathbf{p}_1$, and the spinor kinematic function is defined as 
\begin{align}
    \mathcal{F}(\hat{p_1},\hat{p}_2,\lambda,\lambda')&\equiv\left[\epsilon^{\alpha\delta}h^{\lambda}_{\delta}(\hat{p}_1)h^{\lambda'}_{\alpha}(\hat{p}_2)\right]\left[h^{\lambda\dagger}_{\dot{{\gamma}}}(\hat{p}_1)\bar{\sigma}^{0\dot{\gamma}\beta}h^{\dagger\lambda'}_{\dot{\delta}}(\hat{p}_2)\bar{\sigma}^{0\dot{\delta}\eta}\epsilon_{\beta\eta}\right]\nonumber\\
    &=\frac{1}{4}\Tr\Big[\left(\sigma^0+\lambda\,\hat{p}_1\cdot \vec{\sigma}\right) \left(\sigma^0-\lambda'\,\hat{p}_2\cdot \vec{\sigma}\right)\Big]\nonumber\\
    &=\frac{1}{2}\left(1-\lambda\lambda' \hat{p}_1\cdot \hat{p}_2\right)\,,
\end{align}
in the second line we first apply the identity (\ref{eq: Product_Helicity}) to replace the helicity bases in terms of Pauli matrices, and  then we used $\epsilon^{\a\b}=-i\sigma^2$ as well as the properties of Pauli matrix  i.e. $\sigma^2 (\sigma^i)^T \sigma^2=-\sigma^i$. In the final line, we make use of the trace formula for Pauli matrices that
$\Tr\left(\sigma_a \sigma_b\right)=2\delta_{ab}$. This kinematic factor $\mathcal{F}(\hat{p_1},\hat{p}_2,\lambda,\lambda')$ is symmetric with respect to internal momenta $\hat{p}_1$, $\hat{p}_2$ and helicity variables $\lambda,\lambda'$.

Similarly, using the other type of propagator (\ref{eq:PropagatorDmp2}), we can get 
\begin{align}
    &\int \mathrm{d}^3 x e^{-i \mathbf{s}\cdot \mathbf{x}}\bigl\langle\psi_1^\alpha(x_1)\psi^\dagger_{1\dot{\beta}}(x_2)\bigr\rangle\bigl\langle\psi_{2\alpha}(x_1)\psi_{2}^{\dagger\dot{\beta}}(x_2)\bigr\rangle\nonumber\\
    &=-\left(\tau_1\tau_2\right)^3\sum_{\lambda,\lambda'=\pm}\int\frac{\mathrm{d}^3 \mathbf{p}_1}{(2\pi)^3}\,u_{1\lambda}(p_1,\tau_1)u^*_{1\lambda}(p_1,\tau_2)u_{2\lambda'}(p_2,\tau_1)u^*_{2\lambda'}(p_2,\tau_2)\times\mathcal{F}(\hat{p_1},\hat{p}_2,\lambda,\lambda')\,. \label{eq:Bubble_term2}
\end{align}
We note that both~(\ref{eq:Bubble_term1}) and~(\ref{eq:Bubble_term2}) are invariant under taking the Hermitian conjugate while simultaneously exchanging $\tau_1\leftrightarrow\tau_2$ and $\mathbf{p}_1\leftrightarrow\mathbf{p}_2$, once the sum over helicities is performed. This follows from the relation $u_{\pm}=v_{\mp}$ satisfied by the mode functions, and the fact that $\lambda$ and $\lambda'$ are dummy variables. Adding both contributions, the bubble function is given by
\begin{align}
   \mathcal{Q}_{-+}(s,\tau_1,\tau_2)=2\left(\tau_1\tau_2\right)^3\sum_{\lambda,\lambda'=\pm}\int\frac{\mathrm{d}^3\mathbf{p}_1}{(2\pi)^3}\,&\mathcal{F}(\hat{p_1},\hat{p}_2,\lambda,\lambda') \times u_{1\lambda}(p_1,\tau_1) u_{2\lambda'}(p_2,\tau_1)\nonumber\\
   &\times\Big[u^*_{1\lambda}(p_1,\tau_2)u^*_{2\lambda'}(p_2,\tau_2)-v^*_{1\lambda}(p_1,\tau_2)v^*_{2\lambda'}(p_2,\tau_2)\Big]\,.
\end{align}
Finally, by plugging into the mode functions (\ref{eq:Fermion_Modefunction}) and applying the transformation between Whittaker-$W$ and Hankel function (\ref{eq: WhittakerToHankel}), the explicit form of the loop bubble function in the momentum space reads
\begin{keyeqn}
\begin{align}
    &\mathcal{Q}_{-+}(s,\tau_1,\tau_2)\nonumber\\
    &= \,\frac{\pi^2}{8}(\tau_1 \tau_2)^4 \sum_{\lambda=\pm} \int\frac{\mathrm{d}^3\mathbf{p}_1}{(2\pi)^3}  \Bigg[ p_1 p_2\,\,e^{(\nu_{1\lambda}-\nu_{2\lambda})i\pi}H^{(1)}_{\nu_{1\lambda}}(-p_1\tau_1) H^{(2)}_{\nu^*_{1\lambda}}(-p_1\tau_2) H^{(1)}_{\nu^*_{2\lambda}}(-p_2\tau_1) H^{(2)}_{\nu_{2\lambda}}(-p_2\tau_2)\nonumber\\
    &\qquad\qquad\qquad\qquad- (\mathbf{p}_1 \cdot \mathbf{p}_2)\,H^{(1)}_{\nu^*_{1\lambda}}(-p_1\tau_1) H^{(2)}_{\nu^*_{1\lambda}}(-p_1\tau_2) H^{(1)}_{\nu_{2\lambda}}(-p_2\tau_1) H^{(2)}_{\nu_{2\lambda}}(-p_2\tau_2)\Bigg]\,, \label{eq:BubbleFunction_MomentumSpace}
\end{align}
\end{keyeqn}
where the parameters $\nu_i$ are  complex numbers and are related to the dimensionless mass through
\begin{align}
    \nu_{i\lambda}=\frac{1}{2}-i\,\lambda\,m_i\,,
\end{align}
so the flipping of helicity is equivalent to taking the complex conjugate, i.e. $\nu^*_{i\pm}\equiv\nu_{i\mp}$\,. This is quite different from the scalar case, in which the parameter  of the Hankel function is either purely real (in the complementary series) or purely imaginary (in the principal series). The bubble function is a scalar quantity and parity-even as expected, then the other type time factorised bubble $\mathcal{Q}_{+-}(s,\tau_1,\tau_2)$ can be directly found through the relation
\begin{align}
    \mathcal{Q}_{+-}(s,\tau_1,\tau_2)= \Big[\mathcal{Q}_{-+}(s,\tau_1,\tau_2)\Big]^*=\mathcal{Q}_{-+}(s,\tau_2,\tau_1)\,,
\end{align}
as for the time-ordered and anti-time-ordered, it follows the usual relation as 
\begin{align}
    &\mathcal{Q}_{++}(s,\tau_1,\tau_2)=\mathcal{Q}_{-+}(s,\tau_1,\tau_2)\theta(\tau_1-\tau_2)+\mathcal{Q}_{+-}(s,\tau_1,\tau_2)\theta(\tau_2-\tau_1)\,,\\
    &\mathcal{Q}_{--}(s,\tau_1,\tau_2)=\mathcal{Q}_{+-}(s,\tau_1,\tau_2)\theta(\tau_1-\tau_2)+\mathcal{Q}_{-+}(s,\tau_1,\tau_2)\theta(\tau_2-\tau_1)\,.
\end{align}
In previous studies~\cite{Chen:2018xck, Cui:2021iie}, the fermionic bubble function was presented in position space. To show equivalence with our result \eqref{eq:BubbleFunction_MomentumSpace}, one needs to perform a Fourier transform which  is very non-trivial, due to the presence of special functions. We carry out this calculation in detail in Appendix~\ref{App: derivations}, verifying that our expression indeed reproduces the previous results.
\paragraph{Discussion.}
Before moving to the next section, let us make a few remarks concerning this fermionic bubble loop. Although the internal exchanges we consider involve fermions with spinor structures, the bubble loop itself as a combination of fermions is essentially a scalar quantity.  This in turn raises the following natural questions:
\begin{itemize}
    \item \textit{The relation between scalar and fermionic bubbles}: can we establish some relation between this fermionic bubble and the corresponding scalar bubble? In fact, the fermionic propagators can be expressed as derivatives acting on scalar propagators, with the appropriate insertion of gamma matrices $\gamma^a$\cite{Ema:2023dxm}. Using several identities for Hankel functions, the fermionic bubble can  be written as
    \begin{align}
    &\mathcal{Q}_{-+}(s,\tau_1,\tau_2)= \, \sum_{\lambda=\pm}\int\frac{\mathrm{d}^3\mathbf{p}_1}{(2\pi)^3}\bigg\{(\tau_1 \tau_2)^4\,\mathcal{D}_{\lambda,\tau_1}\mathcal{D}_{\lambda,\tau_2}\left[\left(\tau_1\tau_2\right)^{-3}\mathcal{G}_{\nu_{1\lambda}}(p_1,\tau_1,\tau_2)\mathcal{G}_{\nu_{2\lambda}}(p_2,\tau_1,\tau_2)\right]\nonumber\\
    &\qquad\qquad\qquad\qquad\qquad-(s^2\tau_1\tau_2)\,\mathcal{G}_{\nu^*_{1\lambda}}(p_1,\tau_1,\tau_2)\mathcal{G}_{\nu_{2\lambda}}(p_2,\tau_1,\tau_2)\bigg\}\,, \label{eq:Q_GG}
    \end{align}
    where we have defined the time-operator as
    \begin{align}
    \mathcal{D}_{\lambda,\tau}\equiv\partial_{\tau}+\frac{1-i\lambda (m_1+m_2)}{\tau}\,, \label{def: Doperator}
    \end{align}
    and the scalar two-point function (Wightman function) is 
    \begin{align}
    \mathcal{G}_{\nu}(k,\tau_1,\tau_2)\equiv\langle 0| \sigma_{{k}}(\tau_1)\sigma_{{-k}}(\tau_2)| 0\rangle'=\frac{\pi}{4} (\tau_1\tau_2)^{3/2} H^{(1)}_{\nu}(-k\tau_1) H^{(2)}_{\nu^*}(-k\tau_2)\,. \label{eq:scalar_wightman}
    \end{align}
    The product of two scalar propagators in~\eqref{eq:Q_GG} is nothing but the scalar bubble function, defined as
    \begin{align}
        \mathcal{Q}^{\text{scalar}}_{-+}(s,\tau_1,\tau_2; \mu_1,\mu_2)\equiv\int\frac{\mathrm{d}^3\mathbf{p}_1}{(2\pi)^3}\mathcal{G}_{\mu_1}(p_1,\tau_1,\tau_2)\,\mathcal{G}_{\mu_2}(p_2,\tau_1,\tau_2)\,. \label{eq: scalar_momentum}
    \end{align}
    We thus find that the fermionic bubble can indeed be expressed as some differential operators acting on the scalar bubble:
    \begin{keyeqn}
    \begin{align}
    &\mathcal{Q}_{-+}(s,\tau_1,\tau_2)= \, \sum_{\lambda=\pm}\bigg\{(\tau_1 \tau_2)^4\,\mathcal{D}_{\lambda,\tau_1}\mathcal{D}_{\lambda,\tau_2}\left[\left(\tau_1\tau_2\right)^{-3}\mathcal{Q}^{\text{scalar}}_{-+}(s,\tau_1,\tau_2; \nu_{1\lambda},\nu_{2\lambda})\right]\nonumber\\
    &\qquad\qquad\qquad\qquad\qquad-(s^2\tau_1\tau_2)\,\mathcal{Q}^{\text{scalar}}_{-+}(s,\tau_1,\tau_2; \nu^*_{1\lambda},\nu_{2\lambda})\bigg\}\,, \label{eq: fermion_scalar_relation}
    \end{align}
    \end{keyeqn}
    the detailed derivation of the relations can be found in Appendix \ref{app: relations}. Since the scalar bubble has already been evaluated via spectral decomposition in~\cite{Xianyu:2022jwk}, one approach is to directly apply these operators to the known scalar results to obtain the fermionic bubble, and we will return to this point in the following sections.
    
    \item \textit{Distinguish the fermionic bubble from the scalar \footnote{We thank Xi Tong for related discussions.}}: The bubble function~(\ref{eq:BubbleFunction_MomentumSpace}) is itself a scalar quantity, so how can one distinguish the signatures generated by fermions from those of scalars?  There are several possible approaches to consider. First, the \textit{dynamics} of fermions differ from those of scalars. As a result, the mass parameters in the mode function (the Hankel type) are complex, unlike the purely real or purely imaginary cases for scalars. Another crucial distinction lies in the \textit{statistics}, as the anti-commuting nature of spinors introduces an additional overall minus sign in the fermionic loop, and the particle production rates also differ. A comprehensive discussion of these distinguishing features is beyond the scope of this work and we leave it for the future.
\end{itemize}

\section{Correlators with fermionic bubble loop}

In this section, we present  detailed computations of the fermionic bubble loop contributions to cosmological correlators, with particular emphasis on the non-analytic terms associated with the cosmological collider signals. We will follow two different paths, each with distinct advantages.
$(i)$ The first method is akin to the spectral decomposition and leads to a compact final expression, but it crucially relies on dS symmetry. $(ii)$ The second method is based on the MB representation. While this approach is typically more involved and often leads to nested series, it has the benefit of being more readily generalisable to a broader range of setups, including cases where full de Sitter symmetry is absent.\footnote{More precisely, we are referring to cases where dS symmetry is broken in a non-trivial way. This occurs, for example, when a chemical potential is introduced in the quadratic dispersion relation, induced by the inflaton's rolling background. If the breaking instead comes from interactions between the inflaton and massive sectors, the first method still applies because the bubble function remains dS invariant and the interactions can be treated perturbatively. Likewise, if all fields in the loop share the same sound speed, the first method still works, as the effect of the sound speed can be absorbed by a rescaling of the time variables. }
\subsection{Signals from the bubble loop} \label{sec: signals_from_bubble}
Before moving to the technical details, let us first classify the bubble loop structures. These can be categorised into three distinct structures based on their kinematic dependence, which we label as  non-local signals $\mathcal{I}_{\text{NS}}$, local signals $\mathcal{I}_{\text{LS}}$, and background $\mathcal{I}_{\text{BG}}$:
\begin{equation}
\mathcal I^{q_1q_2}(r_1,r_2) = \mathcal I_{\rm{NS}}^{q_1q_2}(r_1,r_2) + \mathcal I_{\rm{LS}}^{q_1q_2}(r_1,r_2) +  \mathcal I_{\rm{BG}}^{q_1q_2}(r_1,r_2)\,,
\end{equation}
where we have introduced the dimensionless momentum ratios as
\begin{align}
r_1 \equiv \frac{s}{k_{12}}\,,\qquad r_2 \equiv \frac{s}{k_{34}}\,.
\end{align}
This classification follows from the analyticity analysis of the loop integral discussed in \cite{Qin:2023bjk}. Specifically, the non-local signals are non-analytic with respect to the internal momentum, defined in our case as ${\bf{s}}\equiv{\bfk}_1+{\bfk}_2$ and $s\equiv |\bf{s}|$. In contrast, the local signals are analytic in $s$, but are non-analytic with respect to the external variables $k_{12}$ and $k_{34}$. The remaining background piece is a meromorphic function of these momentum ratios. In this section, we will mainly focus on the signals part, namely $\mathcal{I}_{\text{NS}}$ and $\mathcal{I}_{\text{LS}}$. 

To extract the signal parts only, one can use a shortcut known as the CC cutting rules \cite{Tong:2021wai, Qin:2023bjk,Ema:2024hkj}. For instance, in the kinematic region $k_{12}>k_{34}$ or $r_1<r_2$, we can use the identity $\theta(\tau_1 - \tau_2) = 1 - \theta(\tau_2 - \tau_1)$ to decompose the (anti-)time-ordering bubble function as:
\begin{align}
\mathcal{Q}_{\pm\pm}(s,\tau_1,\tau_2)
= \mathcal{Q}_{\mp\pm}(s,\tau_1,\tau_2)
+ \theta(\tau_2-\tau_1)\Big[
\mathcal{Q}_{\pm\mp}(s,\tau_1,\tau_2)
- \mathcal{Q}_{\mp\pm}(s,\tau_1,\tau_2)
\Big]\,,
\end{align}
accordingly, the time integrals are split into two parts, 
\begin{align}
&\mathcal{I}_{\pm\pm, {\rm F}}^{q_1 q_2}
\equiv  -{s^{5 + q_{12}}}
\int_{-\infty}^0 \mathrm{d} \tau_1 \mathrm{d} \tau_2\,
(-\tau_1)^{q_1} (-\tau_2)^{q_2}
e^{\pm i k_{12} \tau_1 \pm i k_{34} \tau_2}
\mathcal{Q}_{\mp\pm}(s,\tau_1,\tau_2),\\
&\mathcal{I}_{\pm\pm, {\rm TO}}^{q_1 q_2}
\equiv - {s^{5 + q_{12}}}
\int_{-\infty}^0 \mathrm{d} \tau_1 \mathrm{d} \tau_2\,
(-\tau_1)^{q_1} (-\tau_2)^{q_2}
e^{\pm i k_{12} \tau_1 \pm i k_{34} \tau_2}
\theta(\tau_2-\tau_1)\Big[
\mathcal{Q}_{\pm\mp}
- \mathcal{Q}_{\mp\pm}
\Big].\label{TO}
\end{align}
By combining the time-factorised component with the opposite-sign integrals $\mathcal{I}_{\pm \mp}^{q_1 q_2}$, we define the total factorised contribution to the seed function as 
\begin{align}
\mathcal{I}_{\text{F}}^{q_1q_2}&\equiv \mathcal{I}^{q_1q_2}_{-+}+\mathcal{I}^{q_1q_2}_{++,\text{F}}+{\rm{c.c.}}\,.\label{I_F} 
\end{align}
From the cutting rules, the nested-time piece~\eqref{TO} contains no cosmological collider signals and contributes only to the background. When discussing the signals, we may therefore restrict attention to the factorised piece~\eqref{I_F}. Strictly speaking, however, the factorised contribution~\eqref{I_F} also contains a portion of the background, as will become clear from the explicit calculations below.

Another appealing feature of the fully factorised contribution is that it is better behaved in the UV than the nested-time piece \eqref{TO}. In fact, it is \textit{UV finite}\footnote{A naive way to see this is that if the signal part were UV divergent, no local tree-level counterterms could cancel that divergence. Moreover, the background contributions from the factorised piece contain no total-energy poles, so they cannot be removed by contact terms either.}, and can therefore be evaluated without the subtleties associated with regularisation and renormalisation. We now turn to the analytical calculation of the bubble.

\subsection{Spectral decomposition} \label{sec: spectral}
The first method for evaluating the bubble loop integral is based on a spectral decomposition. The key step is to obtain the spectral density and then pick up the relevant poles in the complex mass plane. This approach was first applied to the scalar bubble \cite{Xianyu:2022jwk} and has recently been extended to the vector bubble case \cite{Zhang:2025nzd}, with other applications such as bubble chain resummation \cite{Grafe:2026qsm}. In this section, rather than working directly with the spectral density, we adopt a slightly different but essentially equivalent strategy. It is more straightforward and readily generalises to the fermion case, as it can be carried out without first knowing the spectral density. Specifically, we will compute the loop integrals using several useful identities established in earlier works on spectral representations \cite{ProductofHyper,Bros:2011vh}
\begin{keyeqn}
    \begin{align}
    \mathcal{D}(x,\eta,u)\cdot\mathcal{D}(y,\eta,u)=\sum_{k=0}^{\infty}4^{-k} f_k (x,y,\eta) u^k\,\mathcal{D}(x+y+2k,\eta,u)\,, \label{eq: product_2F1}
\end{align}
\end{keyeqn}
here  the function $f_k$ is defined as 
\begin{align}
    f_k(x,y,\eta)=\frac{(x)_k(y)_k (x+y+2\eta-1+k)_k (1-\eta)_k}{(x+\eta)_k(y+\eta)_k(x+y+\eta-1+k)_k k!}\,.
\end{align}
and for convenience, we introduce the shorthand notation for the frequently used hypergeometric function 
\begin{align}
    \mathcal{D}(x,\eta,u)= \pFq{2}{1}{\frac{x}{2}, \frac{x+1}{2}}{x+\eta}{u}\,,
\end{align}
where the important requirement is that the first and second parameters of this hypergeometric function differ by {\bf{one half}}.

\subsubsection{Scalar decomposition}
Let us first revisit the result of the scalar bubble \cite{Xianyu:2022jwk} using the useful identity (\ref{eq: product_2F1}), and we will show the calculation step by step in this subsection. To begin with, the scalar bubble in position space can be expressed as 
\begin{align}
    \mathcal{Q}_{\text{scalar}}(x_1,x_2; i\mu_1,i\mu_2)&=\mathcal{G}_{i\mu_1}(x_1,x_2)\times\mathcal{G}_{i\mu_2}(x_1,x_2)\,,
\end{align}
which is actually the Fourier transform of \eqref{eq: scalar_momentum}. The propagator in the position space reads 
\begin{align}
    \mathcal{G}_{i\mu}(x_1,x_2)=\frac{\Gamma\left(\frac{3}{2}-i\mu\right)\Gamma\left(\frac{3}{2}+i\mu\right)}{16\pi^2}\,\pFq{2}{1}{\frac{3}{2}-i\mu, \frac{3}{2}+i\mu}{2}{\frac{1+Z}{2}}\,,
\end{align}
here we consider the case where the two propagators have different masses and focus on the $\mathcal{G}_{-+}$ type, omitting the SK indices for brevity. The other types can be treated using the same method. $Z$ is the dimensionless embedding distance (\ref{eq: embedding_distance}), and the detailed derivation from the Fourier transform of the mode function in momentum space is provided in Appendix.~\ref{App: Fourier_Trans}. 
At this point, the scalar bubble is expressed as a product of hypergeometric functions, but not yet in the form required by \eqref{eq: product_2F1}. To apply this identity, we first use the transformation (\ref{eq:hyperTrans}) to recast each hypergeometric function into the desired form, after which the scalar bubble can be rewritten as
\begin{align}
\mathcal{Q}_{\text{scalar}}(x_1,x_2;i\mu_1,i\mu_2)=-\prod_{n=1}^2\left[\sum_{\sa=\pm}\frac{(-2Z)^{-i\sa\mu_n-\frac{3}{2}}}{4\pi^{{3}/{2}}}\frac{\Gamma\left(\frac{3}{2}+i\sa\mu_n\right)}{\sinh(\sa\pi\mu_n) }\frac{\mathcal{D}\left(\frac{3}{2}+i\sa\mu_n,-\frac{1}{2},\frac{1}{Z^2}\right)}{\Gamma\left(1+i\sa\mu_n\right)}\right]\,.
\end{align}
The next step is then to apply the product formula \eqref{eq: product_2F1} to expand the bubble function as a series
\begin{align}
    \mathcal{Q}_{\text{scalar}}&=-\sum_{k=0}^{\infty}\sum_{\sa,\sb=\pm}\frac{(-2Z)^{-3}}{16\pi^3}\,\frac{(-2Z)^{-i\sa\mu_1-k}}{\sinh(\sa\pi\mu_1)}\,\frac{(-2Z)^{-i\sb\mu_2-k}}{\sinh(\sb\pi\mu_2)}\times\Gamma \left[\bgm  3/2+i\sa\mu_1, 3/2+i\sb\mu_2\\ 1+i\sa\mu_1\,,1+i\sb\mu_2\edm\right]\nonumber\\
    &\times\,{f_k\left(\frac{3}{2}+ i\sa\mu_1,\frac{3}{2}+i\sb\mu_2,-\frac{1}{2}\right)}\,\,{\mathcal{D}\left(2k+3+i\sa\mu_1+i\sb\mu_2,-\frac{1}{2},\frac{1}{Z^2}\right)}\,,
\end{align}
and we have numerically checked the convergence of the above series.

We now express the product of propagators in terms of a single hypergeometric function, written as an infinite series. In this way, the scalar bubble appears as a sum over an infinite sum of tree-level-type contributions. For later use, the result can also be reformulated in terms of the associate Legendre-$Q$ function using the identity: 
\begin{align}
        \mathcal{D}\left(x,\eta,\frac{1}{Z^2}\right)=-i\,\frac{2^{\eta}e^{i\pi\eta}}{\sqrt{2\pi}}\,\frac{(-2Z)^{x}}{(Z^2-1)^{\frac{1}{4}-\frac{\eta}{2}}}\,\frac{\Gamma(x+\eta)}{\Gamma(x)}\,Q_{-3/2+x+\eta}^{1/2-\eta}(-Z)\,,\label{eq: hypertoQ}
\end{align}
now the bubble function reads 
\begin{align}
  \mathcal{Q}_{\text{scalar}}=\sum_{k=0}^{\infty} \sum_{\sa,\sb=\pm}&\frac{\csch(\sa\pi\mu_1)\csch(\sb\pi\mu_2)}{32\pi^{{7}/{2}}\left(Z^2-1\right)^{1/2}}\times\Gamma \left[\bgm \frac{3}{2}+i\sa\mu_1,\frac{3}{2}+i\sb\mu_2,\frac{5}{2}+2k+i\sa\mu_1+i\sb\mu_2\\1+i\sa\mu_1,1+i\sb\mu_2,3+2k+i\sa\mu_1+i\sb\mu_2 \edm\right]\,\nonumber\\
  &\times f_k\left(\frac{3}{2}+i\sa\mu_1,\frac{3}{2}+i\sb\mu_2,-\frac{1}{2}\right)\,{Q_{1+2k+i\sa\mu_1+i\sb\mu_2}^1\left(-Z\right)}\,.
\end{align}
To make it more transparent, we can perform the Fourier transform using the same method as in Appendix \ref{App: Fourier_Trans}, by using the integration formula (\ref{intFormula2}). Then we get
\begin{align}
    \int \frac{\dd^3 k}{(2\pi)^3}e^{i \mathbf{k}\cdot \mathbf{x}}  J_{i\mu}(-k\tau_1) H^{(2)}_{-i\mu}(-k\tau_2)&= -\frac{i}{2\pi^3}\frac{1}{(\tau_1\tau_2)^{{3}/{2}}}\frac{Q_{i\mu-1/2}^{1}{(-Z)}}{(Z^2-1)^{{1}/{2}}}\,. \label{eq: JH_Fourier}
\end{align}
Finally, we obtain the desired decomposition-like expression, the scalar bubble function in momentum space can be written as an infinite sum of tree-level diagrams: 
\begin{keyeqn}
\begin{align}
    \mathcal{Q}_{\text{scalar}}(s,\tau_1,\tau_2;i\mu_1,i\mu_2)&=\sum_{k=0}^{\infty} \sum_{\sa,\sb=\pm}\rho(k,i\sa\mu_1,i\sb\mu_2)\times\widetilde{\mathcal{G}}_{2k+\frac{3}{2}+i\sa\mu_1+i\sb\mu_2}\left(s,\tau_1,\tau_2\right)\,,\label{eq:scalar_decomps}
\end{align}
\end{keyeqn}
where the new propagator $\widetilde{\mathcal{G}}_{i\mu}$ is defined as
\begin{align}
    \widetilde{\mathcal{G}}_{i\mu}(s,\tau_1,\tau_2)\equiv (\tau_1\tau_2)^{3/2} J_{i\mu}(-s\tau_1) H^{(2)}_{-i\mu}(-s\tau_2)\,,\label{eq:gtil}
\end{align}
and the density function reads 
\begin{align}
    &\rho(k,i\sa\mu_1,i\sb\mu_2)\nonumber\\
    &\equiv \frac{i}{16\sqrt{\pi}}\,\frac{f_k\left(\frac{3}{2}+i\sa\mu_1,\frac{3}{2}+i\sb\mu_2,-\frac{1}{2}\right)}{\sinh(\sa\pi\mu_1)\sinh(\sb\pi\mu_2)}\,\times\Gamma \left[\bgm \frac{3}{2}+i\sa\mu_1,\frac{3}{2}+i\sb\mu_2,\frac{5}{2}+2k+i\sa\mu_1+i\sb\mu_2\\1+i\sa\mu_1,1+i\sb\mu_2,3+2k+i\sa\mu_1+i\sb\mu_2 \edm\right]\,.
\end{align}
To summarise, in the scalar case, the relation above can be represented schematically as follows
\begin{align}
\raisebox{2.0 pt}{\begin{tikzpicture}[baseline={(current bounding box.center)}, line width=1. pt, scale=3.0]
    \draw[red3, thick] (0.5,0) arc[start angle=0, end angle=180, radius=0.25];
    \draw[blue3, thick] (0.5,0) arc[start angle=360, end angle=180, radius=0.25];
    \draw[draw=lightgray2, fill=lightgray2] (0, 0) circle (.035cm);
    \draw[draw=lightgray2, fill=lightgray2] (0.5, 0) circle (.035cm);
    \node at (0.25,0.32){\textcolor{red3}{$\mu_1$}};
    \node at (0.25,-0.32){\textcolor{blue3}{$\mu_2 $}};
    \node at (0.25,0){scalar};
\end{tikzpicture} }
~~=~~ \sum_{\mu_{k}}\, \rho(\mu_k)\,\times\,\bigg(~
\raisebox{1.0 pt}{\begin{tikzpicture}[baseline={(current bounding box.center)}, line width=1. pt, scale=3.0]
    \draw[black, thick] (0.0,0) -- (0.6,0);
    \draw[draw=lightgray2, fill=lightgray2] (0, 0) circle (.035cm);
    \draw[draw=lightgray2, fill=lightgray2] (0.6, 0) circle (.035cm);
    \node at (0.3,0.08){$\mu_k$};
  \node[scale=0.25, transform shape] at (0,-0.11) {$J$};
  \node[scale=0.25, transform shape] at (0.6,-0.11) {$H$};
\end{tikzpicture}}
~
\bigg)\,,
\end{align}
with the sum over $\mu_k=2k+3/2+i\sa\mu_1+i\sb\mu_2$. The final step in evaluating the corresponding diagram is to attach the external legs and carry out the standard time integrals, which are familiar and straightforward. We have checked the cosmological collider signals at the one loop level, including both local and non-local contributions, and find exact agreement \footnote{Up to the overall constant prefactor that may come from different conventions.} with previous work \cite{Xianyu:2022jwk} after taking the equal-mass limit $\mu_1=\mu_2$.
\subsubsection{Fermion decomposition}
After reproducing the scalar result as a warm-up, we now turn to the fermionic case. A straightforward approach is to start from the final scalar expression and apply the time-derivative operators as in \eqref{eq: fermion_scalar_relation}, which yields the corresponding fermionic result. However, this route requires additional efforts to reorganise the different terms into a compact form. Instead, we find it more transparent to work directly with the fermionic bubble in position space and follow steps analogous to those in the previous section.

First, the fermionic bubble in the position space is \cite{Allen:1986qj,Chen:2018xck,Schaub:2023scu}
\begin{align}
    \mathcal{Q}_{\text{fermion}}(x_1,x_2)&\equiv\int\frac{\dd^3 s}{(2\pi)^3}\,e^{i\mathbf{s}\cdot \bfx}\,\mathcal{Q}_{-+}(s,\tau_1,\tau_2)=-4\Big[f_{m_1}(Z)f_{m_2}(Z)+g_{m_1}(Z)g_{m_2}(Z)\Big]\,, \label{eq: Q_position1}
\end{align}
where $f_m$ and $g_m$ are defined as 
\begin{align}
    f_m(Z)&=-i\,\frac{|\Gamma(2-i m)|^2}{16\sqrt{2}\pi^2}\sqrt{1-Z}\times\pFq{2}{1}{2-i m,2+i m}{2}{\frac{1+Z}{2}}\,,\label{def: fm}\\
    g_m(Z)&=\frac{m\,|\Gamma(2-i m)|^2}{32\sqrt{2}\pi^2}\sqrt{1+Z}\times\pFq{2}{1}{2-i m,2+i m}{3}{\frac{1+Z}{2}}\,, \label{def: gm}
\end{align}
the derivation of the above formula is given in Appendix~\ref{App: derivations}. For simplicity, we introduce another shorthand notation
\begin{align}
    {\mathcal{K}}_{\nu}({Z})\equiv\frac{(-2{Z})^{-\frac{3}{2}-\nu}}{2\pi^{5/2}}\,\,\Gamma\left[-\nu,\frac{3}{2}+\nu\right]\times\mathcal{D}\left(\frac{3}{2}+\nu,-\frac{1}{2},\frac{1}{{Z}^2}\right)\,,\label{eq: propagatorK}
\end{align}
which is related to the scalar propagator via 
\begin{align}
    \mathcal{G}_{i\mu}({Z})= \frac{1}{2}\big[{\mathcal{K}}_{i\mu}({Z})+{\mathcal{K}}_{-i\mu}({Z})\big]\,.
\end{align}
It is also related to the propagator \eqref{eq:gtil} through the simple relation as 
\begin{align}
    \mathcal{K}_{i\mu}(Z)=\frac{\pi}{2}\csch(\pi\mu)\int \frac{\mathrm{d}^3k}{(2\pi)^3}\,e^{i\bfk\cdot\bfx}\,\widetilde{\mathcal{G}}_{i\mu}(k,\tau_1,\tau_2)\,,
\end{align}
follows immediately by combining \eqref{eq: hypertoQ} and \eqref{eq: JH_Fourier}. This fermionic bubble in position space can then be written as 
\begin{align}
    \mathcal{Q}_{\text{fermion}}(x_1,x_2)&=\sum_{\lambda,\sa,\sb=\pm} \Big[\partial_Z{\mathcal{K}}_{\sa\nu_{1\lambda}}({Z})\,\partial_Z{\mathcal{K}}_{\sb\nu_{2\lambda}}({Z})-{Z}\,\partial_Z{\mathcal{K}}_{\sa\nu_{1\lambda}^*}({Z})\,\partial_Z{\mathcal{K}}_{\sb\nu_{2\lambda}}({Z})\Big]\,,
\end{align}
and one can numerically check that this is equivalent to \eqref{eq: Q_position1}. We can also expand each $\nu_{\lambda}$ and sum over helicities. Then in a more compact form, equivalently, this reads
\begin{align}
\mathcal{Q}_{\text{fermion}}(x_1,x_2)=\sum_{\sa,\sb,\sc,\mathsf{d}=\pm1}\left(\delta_{\sa\sb,\sc\mathsf{d}}-Z\delta_{\sa\sb,-\sc\mathsf{d}}\right){\partial_Z\mathcal{K}}_{\sc/2+i\sa m_1}(Z)\,{\partial_Z\mathcal{K}}_{\mathsf{d}/2+i\sb m_2}(Z)\,.
\end{align}
Depending on the relative sign between $\sa$ and $\sb$, there are two possible frequencies, set by either $M\equiv m_1+m_2$ or $\delta \equiv m_1-m_2$. Accordingly, we can separate the result into two parts:
\begin{align}
    \mathcal{Q}_{\text{fermion}}(x_1,x_2)=\mathcal{Q}_{M}(x_1,x_2)+\mathcal{Q}_{\delta}(x_1,x_2)\,,
\end{align}
where the first $\mathcal{Q}_M$ comes from $\sa=\sb$, while the other $\mathcal{Q}_{\delta}$ corresponds to the opposite choice $\sa=-\sb$. Let us begin with the equal sign case, using \eqref{eq: propagatorK}, it becomes
\begin{align}
    \mathcal{Q}_{M}=&\sum_{\sa=\pm}\,\frac{\left(-2Z\right)^{-4-i\sa M}}{\pi^3\cosh(\pi m_1)\cosh(\pi m_2)}\,\Gamma\left[\bgm2+i\sa m_1,2+i\sa m_2\\\frac{1}{2}+i\sa m_1,\frac{1}{2}+i\sa m_2\edm\right]\nonumber\\
    &\times\Bigg[\prod_{j=1}^2\,Z\,\mathcal{D}\bigg(1+i\sa m_j,-\frac{1}{2},\frac{1}{Z^2}\bigg)-\prod_{j=1}^2\left(Z^2-1\right)^{1/2}\mathcal{D}\bigg(2+i\sa m_j,-\frac{3}{2},\frac{1}{Z^2}\bigg)\Bigg]\,.
\end{align}
It already has the desired form, so we can directly apply the decomposition formula \eqref{eq: product_2F1} to both products in the second line, yielding
\begin{align}
    \mathcal{Q}_{M}=&\sum_{k=0}^{\infty}\sum_{\sa=\pm}\frac{(-2Z)^{-4-2k-i\sa M}}{\pi^3\cosh(\pi m_1)\cosh(\pi m_2)}\times\Gamma\left[\bgm 2+i\sa m_1, 2+i\sa m_2\\ 1/2+i\sa m_1, 1/2+i \sa m_2\edm\right]\nonumber\\
    &\Bigg[-f_{k,2}(Z^2-1) \mathcal{D}\left(4+2k+i\sa M,-\frac{3}{2},\frac{1}{Z^2}\right)+f_{k,1}Z^2\,\mathcal{D}\left(2+2k+i\sa M,-\frac{1}{2},\frac{1}{Z^2}\right)\Bigg]\,,
\end{align}
and the prefactors are
\begin{align}
    &f_{k,1}\equiv f_k\left(1+i\sa m_1,1+i \sa m_2,-\frac{1}{2}\right)\,,\\
    &f_{k,2}\equiv f_k \left(2+i\sa m_1,2+i\sa m_2,-\frac{3}{2}\right)\,.
\end{align}
As in the scalar case, to transform back to momentum space, it is more convenient to work with the Legendre-$Q$. Using the relation between the hypergeometric function and the Legendre-$Q$ function \eqref{eq: hypertoQ}, we obtain
\begin{align}
    Q_M&=-\sum_{k=0}^{\infty}\sum_{\sa=\pm}\frac{\sech (\pi m_1)\sech (\pi m_2)}{8\pi^{7/2}}\times\Gamma\left[\bgm 2+i\sa m_1, 2+i\sa m_2,5/2+2k+i\sa M\\ 1/2+i\sa m_1, 1/2+i \sa m_2,2+2k+i\sa M\edm\right]\nonumber\\
       &~~~\Bigg\{\frac{2f_{k,2}}{(3+2k+i\sa M)(2+2k+i\sa M)}\,Q_{1+2k+i\sa  M}^2(-Z)+\frac{f_{k,1}}{(3/2+2k+i\sa M)}\frac{Q_{2k+i\sa M}^1(-Z)}{\sqrt{Z^2-1}}\Bigg\}\,. \label{eq: fermionStep1}
\end{align}
To go back to the momentum space using \eqref{eq: JH_Fourier}, we need $Q^1_{\nu}$ instead of the higher order $Q^2_{\nu}$. The next step is therefore to apply the recurrence relation \eqref{eq: Q2toQ1}, after which the expression above becomes
\begin{align}
     Q_M=&\sum_{k=0}^{\infty}\sum_{\sa=\pm}\frac{\sech (\pi m_1)\sech (\pi m_2)}{8\pi^{7/2}\sqrt{Z^2-1}}\,\Gamma\left[\bgm 2+i\sa m_1, 2+i\sa m_2,7/2+2k+i\sa M\\ 3/2+i\sa m_1, 3/2+i \sa m_2,4+2k+i\sa M\edm\right]\nonumber\\
     &\times (3+2k)(k+1+i \sa M) \,f_k\left(2+i\sa m_1,2+i\sa m_2,-\frac{1}{2}\right)\,Q_{2+2k+i\sa M}^1(-Z)\,.\label{eq: fermionStep2}
\end{align}
At this point, only a single term involving $Q^1_{\nu}$ remains. The above expression follows from rearranging and combining two infinite series, together with some properties about their coefficients, more detailed steps are provided in the Appendix \ref{sec: spectral_app}. Finally, applying the transformation \eqref{eq: JH_Fourier} yields the decomposition result which is
\begin{keyeqn}
    \begin{align}
        \mathcal{Q}_{M}(s,\tau_1,\tau_2;m_1,m_2)=\sum_{\sa=\pm}\sum_{k=0}^{\infty} \, \rho_M(k,i\sa m_1,i\sa m_2)\times \widetilde{\mathcal{G}}_{2k+\frac{5}{2}+i\sa M}(s,\tau_1,\tau_2)\,, \label{eq: fermion_M}
    \end{align}
    \text{with the spectral density:}
    \begin{align}
       \rho_M(k,i\sa m_1,i\sa m_2)&\equiv \frac{i}{4\sqrt{\pi}}\frac{(3+2k)(k+1+i\sa M)}{\cosh{(\pi m_1)}\cosh{(\pi m_2)}}\, f_k\left(2+i\sa m_1,2+i\sa m_2,-\frac{1}{2}\right)\nonumber\\
       &\times  \Gamma\left[\bgm 2+i\sa m_1, 2+i\sa m_2,7/2+2k+i\sa M\\ 3/2+i\sa m_1, 3/2+i \sa m_2,4+2k+i\sa M\edm\right]\,.
    \end{align}
\end{keyeqn}
The other case, with opposite signs $\sa=-\sb$ and frequency $\delta\equiv m_1-m_2$, follows similar steps. We  quote the final result below, and provide the derivation in Appendix~\ref{sec: spectral_app}:
\begin{keyeqn}
    \begin{align}
        \mathcal{Q}_{\delta}(s,\tau_1,\tau_2;m_1,m_2)=\sum_{\sa=\pm}\sum_{k=0}^{\infty} \, \rho_\delta(k,i\sa m_1,i\sa m_2)\times \widetilde{\mathcal{G}}_{2k+\frac{3}{2}+i\sa \delta}(s,\tau_1,\tau_2)\,, \label{eq: fermion_delta}
    \end{align}
    \text{with the spectral density:}
    \begin{align}
       \rho_\delta(k,i\sa m_1,i\sa m_2)&\equiv -\frac{3i}{2\sqrt{\pi}}\frac{(2k+i\sa \delta)}{(3+2k)(k+i\sa\delta)}\, \frac{f_k\left(2+i\sa m_1,2-i\sa m_2,-\frac{3}{2}\right)}{{\cosh{(\pi m_1)}\cosh{(\pi m_2)}}}\nonumber\\
       &\times  \Gamma\left[\bgm 2+i\sa m_1, 2-i\sa m_2,5/2+2k+i\sa \delta\\ 1/2+i\sa m_1, 1/2-i \sa m_2,3+2k+i\sa \delta\edm\right]\,.\label{eq: density_delta}
    \end{align}
\end{keyeqn}
Combining both, we obtain the total contribution to the signals from the fermionic bubble. It can also be expressed schematically as
\begin{align}
\raisebox{2.0 pt}{\begin{tikzpicture}[baseline={(current bounding box.center)}, line width=1. pt, scale=3.2]
    \draw[red3, thick] (0.5,0) arc[start angle=0, end angle=180, radius=0.25];
    \draw[blue3, thick] (0.5,0) arc[start angle=360, end angle=180, radius=0.25];
    \draw[draw=lightgray2, fill=lightgray2] (0, 0) circle (.035cm);
    \draw[draw=lightgray2, fill=lightgray2] (0.5, 0) circle (.035cm);
    \node at (0.25,0.32){\textcolor{red3}{$m_1$}};
    \node at (0.25,-0.32){\textcolor{blue3}{$m_2 $}};
    \node at (0.25,0){fermion};
\end{tikzpicture} }
~~=~~ \sum_{k,\sa}\, \rho_M\,\times\,\bigg(~
\raisebox{1.0 pt}{\begin{tikzpicture}[baseline={(current bounding box.center)}, line width=1. pt, scale=3.0]
    \draw[black, thick] (0.0,0) -- (0.6,0);
    \draw[draw=lightgray2, fill=lightgray2] (0, 0) circle (.035cm);
    \draw[draw=lightgray2, fill=lightgray2] (0.6, 0) circle (.035cm);
    \node[scale=0.25, transform shape] at (0.3,0.12){$2k+\frac{5}{2}+i\sa M$};
  \node[scale=0.25, transform shape] at (0,-0.11) {$J$};
  \node[scale=0.25, transform shape] at (0.6,-0.11) {$H$};
\end{tikzpicture}}
~
\bigg)+
\sum_{k,\sa}\, \rho_\delta\,\times\,\bigg(~
\raisebox{1.0 pt}{\begin{tikzpicture}[baseline={(current bounding box.center)}, line width=1. pt, scale=3.2]
    \draw[black, thick] (0.0,0) -- (0.6,0);
    \draw[draw=lightgray2, fill=lightgray2] (0, 0) circle (.035cm);
    \draw[draw=lightgray2, fill=lightgray2] (0.6, 0) circle (.035cm);
    \node[scale=0.25, transform shape] at (0.3,0.12){$2k+\frac{3}{2}+i\sa\delta$};
  \node[scale=0.25, transform shape] at (0,-0.11) {$J$};
  \node[scale=0.25, transform shape] at (0.6,-0.11) {$H$};
\end{tikzpicture}}
~
\bigg)\,.\label{fermionDecom}
\end{align}
These two different series, \eqref{eq: fermion_M} and \eqref{eq: fermion_delta}, can be regrouped into a single spectral integral by taking the corresponding residues,
\begin{align}
    Q_{-+}(s,\tau_1,\tau_2)\equiv \int_{-\infty}^{+\infty}\mathrm{d}\nu\,\nu\,\rho_{\text{fermion}}(\nu)\times\widetilde{\mathcal{G}}_{i\nu}(s,\tau_1,\tau_2)\,, \label{eq: spectral_integral}
\end{align}
with the fermion spectral density
\begin{align}
    \rho_{\text{fermion}}(\nu)\equiv\frac{\prod_{\sa,\sb,\sc,\mathsf{d}=\pm1}\Gamma\left(\frac{1}{2}+\frac{\frac{3}{2}+ i\sa M+i\sb\nu}{2}\right)\,\Gamma\left(\frac{\frac{3}{2}+i\sc\delta+i\mathsf{d} \nu}{2}\right)}{4\pi^4\times\Gamma\left(\frac{3}{2}+ i\nu\right)\Gamma\left(\frac{3}{2}- i\nu\right)}\,,
\end{align}
so there are four Gamma functions that involve $M$ and four associated with $\delta$, and we have checked that it is consistent with the result reported in the first related work \cite{Altshuler:2025qmk}.
With this spectral density, it would also be interesting to generalise various bounds \cite{deRham:2025mjh,Lee:2025kgs} on cosmological correlators involving fermionic states.

Now the bubble functions have been expressed as a series summation of tree-level propagators, the final step is to complete the time integrals,
\begin{align}
&-\sum_{\sa=\pm}\sa\,{s^{5+q_{12}}}\int_{-\infty}^0 \mathrm{~d} \tau_1 \mathrm{~d} \tau_2  \left(-\tau_1\right)^{q_1}\left(-\tau_2\right)^{q_2} e^{i\sa k_{12} \tau_1+i k_{34} \tau_2}\times \widetilde{\mathcal{G}}_{i\nu}(s,\tau_1,\tau_2)\nonumber\\
&\equiv \mathcal{T}^{\text{LS}}_{i\nu}(r_1,r_2)+\mathcal{T}^{\text{NS}}_{i\nu}(r_1,r_2)\,, \label{eq: spectral_timeInt}
\end{align}
where we have decomposed it into two parts, corresponding to the non-local and local CC signals, respectively, with the explicit forms given by
\begin{align}
    \mathcal{T}^{\rm{LS}}_{i\nu}(r_1,r_2)&=r_1^{\frac{5}{2}+q_1}r_2^{\frac{5}{2}+q_2}\left(\frac{r_1}{r_2}\right)^{i\nu}\frac{2(e^{i\pi(q_1+i\nu)}+i)}{e^{i\pi q_{12}/2}(e^{-2\pi\nu}-1)}\,\Gamma\left[\frac{5}{2}+q_1+i\nu, \frac{5}{2}+q_2-i\nu\right]\nonumber\\
    &\times\pregFq{2}{1}{\frac{5}{4}+\frac{q_2}{2}-\frac{i\nu}{2}, {\frac{7}{4}+\frac{q_2}{2}-\frac{i\nu}{2}}}{1-i\nu}{r_2^2}\times\pregFq{2}{1}{\frac{5}{4}+\frac{q_1}{2}+\frac{i\nu}{2},{\frac{7}{4}+\frac{q_1}{2}+\frac{i\nu}{2}}}{1+i\nu}{r_1^2}\,, \label{eq: T_LS}
\end{align}
and the non-local type as 
\begin{align}
    \mathcal{T}^{\rm{NS}}_{i\nu}(r_1,r_2)&=r_1^{q_1}r_2^{q_2}\left({r_1}{r_2}\right)^{\frac{5}{2}+i\nu}\,\frac{i+e^{i\pi(q_1+i\nu)}}{e^{-2\pi\nu}-1}\frac{e^{-i\pi\bar{q}_{12}/2}}{2^{2i\nu-1}}\Gamma\left[\frac{5}{2}+q_1+i\nu, \frac{5}{2}+q_2+i\nu\right]\nonumber\\
    &\times\pregFq{2}{1}{\frac{5}{4}+\frac{q_2}{2}+\frac{i\nu}{2}, {\frac{7}{4}+\frac{q_2}{2}+\frac{i\nu}{2}}}{1+i\nu}{r_2^2}\times\pregFq{2}{1}{\frac{5}{4}+\frac{q_1}{2}+\frac{i\nu}{2},{\frac{7}{4}+\frac{q_1}{2}+\frac{i\nu}{2}}}{1+i\nu}{r_1^2}\,. \label{eq: T_NS}
\end{align}
Finally, the complete expressions for the two parts of signals are
\begin{keyeqn}
\begin{align}
    &\mathcal{I}^{q_1 q_2}_{\rm{NS}(LS)}\nonumber\\
    &=\sum_{\sa=\pm}\sum_{k=0}^{\infty}\left[\rho_{M}(k,i\sa m_1,i\sa m_2)\,\mathcal{T}^{\rm{\rm{NS(LS)}}}_{2k+\frac{5}{2}+i\sa M}(r_1,r_2)+\rho_{\delta}(k,i\sa m_1,i\sa m_2)\,\mathcal{T}^{\rm{NS(LS)}}_{2k+\frac{3}{2}+i\sa \delta}(r_1,r_2)\right]+\text{c.c.}\,\,. \label{spectral_Signals}
\end{align}
\end{keyeqn}
\subsection{Mellin-Barnes transformation} 
\label{sec: MBmethod}
In this section, we follow the second route to evaluate the bubble diagram, extending the scalar analysis of \cite{Qin:2024gtr} to the fermionic case. The key ingredient is to employ the MB representation of the Hankel function: 
\begin{align}
H_{\nu}^{(j)}(-k \tau)
= \int_{-i \infty}^{i \infty} \frac{\mathrm{d}s}{2 \pi i} 
\frac{(-k \tau / 2)^{-2s}}{\pi}\, 
e^{(-1)^{j+1} i \frac{\pi}{2}(2s - \nu - 1)} 
\times\Gamma\left[s - \frac{\nu}{2},\, s + \frac{\nu}{2} \right],
\quad (j = 1, 2).\label{MB}
\end{align}
As we will see later, this representation has the advantage of making both the momentum and time integrals essentially straightforward. It also provides a transparent way to track both non-local and local CC signals by picking up the corresponding poles.
We begin with the opposite-sign seed function $\mathcal{I}_{-+}$ as an example, its counterpart $\mathcal{I}_{+-}$ can be obtained simply by taking the complex conjugate. Substituting the explicit expression for the bubble function $\mathcal{Q}_{-+}$ from \eqref{eq:BubbleFunction_MomentumSpace} into the seed function \eqref{eq:SeedIntegral} with $(\sa, \sb) = (-, +)$, and applying the MB transformation \eqref{MB} to each Hankel function inside the integrand, we find that
\begin{align}
\nonumber \mathcal{I}_{-+}^{q_1q_2}&=\ \frac{ s^{5+q_{12}}}{8\pi^2} \int_{-i \infty}^{+i \infty}\prod_{i=1}^4 \frac{\mathrm{~d} s_i}{2 \pi i}\int_{-\infty}^0 \mathrm{~d} \tau_1 \mathrm{~d} \tau_2  \left(-\tau_1\right)^{q_1+4-2s_{13}}\left(-\tau_2\right)^{q_2+4-2s_{24}} e^{-i k_{12} \tau_1+i k_{34} \tau_2} \\
& \times e^{ i\pi (s_{13}-  s_{24})} \int \frac{\mathrm{d}^3 \mathbf{p}_1}{(2 \pi)^3}  \left(\frac{p_1}{2}\right)^{-2s_{12}}\left(\frac{p_2}{2}\right)^{-2s_{34}}\sum_{\lambda=\pm}\bigg[ p_1p_2\, \widetilde{\Gamma}_{1\lambda}\big(\{s_i\}\big)  -(\mathbf{p}_1 \cdot \mathbf{p}_2)\,\widetilde{\Gamma}_{2\lambda}\big(\{s_i\}\big)\bigg]\,,
\end{align}
where the shorthand notations for products of Gamma functions are defined through
\begin{align}
&\widetilde{\Gamma}_{1\lambda}\big(\{s_i\}\big)
\equiv \Gamma\left[s_1+\frac{\nu_{1\lambda}}{2}, s_1-\frac{\nu_{ 1\lambda}}{2},s_2+\frac{ \nu_{1\lambda}^*}{2}, s_2-\frac{\nu_{1\lambda}^*}{2},s_3+\frac{ \nu_{2\lambda}^*}{2}, s_3-\frac{\nu_{2\lambda}^*}{2},s_4+\frac{ \nu_{ 2\lambda}}{2}, s_4-\frac{ \nu_{2\lambda}}{2}\right],\label{til_G_1}\\
&\widetilde{\Gamma}_{2\lambda}\big(\{s_i\}\big)   
\equiv\Gamma\left[s_1+\frac{\nu_{1\lambda}^*}{2}, s_1-\frac{\nu_{ 1\lambda}^*}{2},s_2+\frac{ \nu_{1\lambda}^*}{2}, s_2-\frac{\nu_{1\lambda}^*}{2}, s_3+\frac{ \nu_{2\lambda}}{2}, s_3-\frac{\nu_{2\lambda}}{2}, s_4+\frac{ \nu_{ 2\lambda}}{2}, s_4-\frac{ \nu_{2\lambda}}{2}\right]\,,\label{til_G_2}
\end{align}
the product of the four Hankel functions in the bubble loop yields these products of eight Gamma functions. Here the $s_2$ and $s_4$ sectors are identical in $\widetilde{\Gamma}_{1\lambda}$ and $\widetilde{\Gamma}_{2\lambda}$, while the $s_1$ and $s_3$ parts are complex conjugates of each other.
Now the seed function involves three types of integrations: over time, loop momentum, and the MB variables $s_i$. The time and momentum integrands are elementary and can be handled straightforwardly. We now proceed by evaluating each integral in sequence.
\paragraph{- Step I: \textit{Time integral}}~~
\\
The time integrands consist of power laws and exponential factors, so they can be evaluated as
\begin{align}
&\int_{-\infty}^0 \mathrm{~d} \tau_1 \mathrm{~d} \tau_2  \left(-\tau_1\right)^{q_1+4-2s_{13}}\left(-\tau_2\right)^{q_2+4-2s_{24}} e^{- i k_{12} \tau_1 +i k_{34} \tau_2} \nonumber\\
&=e^{i{\pi}\bar{q}_{12}/2}\,e^{i\pi(s_{24}-s_{13})}\,k_{12}^{\left(2 s_{13}-q_1-5\right)}\,k_{34}^{\left(2 s_{24}-q_2-5\right)}\times\Gamma\Big[q_1+5-2 s_{13}, q_2+5-2 s_{24}\Big]\,,
\end{align}
where $\bar{q}_{12}\equiv q_1-q_2$, and this time integral introduces two more Gamma functions.
\paragraph{- Step II: \textit{Loop integral}}
~\\  
The loop momentum integral is evaluated by \eqref{eq_loopint1} and \eqref{eq_loopint2} which can be derived using the Schwinger parametrisation, resulting in
\begin{align}\label{eq_loop1}
\int \frac{\mathrm{d}^3 \mathbf{p}_1}{(2 \pi)^3}\, p_1p_2 \left(\frac{p_1}{2}\right)^{-2s_{12}} \left(\frac{p_2}{2}\right)^{-2s_{34}}=\frac{2^{2s_{1234}}s^{5-2s_{1234}}}{(4\pi)^{3/2}}\,
 \Gamma\left[
 \begin{matrix}
     s_{1234}-\frac{5}{2},-s_{12}+2,-s_{34}+2\\
     -s_{1234}+4,s_{12}-\frac{1}{2},s_{34}-\frac{1}{2}
 \end{matrix}
 \right],
\end{align}
and the other one with the vector dot product is
\begin{align}\label{eq_loop2}
\int \frac{\mathrm{d}^3 \mathbf{p}_1}{(2 \pi)^3}\,(\mathbf{p}_1 \cdot \mathbf{p}_2)\left(\frac{p_1}{2}\right)^{-2s_{12}} \left(\frac{p_2}{2}\right)^{-2s_{34}} =-\frac{2^{2s_{1234}}s^{5-2s_{1234}}}{(4\pi)^{3/2}}
 \Gamma\left[
 \begin{matrix}
     s_{1234}-\frac{5}{2},-s_{12}+\frac{5}{2},-s_{34}+\frac{5}{2}\\
     -s_{1234}+4,s_{12},s_{34}
 \end{matrix}
 \right]\,.
\end{align}
Here we recall that $s \equiv |{\bf k}_1 + {\bf k}_2|$ (not to be confused with the MB integration variables).
After finishing both the time and loop integrals, now the seed function $\mathcal{I}_{-+}^{q_1q_2}$ can be summarised as
\begin{align}
&\nonumber \mathcal{I}_{-+}^{q_1q_2}=\ \frac{e^{ i{\pi}\bar{q}_{12}/2}\,r_1^{5+q_1}r_2^{5+q_2} }{64\pi^{7/2}} \sum_{\lambda= \pm}\int_{-i \infty}^{+i \infty}\prod_{i=1}^4 \frac{\mathrm{~d} s_i}{2 \pi i}\,\,\Gamma\left[
 \begin{matrix}
     s_{1234}-\frac{5}{2},q_1+5-2 s_{13}, q_2+5-2 s_{24}\\
     -s_{1234}+4
 \end{matrix}
 \right]\\
&\times\left(\frac{r_1}{2}\right)^{-2 s_{13}}\left(\frac{r_2}{2}\right)^{-2 s_{24}}\Bigg\{  \widetilde{\Gamma}_{1\lambda}\big(\{s_i\}\big)\,\Gamma\left[\begin{array}{c}
2-s_{12},2-s_{34} \\
s_{12}-\frac{1}{2}, s_{34}-\frac{1}{2}
\end{array}\right]+\widetilde{\Gamma}_{2\lambda}\big(\{s_i\}\big)\,\Gamma\left[\begin{array}{c}
\frac{5}{2}-s_{12},\frac{5}{2}-s_{34} \\
s_{12}, s_{34}
\end{array}\right]\Bigg\}\,,
\end{align}
where only MB-$s_i$ integrals remain and it depends only on the momentum ratios: 
\begin{align}
r_1 \equiv \frac{s}{k_{12}}\,,\qquad r_2 \equiv \frac{s}{k_{34}}\,.
\end{align}
The evaluation of $\mathcal{I}_{\pm \pm, {\rm F}}$ is completely parallel to the above steps. The only modification comes from the sign assignment on the external legs. For instance, replacing  $e^{-i k_{12}\tau_1}$ by $e^{+i k_{12}\tau_1}$ turns $\mathcal{I}_{-+}$ into  $\mathcal{I}_{++,{\rm{F}}}$. Combining all contributions, the total factorised part of the seed function \eqref{I_F} is then given by
\begin{align}
\nonumber \mathcal{I}_{\text{F}}^{q_1q_2}&=\ \frac{r_1^{5+q_1}r_2^{5+q_2} }{32\pi^{7/2}}\int_{-i \infty}^{+i \infty}\left[\prod_{i=1}^4 \frac{\mathrm{~d} s_i}{2 \pi i}\right]\left(\frac{r_1}{2}\right)^{-2 s_{13}}\left(\frac{r_2}{2}\right)^{-2 s_{24}}\left[\cos \frac{\pi \bar{q}_{12}}{2}+\cos \left(2 \pi s_{13}-\frac{\pi q_{12}}{2}\right)\right]\\
\nonumber 
&\times \sum_{\lambda=\pm}\Bigg\{  \textcolor{red3}{\widetilde{\Gamma}_{1\lambda}\left\{(s_i)\right\}}\,\Gamma\left[\begin{array}{c}
2-s_{12},2-s_{34} \\
s_{12}-\frac{1}{2}, s_{34}-\frac{1}{2}
\end{array}\right]+\textcolor{red3}{\widetilde{\Gamma}_{2\lambda}\big\{(s_i)\big\}}\, \Gamma\left[\begin{array}{c}
\frac{5}{2}-s_{12},\frac{5}{2}-s_{34} \\
s_{12}, s_{34}
\end{array}\right]\Bigg\}\nonumber\\
&\times \Gamma\left[
 \begin{matrix}
     \textcolor{blue3}{s_{1234}-\frac{5}{2}},q_1+5-2 s_{13}, q_2+5-2 s_{24}\\
     -s_{1234}+4
 \end{matrix}
 \right]\,.\label{I_F2}
\end{align}
\paragraph{- Step III: \textit{Contour integral}}
~\\  
Now, the remaining task is to perform the  $s_i$-integrals, which can be done by closing the contours to the left and picking up the residues at the left-hand poles of all $s_i$. We then find two types of poles: {\it spectrum poles}, originating from $\widetilde{\Gamma}_{1,2\lambda}$, and {\it UV poles} coming from the factor $\Gamma(s_{1234}-5/2)$ in \eqref{I_F2}, as previously noted in \cite{Qin:2024gtr} for the scalar loop case. Explicitly, they are given by
\begin{itemize}
\item[\textcolor{red3}{$\bullet$}]{\textcolor{red3}{Spectrum poles:}}
\begin{align}
s_1=-n_1+\mathsf{c}_1 \frac{\nu_{1\lambda}}{2}, \quad s_2=-n_2+\mathsf{c}_2 \frac{\nu_{1\lambda}^*}{2}, \quad s_3=-n_3+\mathsf{c}_3 \frac{\nu_{2\lambda}^*}{2}, \quad s_4=-n_4+\mathsf{c}_4 \frac{\nu_{2\lambda}}{2}, \label{s_pole_1}
\end{align}
for the term with $\widetilde{\Gamma}_{1\lambda}$ \eqref{til_G_1} and 
\begin{align}
s_1=-n_1+\mathsf{c}_1 \frac{\nu_{1\lambda}^*}{2}, \quad s_2=-n_2+\mathsf{c}_2 \frac{\nu_{1\lambda}^*}{2}, \quad s_3=-n_3+\mathsf{c}_3 \frac{\nu_{2\lambda}}{2}, \quad s_4=-n_4+\mathsf{c}_4 \frac{\nu_{2\lambda}}{2},\label{s_pole_2}
\end{align}
for those with $\widetilde{\Gamma}_{2\lambda}$ \eqref{til_G_2}, where $n_i=0,1, \cdots, \text { and } \mathsf{c}_i= \pm$. The locations of these poles are shown in the left panel of Figure~\ref{fig: poles}. This is different from the scalar case, in which all spectral poles lie in the left half-plane. Here, however, since the mass parameter $\nu_{i}$ has the real part one half, the contour then should be closed so as to include the region ${\rm{Re}}\,s_i\leq 1/4$. Since all poles here encode information about the fermion masses, these spectral poles will contribute to the oscillatory signals.
\item[\textcolor{blue3}{$\bullet$}]{{\textcolor{blue3}{UV poles:}}}
\begin{align}
s_{1234}=\frac{5}{2}-m\,, \quad m=0,1, \cdots,    \label{uv_pole}
\end{align}
the poles are shown in the right panel of Figure~\ref{fig: poles}. These are called UV poles because  the loop integrals \eqref{eq_loop1} and \eqref{eq_loop2} diverge in the UV region $q\to\infty$ with this set of poles. The sum of the $s_i$ variables takes half-integer values and, as we will see shortly, contributes to both the signal and background parts.
\end{itemize}
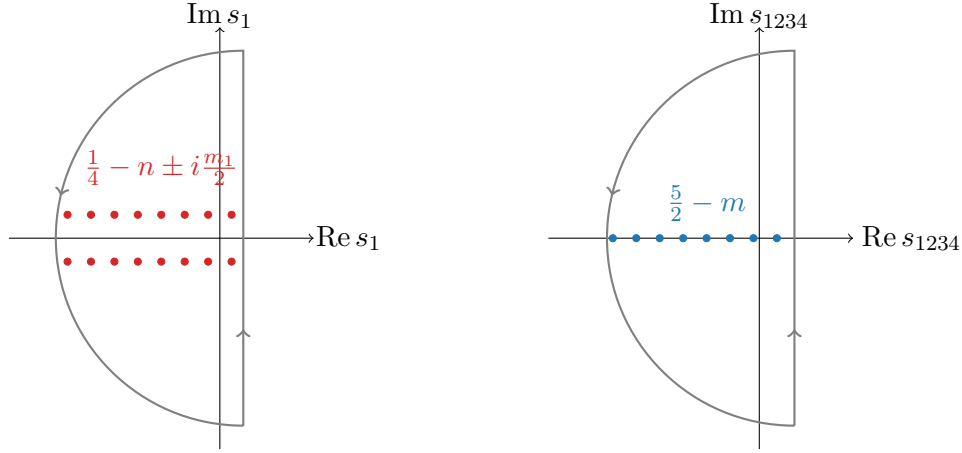
\begin{figure}[h]
    \centering
    \begin{subfigure}{0.45\textwidth}
    \centering
    \centering
    \begin{tikzpicture}[scale = 1.55]
        \draw[black, ->] (-1.8,0) -- (0.8,0) coordinate (xaxis);
        \draw[black, ->] (0,-1.8) -- (0,1.8) coordinate (yaxis);
        \node at (1.1, 0) {$\Re s_1$};
        \node at (0, 1.9) {$\Im s_1$};
		\draw[red3, fill = red3] (0.1, 0.2) circle (.03cm);
        \draw[red3, fill = red3] (-0.1, 0.2) circle (.03cm);
        \draw[red3, fill = red3] (-0.3, 0.2) circle (.03cm);
        \draw[red3, fill = red3] (-0.5, 0.2)  circle (.03cm);
        \draw[red3, fill = red3] (-0.7, 0.2)  circle (.03cm);
        \draw[red3, fill = red3] (-0.9, 0.2) circle (.03cm);
        \draw[red3, fill = red3] (-1.1, 0.2) circle (.03cm);
        \draw[red3, fill = red3] (-1.3, 0.2) circle (.03cm);
        \draw[red3, fill = red3] (0.1, -0.2) circle (.03cm);
        \draw[red3, fill = red3] (-0.1, -0.2) circle (.03cm);
        \draw[red3, fill = red3] (-0.3, -0.2) circle (.03cm);
        \draw[red3, fill = red3] (-0.5, -0.2)  circle (.03cm);
        \draw[red3, fill = red3] (-0.7, -0.2)  circle (.03cm);
        \draw[red3, fill = red3] (-0.9, -0.2)  circle (.03cm);
        \draw[red3, fill = red3] (-1.1, -0.2) circle (.03cm);
        \draw[red3, fill = red3] (-1.3, -0.2) circle (.03cm);
        \node at (-0.5, 0.6) {\textcolor{red3}{ $\frac{1}{4}-n\pm i \frac{m_1}{2}$}};	
	\path[gray, draw, line width = 0.8pt, postaction = decorate, decoration={markings,
			mark=at position 0.1 with {\arrow[line width=1pt]{>}},
			mark=at position 0.65 with {\arrow[line width=1pt]{>}}}] (0.2, -1.6) -- (0.2, 1.6) arc (90:270:1.6) -- (0.2, -1.6);
	\end{tikzpicture}
     \end{subfigure}
    \begin{subfigure}{0.45\textwidth}
    \centering
    \centering
    \begin{tikzpicture}[scale = 1.55]
        \draw[black, ->] (-1.8,0) -- (0.8,0) coordinate (xaxis);
        \draw[black, ->] (0,-1.8) -- (0,1.8) coordinate (yaxis);
        \node at (1.3, 0) {$\Re s_{1234}$};
        \node at (0, 1.9) {$\Im s_{1234}$};
		\draw[blue3, fill = blue3] (0.15, 0) circle (.03cm);
        \draw[blue3, fill = blue3] (-0.05, 0) circle (.03cm);
        \draw[blue3, fill = blue3] (-0.25, 0) circle (.03cm);
        \draw[blue3, fill = blue3] (-0.45, 0)  circle (.03cm);
        \draw[blue3, fill = blue3] (-0.65, 0)  circle (.03cm);
        \draw[blue3, fill = blue3] (-0.85, 0)  
        circle (.03cm);
        \draw[blue3, fill = blue3] (-1.05, 0) circle (.03cm);
        \draw[blue3, fill = blue3] (-1.25, 0) circle (.03cm);
        \node at (-0.45, 0.30) {\textcolor{blue3}{ $\frac{5}{2}-m$}};
	\path[gray, draw, line width = 0.8pt, postaction = decorate, decoration={markings,
			mark=at position 0.1 with {\arrow[line width=1pt]{>}},
			mark=at position 0.65 with {\arrow[line width=1pt]{>}}}] (0.3, -1.6) -- (0.3, 1.6) arc (90:270:1.6) -- (0.3, -1.6);
	\end{tikzpicture}
     \end{subfigure}
     \caption{The location of different poles that contribute to the $s_i$ contour integrals.}
     \label{fig: poles}
\end{figure}
\subsubsection*{Non-local Signal}
Different poles contribute to different parts of the result, according to their distinct analytic structures. Let us begin with the non-local CC signal, which is non-analytic in the internal momentum $|\mathbf{s}| = |\mathbf{k}_1+\mathbf{k}_2|$. In the factorised seed function \eqref{I_F2}, the $s$ dependence appears as $s^{-2s_{1234}}$, arising from the factor $r_1^{-2s_{13}} r_2^{-2s_{24}}$. Because the UV poles in \eqref{uv_pole} are located at half-integer values of $s_{1234}$, they are irrelevant to the non-local signal, which thus receives contributions only from the spectrum poles.
The spectrum poles associated with   $\widetilde{\Gamma}_{1\lambda}$ from \eqref{s_pole_1} satisfy
\begin{align}
    s_{12}=-n_{12}+\frac{\sf c_{12}}{2}-i\lambda m_1\frac{\bar{\sf c}_{12}}{2}\,,\qquad s_{34}=-n_{34}+\frac{\sf c_{34}}{2}+i\lambda m_2\frac{\bar{\sf c}_{34}}{2}\,.
\end{align}
As seen in \eqref{I_F2}, $\widetilde{\Gamma}_{1\lambda}$ appears together with the additional denominator factors $1/\Gamma(s_{12}-1/2)\Gamma(s_{34}-1/2)$, that eliminate the contribution whenever $\bar{\sf c}_{12}=0$ or  $\bar{\sf c}_{34}=0$. Consequently, the surviving terms require $\mathsf{c}_2 = -\mathsf{c}_1$ and $\mathsf{c}_4 = -\mathsf{c}_3$. Likewise, for the integral involving $\widetilde{\Gamma}_{2\lambda}$, the contributing terms obey $\mathsf{c}_2 = \mathsf{c}_1$ and $\mathsf{c}_4 = \mathsf{c}_3$, due to the presence of $\Gamma(s_{12})\Gamma( s_{34})$ in the denominator. Imposing these conditions leaves only two independent dummy variables $\sc_i$ , which we relabel as $\sc_1$ and $\sc_2$. As a result, the final non-local signals take the form
\begin{align}
\mathcal{I}_{{\rm{NS}}}^{q_1q_2}=& \sum_{n_i=0}^{\infty}\sum_{\sc_i,\lambda=\pm}
\left(\frac{r_1 r_2}{4}\right)^{i \lambda(\mathsf{c}_1m_1-\mathsf{c}_2m_2)} \left(\frac{r_1}{2}\right)^{2 n_{13}+q_1+5}\left(\frac{r_2}{2}\right)^{2 n_{24}+q_2+5}\nonumber\\
&\times\left[\mathcal{A}_{1\lambda}(m_i,q_i,\sc_i,n_i)\left(\frac{r_2}{r_1}\right)^{\frac{\sf c_{12}}{2}}+\mathcal{A}_{2\lambda}(m_i,q_i,\sc_i,n_i)\left(\frac{r_1 r_2}{4}\right)^{-\frac{\sf c_{12}}{2}}\right]\,,\label{MB_NLS}
\end{align}
where we keep only the dependence on the kinematic variables explicit. These  coefficients $\mathcal{A}_{i\lambda}$ are products of several Gamma functions and their full expressions are collected in Appendix \ref{app: MB_nonlocal}. Clearly, in the soft limit, the leading terms are
\begin{align}
    \lim_{s\to0}\mathcal{I}_{\text{NS}}^{q_1q_2}\sim r_1^{q_1} r_2^{q_2}(r_1 r_2)^{i\lambda\delta+4}\,,
\end{align}
where the  characteristic frequency is $\delta=m_1-m_2$, thus the CC signal is dominated by the mass difference mode, which disappears in the equal-mass limit.
\subsubsection*{Local Signal}
Next, let us turn to the local type CC signals, which by definition should be analytic in the internal energy $|\bf{s}|$. The only possible contributions therefore come from the UV poles \eqref{uv_pole}. Evaluating the contour integrals is more subtle in this case, since the UV poles entangle all integration variables, and the residues have to be taken iteratively.
 The calculation is straightforward at the first layer, any choice of integration variable is equivalent. For definiteness, we carry out the $s_4$-integral first by taking the residue at $s_4 = 5/2 - m - s_{123}$, which yields
\begin{align}
\nonumber\mathcal{I}_{\mathrm{F}, {\rm{UV}}}^{q_1 q_2}&=\ \frac{r_1^{5+q_1} r_2^{q_2}}{ \pi^{7 / 2}} \int_{-i \infty}^{+i \infty}\frac{\mathrm{~d} s_1}{2 \pi i}\frac{\mathrm{~d} s_3}{2 \pi i}\left[\cos \frac{\pi \bar{q}_{12}}{2}+\cos \left(2 \pi s_{13}-\frac{\pi q_{12}}{2}\right)\right]\left(\frac{r_1}{r_2}\right)^{-2 s_{13}}  \\
&\times \sum_{m=0}^{\infty} \frac{(-1)^m}{m!}\left(\frac{r_2}{2}\right)^{2 m}\,\Gamma\left[\begin{array}{c}
q_1+5-2 s_{13}, q_2+2m+2 s_{13} \\
\frac{3}{2}+m
\end{array}\right] \sum_{\lambda=\pm}f_\lambda\left(s_1, s_3;m\right)\,,\label{I_FIII_2}
\end{align}
here the subscript on $\mathcal{I}_{\mathrm{F}, {\rm{UV}}}^{q_1 q_2}$ indicates that this is the contribution from the UV poles, and we have grouped all terms that involve $s_2$ into
 \begin{align}
\nonumber  f_\lambda(s_1,s_3;m)&\equiv \int_{-i \infty}^{+i \infty} \frac{\mathrm{~d} s_2}{2 \pi i}\,\Gamma\left[s_2+\frac{ \nu_{1\lambda}^*}{2}, s_2-\frac{\nu_{1\lambda}^*}{2},\frac{5}{2}-m-s_{123}+\frac{\nu_{2\lambda}}{2}, \frac{5}{2}-m-s_{123}-\frac{\nu_{2\lambda}}{2}\right]\\
&\times\left[\left(s_{12}-\frac{1}{2}\right)_m\left(2-m-s_{12}\right)_m\hat{\Gamma}_{\lambda}(s_1,s_3)+\left(s_{12}\right)_m\left(\frac{5}{2}-m-s_{12}\right)_m\hat{\Gamma}_{-\lambda}(s_1,s_3)\right] ,\label{def_fm_2}
\end{align}   
where $(z)_m\equiv \Gamma(z+m)/ \Gamma(z)$ is the Pochhammer symbol and we have defined another shorthand notation for Gamma functions as 
\begin{align}
&\hat{\Gamma}_{\lambda}(s_1,s_3)\equiv {\Gamma}\left[s_1+\frac{\nu_{1\lambda}}{2}, s_1-\frac{ \nu_{1\lambda}}{2}, s_3+\frac{ \nu_{2\lambda}^*}{2}, s_3-\frac{ \nu_{2\lambda}^*}{2}\right]\,. 
\end{align}
Now we still have three-layer integrals, we then proceed to the $s_2$-integral, which can be evaluated using Barnes’ lemma~\cite{BarnesLemma1}. The details of the calculation are provided in Appendix~\ref{app: local_steps}, and the final result is shown below as:
\begin{align}
\nonumber  f_\lambda(s_1,s_3;m)=&\ \sum_{t_1, t_2=0}^m(-1)^{t_{12}}\binom{m}{t_1}\binom{m}{t_2}\Gamma\left[\frac{-\nu_{1\lambda}^*+\nu_{2\lambda}+5}{2}-s_{13}-t_1,\frac{\nu_{1\lambda}^*-\nu_{2\lambda}+5}{2}-s_{13}-t_2\right] \\
\nonumber &\times\left[\left(\frac{3}{2}-s_1-\frac{\nu_{1\lambda}^*}{2}-t_1\right)_{t_1}\left(\frac{3}{2}-s_3-\frac{\nu_{2\lambda}}{2}-t_2\right)_{t_2}\hat{\Gamma}_{\lambda}(s_1,s_3)\right.\\
\nonumber &\left.~~+\left(1-s_1-\frac{\nu_{1\lambda}^*}{2}-t_1\right)_{t_1}\left(1-s_3-\frac{\nu_{2\lambda}}{2}-t_2\right)_{t_2}\hat{\Gamma}_{-\lambda}(s_1,s_3)\right]\\
&\times \Gamma\left[\begin{array}{c}
\frac{\nu_{1\lambda}^*+\nu_{2\lambda}+5}{2}-m-s_{13}, \frac{-\nu_{1\lambda}^*-\nu_{2\lambda}+5}{2}+m-s_{13}-t_{12} \\
5-t_{12}-2s_{13}
\end{array}\right]\,.\label{f_m_2}
\end{align}
To perform the remaining $s_1$ and $s_3$ integrals, we need to sum over residues from several sets of poles, one such set comes from $\hat{\Gamma}_{\lambda}$, with poles at 
\begin{align}
\hat{\Gamma}_{\lambda}(s_1,s_3):\quad s_1 = -n_1 + \mathsf{c}_1 \frac{\nu_{1\lambda}}{2}, \quad 
s_3 = -n_2 + \mathsf{c}_2 \frac{\nu_{2\lambda}^*}{2}\,,\label{eq:Local1}
\end{align}
similarly, in the integral involving $\widetilde{\Gamma}_{-\lambda}(s_1, s_3)$, the poles are 
\begin{align}
\hat{\Gamma}_{-\lambda}(s_1,s_3):\quad s_1 = -n_1 + \mathsf{c}_1 \frac{\nu_{1\lambda}^*}{2}, \quad 
s_3 = -n_2 + \mathsf{c}_2 \frac{\nu_{2\lambda}}{2}\,.\label{eq:Local2}
\end{align}
A final possibility \footnote{There are also other Gamma functions of the form $\Gamma(...-s_{13})$ which might suggest additional poles in the left half-plane of variables $s_{1,3}$. However, in the MB representation, the ``left" poles we refer to are, strictly speaking, those arising from factors of the type $\Gamma(...+s_i)$, i.e. Gamma functions with a positive coefficient in front of the MB variable\cite{Smirnov:2006ry,Smirnov:2012gma}. } is to pick up the poles of the factor $\Gamma(q_2+2m+2s_{13})$ coming from the time integral, located at
\begin{align}
\Gamma(q_2+2m+2s_{13}):\quad 2 s_{13} = -q_2 - 2m - k,\,\,\text{with}\quad k = 0, 1, \cdots. \label{UV2} 
\end{align}
However, recall the kinematic dependence in \eqref{I_FIII_2}, where it appears as $(r_1/r_2)^{-2s_{13}}$, the set of poles \eqref{UV2} above generate only analytic powers, and therefore do not contribute to the signal part. As a result, the signal receives contributions only from the poles in \eqref{eq:Local1} and \eqref{eq:Local2}. Summing over both sets then yields the final local signal as,
\begin{align}
\mathcal{I}_{\rm{LS}}^{q_1q_2}=\sum_{m,n_i=0}^{\infty}\sum_{\sc_i,\lambda=\pm}\mathcal{B}_{m,\lambda}(m_i,n_i,q_i,\sc_i)\,r_1^{5+q_1} r_2^{q_2+2m}\left(\frac{r_1}{r_2}\right)^{-\frac{\mathsf{c}_{12}}{2}+i\lambda(\mathsf{c}_1{m}_1-\mathsf{c}_2{m}_2)+2n_{12}}\,,\label{MB_LS}
\end{align}
for brevity again, we keep only the kinematic dependence and absorb all remaining coefficients into $\mathcal{B}_{m,\lambda}$ whose explicit form is given in Appendix \ref{app: local_steps}. We are considering the regions $r_1<r_2$, under the limit $r_i\to 0$ then the leading terms are 
\begin{align}
    \lim_{r_1\ll r_2\ll 1}\mathcal{I}^{q_1q_2}_{\text{LS}}\sim r_1^{5+q_1}r_2^{q_2}\left[\left(\frac{r_1}{r_2}\right)^{-1+i\lambda\delta}+\left(\frac{r_1}{r_2}\right)^{i\lambda M}\right]\,,
\end{align}
as in the non-local signal case, the leading contribution in this hierarchical squeezed limit is dominated by the local signal mode with frequency set by the mass difference $\delta$, which we will discuss in more detail in the later section. Now, combining \eqref{MB_LS} and \eqref{MB_NLS}, we obtain the complete signal parts from the fermionic bubble. On the other hand, even the factorised component still contains background terms coming from the pole set in \eqref{UV2}, we defer their evaluation to Appendix \ref{app: MP_background}.
\subsection{Comparison of different methods}\label{sec: compare}
We have now derived the full expressions for the CC signals from the fermionic loop by two completely different methods. The two results take rather different forms: the spectral method yields a single series involving products of hypergeometric functions, while the MB method gives a four-fold power series. Although some of the summation layers can be performed explicitly, proving the complete equivalence of the two expressions is still not straightforward. As a first check, we therefore study certain kinematic limits, in which the full expressions simplify significantly.
\subsection*{Single squeezed limit}
Let us begin with the single squeezed limit $r_1\to 0$, which is equivalent to the hard energy limit, namely $k_{12}\to \infty$ with $k_{34}$ and $s$ kept fixed. From the MB method, the leading contribution to the non-local signal \eqref{MB_NLS} comes from the term with $n_1=n_3=0$ and ${\mathsf c}_1={\mathsf c}_2=+1$. Then two remaining sums are those over $n_2$ and $n_4$. At this point, it is convenient to introduce the new variable $k=n_2+n_4$, so that the summation variables $(n_2,n_4)$ are replaced by $(n_2,k)$. The sum over $n_2$, with $0\le n_2\le k$, can be carried out first, yielding a product of gamma functions. The final sum over $k$ from $0$ to $\infty$ then gives the closed form as 
\begin{align}
\nonumber \lim_{r_1\to 0} \mathcal I^{q_1q_2}_{\rm{NS}} =&\ \frac{r_1^{4+q_1} r_2^{4+q_2}}{8 \pi^{7 / 2}} \sum_{\lambda= \pm}\left[\cos \frac{\pi \bar{q}_{12}}{2}-\cos \left( \frac{\pi q_{12}}2 + i\lambda \pi \delta\right)\right]\\
\nonumber &\times \left(\frac{r_1r_2}{4}\right)^{i\lambda \delta}
\times\pFq{2}{1}{\frac{4+q_2+i\lambda\delta}2,\frac{5+q_2+i\lambda\delta}2}{\frac52+i\lambda\delta}{r_2^2}
\times \Gamma\left[\frac{1}{2}-i\lambda{m}_1,\frac{1}{2}+i\lambda{m}_2\right] \\
&\times \Gamma\left[\begin{array}{c}
-\frac{3}{2}-i\lambda \delta,2+i\lambda{m}_1, 2-i\lambda{m}_2, 4+q_1+i\lambda \delta , 4+q_2+i\lambda \delta \\
3+i\lambda \delta
\end{array}\right].\label{squeezedNLS}
\end{align}
Similarly, for the local signal, only the terms with $n_1=n_2=0$ and ${\mathsf c}_1={\mathsf c}_2=+1$ in \eqref{MB_LS} contribute at the leading order in the limit $r_1\to 0$. The sums over $t_1$ and $t_2$, both running from $0$ to $m$, can be performed first, followed by the sum over $m$ from $0$ to $\infty$. The result is
\begin{align}
\nonumber \lim_{r_1\to 0} \mathcal I^{q_1q_2}_{\rm{LS}} =&\ \frac{r_1^{4+q_1} r_2^{1+q_2}}{\pi^{7 / 2}} \sum_{\lambda= \pm}\left[\cos \frac{\pi \bar{q}_{12}}{2}-\cos \left( \frac{\pi q_{12}}2 + i\lambda \pi \delta\right)\right]\\
\nonumber &\times \left(\frac{r_1}{r_2}\right)^{i\lambda \delta}
\times \pFq{2}{1}{
\frac{1+q_2-i\lambda\delta}2,\frac{2+q_2-i\lambda\delta}2}
{-\frac12-i\lambda\delta}
{r_2^2}
\times \Gamma\left[\frac{1}{2}-i\lambda{m}_1,\frac{1}{2}+i\lambda{m}_2\right]\\
\nonumber &\times \Gamma\left[\begin{array}{c}
\frac{3}{2} + i\lambda \delta, 2+i\lambda{m}_1, 2-i\lambda{m}_2, 4+q_1+i\lambda \delta , 1+q_2-i\lambda \delta\\
3+i\lambda \delta 
\end{array}\right].\label{squeezedLS}
\end{align}
In the spectral decomposition method, if we keep only the leading term in the series ($k=0$) and expand about $r_1=0$, we obtain exactly the same expressions as above, which then confirms the consistency of the two methods. Clearly, both the local and non-local signals are dominated by the frequency mode $\delta$, with the leading order behaviour in $r_1$ being $\mathcal{O}(r_1^{4+q_1})$.
By comparison, the background piece scales as (see Appendix \ref{app: MP_background}) 
\begin{align}
\lim_{r_1\to0}\mathcal I^{q_1q_2}_{\rm BG} = \mathcal O\left(r_1^{5+q_{12}}\right)\,,
\end{align}
so the background is always subdominant relative to both the non-local and local signals, provided that $q_2>-1$, which is typically the case in realistic models.
\subsection*{Hierarchical double squeezed limit}
We can further take $r_2 \to 0$ while keeping $r_1 \ll r_2$. In this hierarchically double squeezed limit, the signal expressions simplify further as
\begin{align}
\nonumber \lim_{r_1 \ll r_2 \to 0} \mathcal I^{q_1q_2}_{\rm{NS}} =&\ \frac{r_1^{4+q_1} r_2^{4+q_2}}{8 \pi^{7 / 2}} \sum_{\lambda= \pm}\left(\frac{r_1r_2}{4}\right)^{i\lambda \delta}\left[\cos \frac{\pi \bar{q}_{12}}{2}-\cos \left( \frac{\pi q_{12}}2 + i\lambda \pi \delta\right)\right]\times\Gamma\left[\frac{1}{2}-i\lambda{m}_1,\frac{1}{2}+i\lambda{m}_2\right] \\
&\times \Gamma\left[\begin{array}{c}
-\frac{3}{2}-i\lambda \delta,2+i\lambda{m}_1, 2-i\lambda{m}_2, 4+q_1+i\lambda \delta, 4+q_2+i\lambda \delta\\
3+i\lambda \delta
\end{array}\right],\\
\lim_{r_1\ll r_2 \to 0} \mathcal I^{q_1q_2}_{\rm{LS}} =&\ \frac{r_1^{4+q_1} r_2^{1+q_2}}{\pi^{7 / 2}} \sum_{\lambda= \pm} \left(\frac{r_1}{r_2}\right)^{i\lambda \delta}\left[\cos \frac{\pi \bar{q}_{12}}{2}-\cos \left( \frac{\pi q_{12}}2 + i\lambda \pi \delta\right)\right]
\times \Gamma\left[\frac{1}{2}-i\lambda{m}_1,\frac{1}{2}+i\lambda{m}_2\right]\nonumber\\
 &\times \Gamma\left[\begin{array}{c}
\frac{3}{2} + i\lambda \delta, 2+i\lambda{m}_1, 2-i\lambda{m}_2, 4+q_1+i\lambda \delta, 1+q_2-i\lambda \delta\\
3+i\lambda \delta
\end{array}\right]\,.
\end{align}
As we can see, the local signal dominates over the non-local signal in this double squeezed limit. Using the asymptotic behaviour of the Gamma functions, both signals are suppressed by the exponential factor $\exp(-\pi M)$ coming from their product, which is the same Boltzmann suppression as in the scalar case \cite{Qin:2024gtr}.
\begin{figure}[htp]
\centering
\hspace{-1cm}
\includegraphics[width=0.98\textwidth]{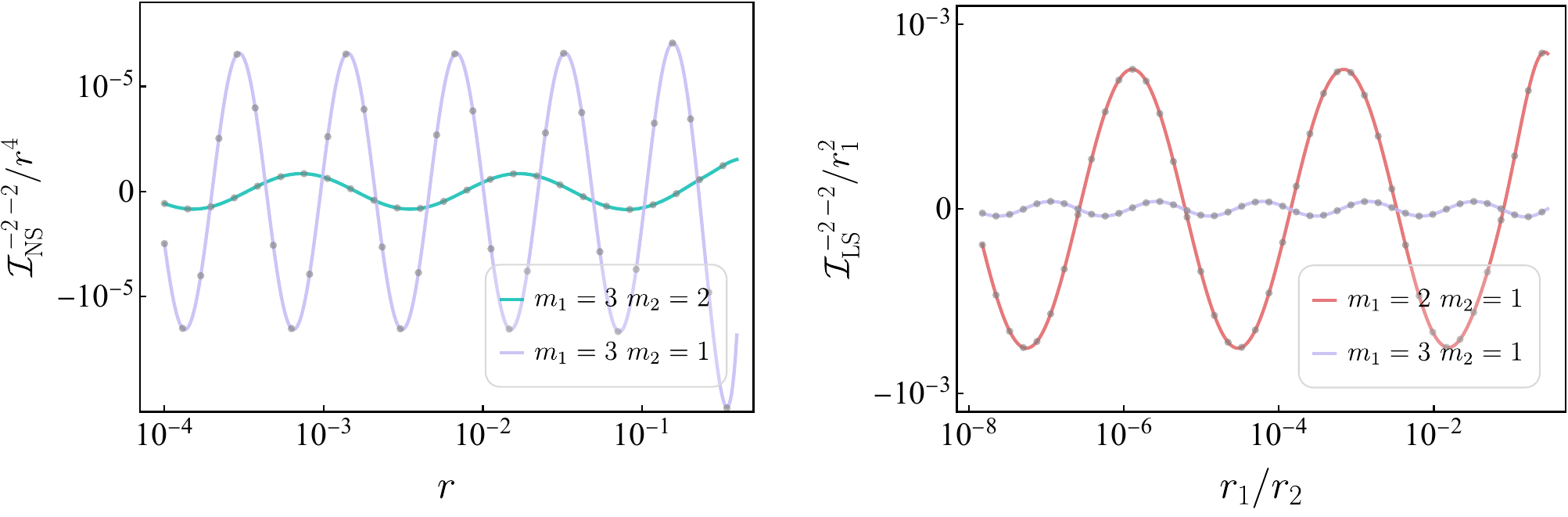}
\caption{The signals from the fermionic bubble loop obtained using two different methods. We take the parameters $q_1=q_2=-2$. \textit{Left panel}: \textbf{The non-local signals}, with the kinematic variables set to $r_1=r_2/2\equiv r$. The solid curves are obtained with the spectral method, shown in \textcolor[RGB]{10,190,180}{green} for $m_1=3,m_2=2$ and in \textcolor[RGB]{197,187,245}{purple} for $m_1=3,m_2=1$. The corresponding MB results are marked by \textcolor[RGB]{178,179,178}{gray dots} with the same mass choices. For the better visualisation, we normalise the vertical axis by $r^{-4}$. 
\textit{Right panel}: \textbf{The local signals}. We have fixed the second variable as $r_2=2/3$. Solid curves again denote the spectral results, with \textcolor[RGB]{230,121,123}{red} for $m_1=2,m_2=1$ and \textcolor[RGB]{197,187,245}{purple} for $m_1=3,m_2=1$. The corresponding MB results are shown as \textcolor[RGB]{178,179,178}{gray dots}. In this case, the vertical axis is scaled by $r_1^{-2}$.}
\label{fig:compare} 
\end{figure}

To further compare the results from the spectral method and the MB method, we plot both in Figure \ref{fig:compare}. The left panel shows the non-local signals, while the right panel shows the local ones. The results from the spectral method in \eqref{spectral_Signals} are plotted as solid curves for several mass choices, while the MB results are shown as gray dots. Clearly, the two sets of results are in excellent agreement. In both panels, we set the parameters to $q_1=q_2=-2$. For the non-local signals, we normalise the result by $r_1^4$ (or equivalently, $(r_1 r_2)^2$). For the local signals, by contrast, we normalise by $r_1^2$, since $r_2$ is kept fixed in that case. Both normalisation choices are consistent with the scaling behaviour in \eqref{squeezedNLS} and \eqref{squeezedLS}. The figure also makes clear that the oscillation frequency is set by the mass difference $\delta$, while the overall amplitude is controlled by the total mass $M$. As the total mass increases, the amplitude becomes more strongly suppressed. This is again  consistent with our previous analysis.
\paragraph{Discussion.} Let us make a few comments about these two methods.
\begin{itemize}
    \item The two methods lead to entirely different series expressions, yet our numerical checks show that they agree perfectly. In practice, however, the spectral method series is much more efficient for evaluating the signal than the MB series, since the latter involves four separate sums, whereas the former contains only a single sum. As we will see in the next section, the spectral method expressions are also more convenient for taking the three-point limit.
    \item In this section, we focus primarily on the signal contributions, all of which come from the factorised seed integral \eqref{I_F}. At the same time, this factorised part also contains a portion of the background contribution. More precisely,
    \begin{align}
    \mathcal{I}^{q_1q_2}_{\mathrm{F}}=\mathcal{I}^{q_1q_2}_{\mathrm{NS}}+\mathcal{I}^{q_1q_2}_{\mathrm{LS}}+(\text{part of}~~\mathcal{I}^{q_1q_2}_{\mathrm{BG}})\,.
    \end{align}
    The detailed derivation of the background contribution from the factorised integral is given in Appendix \ref{app: MP_background}. In the spectral method, however, this background contribution seems to be missing, since the time integrals for the tree-level propagators always involve oscillatory powers \eqref{eq: spectral_timeInt}. To resolve this issue, we need to revisit the derivation more carefully. The apparent inconsistency can be traced to the treatment of the time integral and the spectral series. Take the frequency mode $M$ as an example, for which $\mathcal{Q}_{M}$ is expressed as a series of $\widetilde{\mathcal{G}}_{2k+\frac{5}{2}+i\sa M}$ in \eqref{eq: fermion_M}. Using the definition \eqref{eq:gtil}, under the IR limit $\tau_i\rightarrow0$:
    \begin{align}
        \lim_{\tau_1\to0} \tau_1^{3/2}J_{2k+\frac{5}{2}+i\sa M}(-s\tau_1)&\sim \mathcal{O}\left(\tau_1^{4+2k+i \sa M}\right)\,,\qquad\\
        \lim_{\tau_2\to0} \tau_2^{3/2}H^{(2)}_{-(2k+\frac{5}{2}+i\sa M)}(-s\tau_2)&\sim \mathcal{O}\left(\tau_2^{-1-2k-i \sa M}\right)\,, \label{eq: H_diver}
    \end{align}
    so the $\tau_1$ integral is well convergent, whereas the $\tau_2$ integral is apparently divergent, especially for the terms with larger $k$. For the signal part, this is easy to understand: the oscillatory factors (with $iM$ or $i\delta$) act as regulators, allowing us to obtain finite results. This may be viewed as an analytic continuation, since we know that the signal part itself is free of divergences. For the background contribution, no such regulator is present, and the background information is therefore lost under the continuation. A similar point can also be understood directly from the spectral integral \eqref{eq: spectral_integral}, rather than start from the series formula \eqref{eq: product_2F1}. The spectral integral over $\nu$ in \eqref{eq: spectral_integral} generally does not commute with the time integrals. If the $\nu$ integral is performed first using the residue theorem, with the contour closed in one half-plane, the contribution from the large arc at $|\nu|\to\infty$ vanishes only when $\tau_1>\tau_2$ (with this ordering, the divergence in \eqref{eq: H_diver} is absent). Strictly speaking, one should therefore carry out the time integrals first. This will introduce an additional set of poles in the $\nu$-plane, which gives the background contribution as shown in the previous work \cite{Xianyu:2022jwk}. For the signal part, the result is unaffected, so the spectral and time integrals can be interchanged.
    \item The last part, $\mathcal{I}^{q_1q_2}_{\text{TO}}$ contributes only to the background. There are also several ways to compute the full background contribution. The first is to work directly with the MB representation, where the nested time integrals can still be handled without much difficulty. The second is to use the spectral decomposition, in this case, it is more convenient to start from the spectral integral \eqref{eq: spectral_integral}, perform the time integrals first, and then follow the remaining steps as in the scalar case \cite{Xianyu:2022jwk}. The third is to use the relation we have established between the scalar and fermion cases \eqref{eq: fermion_scalar_relation}. Since the scalar result is already known in \cite{Xianyu:2022jwk}, the fermion result can then be obtained directly by applying the corresponding overall time-derivative operators. Finally, one may also use dispersion relations. Since the full signal expressions are now available, the background contribution can in principle be reconstructed from the signal part through a suitable dispersion relation \cite{Liu:2024xyi}. In this paper, we present the calculation of the background contribution using the first method, namely the MB representation in Appendix \ref{app: MP_background}. The other methods and the comparison between them, are left for future investigation.
\end{itemize}
\section{Towards the three-point function}
We have derived the full expressions for the signals of the four-point functions with a fermionic bubble. In this section, we now move to the corresponding three-point functions, which can be obtained by taking suitable limits, such as $k_4\to0$. These are more directly relevant for phenomenology and observations, especially in the presence of Yukawa type interactions.
\subsection{Cancellation of the divergence} \label{sec: cancell_div}
Let us begin with the expression in \eqref{spectral_Signals}. Taking the three-point limit, $k_4 \to 0$, is equivalent to sending $r_2 \to 1$. However, as can be seen explicitly, both the non-local \eqref{eq: T_NS} and local signals \eqref{eq: T_LS} diverge in this limit, since $r_i = 1$ is the singular point of the hypergeometric functions. 
Quite non-trivially, the divergences cancel exactly once the two different types of signals are combined. For example,
\begin{align}
    \mathcal{T}^{\rm{3pt}}_{\nu}(r_1)&\equiv \lim_ {r_2\to1} \bigg[\mathcal{T}^{\rm{NS}}_{\nu}(r_1,r_2)+ \mathcal{T}^{\rm{LS}}_{\nu}(r_1,r_2)\bigg]\nonumber\\
    &=\frac{\pi^{3/2}e^{-i\pi (q_{12}/2+\nu)}}{2^{1/2+q_2+\nu}}\frac{ie^{i\pi(q_1+\nu)}-1}{\cos(2\pi q_2)+\cos(2\pi\nu)}\,\,\Gamma\left[\begin{array}{c}
\frac{5}{2}+q_1+\nu \\
\frac{3}{2}-q_2-\nu,-\frac{3}{2}-q_2+\nu
\end{array}\right]\nonumber\\
&\times\frac{r_1^{\frac{5}{2}+q_1+\nu}}{\textcolor{red3}{\Gamma(3+q_2)}}\,\,\pregFq{2}{1}{\frac{5}{4}+\frac{q_2}{2}+\frac{\nu}{2},\frac{7}{4}+\frac{q_2}{2}+\frac{\nu}{2}}{1+\nu}{r_1^2}\,, \label{eq:T3pt}
\end{align}
where the divergent terms cancel exactly, leaving this finite expression. The non-local and local CC signals are indistinguishable at the three-point level, as they now depend only on the oscillation with respect to $r_1$, and the final three-point signals are 
\begin{align}
    \mathcal{I}^{q_1q_2}_{\text{signal}}(r_1)&\equiv \lim_ {r_2\to1} \bigg[\mathcal{I}_{\rm{NS}}^{q_1q_2}(r_1,r_2)+ \mathcal{I}_{\rm{LS}}^{q_1q_2}(r_1,r_2)\bigg]\nonumber\\
    &=\sum_{\sa=\pm1}\sum_{k=0}^{\infty}\left[\rho_{M}\cdot\mathcal{T}^{\rm{3pt}}_{2k+\frac{5}{2}+i\sa M}(r_1)+\rho_{\delta}\cdot\mathcal{T}^{\rm{3pt}}_{2k+\frac{3}{2}+i\sa \delta}(r_1)\right]+\text{c.c.}\,.\label{fermion_bispectrum}
\end{align}
We can smoothly reduce the four-point signal to the three-point one. The reason is simple, the correlators should not contain any folded poles assuming the Bunch-Davies initial condition. Since the background part has a different dependence on $r_1$, cancellations must take place between the non-local and local signal parts.
\subsection{Vanishing signals from Yukawa coupling} \label{sec: yukawa}
When it comes to the phenomenology of fermions in correlators, one of the most interesting couplings to consider is the Yukawa-type interaction. This is because the Yukawa coupling is renormalisable and is not suppressed by the cutoff scale, and therefore typically provides the leading contribution compared to higher dimensional operators in the effective field theory (EFT) expansion, unless it is forbidden by symmetries. Furthermore, in the so-called Higgs-modulated reheating scenario~\cite{Cui:2021iie,Lu:2019tjj}, Yukawa-type interactions between the Higgs field and Standard Model (SM) fermions provide a concrete example of fermionic effects in cosmological correlators.

For example, consider the three-point vertex between the scalar $\varphi$ and fermions $\psi_i$
\begin{align}
    \mathcal{L}_{\rm{int,3}}=g a^4\varphi\left(\psi_1\psi_2+\psi_2^\dagger\psi_1^\dagger\right)\,,
\end{align}
it is straightforward to count the powers of $\tau_i$ in the seed integrals associated with this vertex. Here the scale factor $a^4$ contributes a factor of $\tau^{-4}$, while the external propagator contributes a factor of $(1 \pm i k \tau)$. The overall dependence on $\tau$ is either $\tau^{-4}$ or $\tau^{-3}$, corresponding to $q_2 = -4$ or $-3$. However, there is also a factor of $\textcolor{red3}{\Gamma(3+q_2)}$ in the denominator of \eqref{eq:T3pt}, which we have intentionally highlighted in \textcolor{red3}{red}. This factor diverges in both cases $q_2=-3$ and $q_2=-4$, giving rise to a surprising result: 
\textit{No cosmological collider signals in fermionic bubble loops with Yukawa couplings.}\footnote{Although this coupling is IR divergent and some terms depend on the IR cut-off $\tau_0$, by ``vanishing'' we mean that the usual cosmological collider signals are absent, independent of $\tau_0$.}

In fact, this $\Gamma(3+q_2)$ factor already appears at the tree level of scalar exchange through the same reason as in \eqref{eq:T3pt}. For example in the closed-form formulae for the three-point function (see the Summary of \cite{Qin:2023ejc}). To understand why this happens, it is useful to begin with the tree-level example.
\subsubsection*{Tree-level toy example}
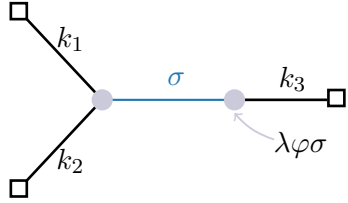
\begin{figure}[htp]
\centering
{\begin{tikzpicture}[baseline={([yshift=-.5ex]current bounding box.center)}, line width=1. pt, scale=3.5]
    \draw[blue3, thick] (0, 0) -- (0.5,0);
    \draw[black] (0, 0) -- (-0.285, 0.305);
    \draw[black] (0, 0) -- (-0.285, -0.305);
    \node[draw, rectangle, minimum width=6.0pt, minimum height=6.0pt, inner sep=0pt, anchor=south east] at (-0.28,0.3) {};
    \node[draw, rectangle, minimum width=6.0pt, minimum height=6.0pt, inner sep=0pt, anchor=north east] at (-0.28,-0.3) {};
    \draw[black] (0.5, 0) -- (0.85, 0);
    \draw[draw=lightgray2, fill=lightgray2] (0, 0) circle (.035cm);
    \draw[draw=lightgray2, fill=lightgray2] (0.5, 0) circle (.035cm);
    \node[draw, rectangle, minimum width=6.0pt, minimum height=6.0pt, inner sep=0pt, anchor=south west] at (0.85,-0.8pt) {};
    \node at (-0.12, 0.23) {$k_1$};
    \node at (-0.12, -0.23) {$k_2$};
    \node at (0.72, 0.08) {$k_3$};
    \draw[->,color=lightgray2,line width=0.7pt] (0.65,-0.15) to[bend left=15] (0.5,-0.05);
    \node at (0.28,0.08){\textcolor{blue3}{$\sigma$}};
    \node at (0.75,-0.17){$\lambda\varphi\sigma$};
\end{tikzpicture} 
    \caption{The tree-level three-point correlator from massive scalar exchange with a $\varphi$$\sigma$ coupling.\label{figure: tree_example}}}
\end{figure}
As shown in the previous section, bubble diagrams can be decomposed into a series of corresponding tree-level diagrams with massive propagators of different masses. For the Yukawa-type fermionic bubble, the relevant tree-level building blocks are those involving the $\lambda\varphi\sigma$ coupling. To understand the disappearance of the CC signals, we consider the general diagram shown in Figure \ref{figure: tree_example}, where the two-point vertex comes from the coupling $\lambda\varphi\sigma$. The detailed form of the remaining interaction is irrelevant for the present discussion, since the factor $\Gamma(3+q_2)$ depends only on the power of this two-point interaction. 

Previous work \cite{Qin:2023ejc} calculated the seed integral for the general tree-level bispectra using bootstrap equations. In the case considered here, the two-point vertex may have an IR divergence, which is not considered in the previous boundary differential equations. Instead, in this work we evaluate the diagram directly from the bulk time integrals.

In the following discussion, we will restore the SK indices explicitly on the massive propagators, with $\mathcal{G}_{1,\sa\sb}$ and $\mathcal{G}_{2,\sa\sb}$ denoting the propagators of masses $m_1$ and $m_2$, respectively. These propagators satisfy the differential equations
\begin{align}
 \mathcal{O}_{n,\tau_j}\,\mathcal{G} _{n,\, \sa\sb}\left(k ,\tau_1, \tau_2\right)=- i\sa \tau_1^2 \tau_2^2 \delta_{\sa\sb}\,\delta\left(\tau_1-\tau_2\right)\,, \label{eq: eom_O}
\end{align}
where we have defined the equation of motion operator as $\mathcal{O}_{n,\tau_j}\equiv  \tau_j^2 \partial_{\tau_j}^2-2 \tau_j \partial_{\tau_j}+k^2 \tau_j^2+m_n^2$, with both $j,n$ taking values in $\{1,2\}$. For the non-time-ordered $\mathcal{G}_{n,-+}$, it is given explicitly in \eqref{eq:scalar_wightman}. These equation of motion operators have the following properties which will be used in the discussion later:
\begin{enumerate}[label=(\roman*), ref=(\roman*)]
    \item  They are symmetric with respect to the two time variables: \label{O_property1}
\begin{align}
    \qquad\big(\mathcal{O}_{n,\tau_1}-\mathcal{O}_{n,\tau_2}\big)\,\mathcal{G}_{n,\sa\sb}(k,\tau_1,\tau_2)=0\,.
\end{align}
    \item The difference between two operators is simply the difference between the masses: \label{O_property2}
\begin{align}
    \qquad\big(\mathcal{O}_{1,\tau_j}-\mathcal{O}_{2,\tau_j}\big)=m_1^2-m_2^2\,.
\end{align}
    \item For well behaved functions $f(\tau)$ and $g(\tau)$ in $\tau\to-\infty$, it follows \label{O_property3}
\begin{align}
    \int^{\tau_0} d\tau a^4(\tau)\,\big[\mathcal{O}_{n,\tau}f(\tau)\big]\, \,g(\tau)= \int^{\tau_0} d\tau a^4(\tau)\big[\mathcal{O}_{n,\tau}\,g(\tau)\big]\, f(\tau)+\frac{1}{\tau^2}\big[f'g-fg'\big]\Big|_{\tau_0}\,,
\end{align}
which follows straightforwardly from integration by parts. By ``well behaved'', we mean that these functions decay sufficiently rapidly so that the boundary term at the far past can be safely neglected.
\end{enumerate}
\paragraph{Mixed propagator identity}
To calculate the bispectrum in Figure \ref{figure: tree_example}, we first derive a useful result for the mixed propagator associated with the quadratic coupling $\lambda\varphi\sigma$, which is
\begin{align}
    \sum_{\sc=\pm}\bigg(
\raisebox{8 pt}{\begin{tikzpicture}[baseline={(current bounding box.center)}, line width=1. pt, scale=3.0]
    \draw[blue3, thick] (0.0,0) -- (0.35,0);
    \draw[red3, thick] (0.35,0) -- (0.7,0);
    \draw[draw=lightgray2, fill=lightgray2] (0, 0) circle (.035cm);
    \draw[draw=lightgray2, fill=lightgray2] (0.35, 0) circle (.035cm);
    \draw[draw=lightgray2, fill=lightgray2] (0.7, 0) circle (.035cm);
   \node[scale=0.3, transform shape] at (0,0.11) {$\sa$};
  \node[scale=0.3, transform shape] at (0.7,0.11) {$\sb$};
   \node[scale=0.3, transform shape] at (0.35,0.11) {$\sc$};
\end{tikzpicture}}
\bigg)
&\equiv\sum_{\sc=\pm}i\sc\int d\tau_3\, \mathcal{G}_{1,\sa\sc}(k,\tau_1,\tau_3)a^4(\tau_3)\mathcal{G}_{2,\sc\sb}\,(k,\tau_3,\tau_2)\,,
\end{align}
then we can apply the property \ref{O_property2} of the equation of motion operator to  both sides, yielding
\begin{align}
&(m_1^2-m_2^2)\,\sum_{\sc=\pm}\bigg(
\raisebox{8 pt}{\begin{tikzpicture}[baseline={(current bounding box.center)}, line width=1. pt, scale=3.0]
    \draw[blue3, thick] (0.0,0) -- (0.35,0);
    \draw[red3, thick] (0.35,0) -- (0.7,0);
    \draw[draw=lightgray2, fill=lightgray2] (0, 0) circle (.035cm);
    \draw[draw=lightgray2, fill=lightgray2] (0.35, 0) circle (.035cm);
    \draw[draw=lightgray2, fill=lightgray2] (0.7, 0) circle (.035cm);
   \node[scale=0.3, transform shape] at (0,0.11) {$\sa$};
  \node[scale=0.3, transform shape] at (0.7,0.11) {$\sb$};
   \node[scale=0.3, transform shape] at (0.35,0.11) {$\sc$};
\end{tikzpicture}}
\bigg)\nonumber\\
&=\sum_{\sc=\pm}i\sc\int d\tau_3\, \mathcal{G}_{1,\sa\sc}(k,\tau_1,\tau_3)a^4(\tau_3)\Big(\mathcal{O}_{1,\tau_3}-\mathcal{O}_{2,\tau_3}\Big)\mathcal{G}_{2,\sc\sb}\,(k,\tau_3,\tau_2)\,,\nonumber\\
    &=\sum_{\sc=\pm}i\sc\int d\tau_3\, a^4(\tau_3)\bigg\{\Big[\mathcal{O}_{1,\tau_3}\mathcal{G}_{1,\sa\sc}(k,\tau_1,\tau_3)\Big]\mathcal{G}_{2,\sc\sb}\,(k,\tau_3,\tau_2)-\mathcal{G}_{1,\sa\sc}(k,\tau_1,\tau_3)\Big[\mathcal{O}_{2,\tau_3}\mathcal{G}_{2,\sc\sb}\,(k,\tau_3,\tau_2)\Big]\bigg\}\nonumber\\
    &=\mathcal{G}_{2,\sa\sb}(k,\tau_1,\tau_2)-\mathcal{G}_{1,\sa\sb}(k,\tau_1,\tau_2)\,.
\end{align}
In the second step, we used the property \ref{O_property3}, while the final step follows from \eqref{eq: eom_O}. Diagrammatically, this relation can be represented as \footnote{This is the dS generalisation of the flat-space result, which simply takes the form $\frac{m_1^2-m_2^2}{(-p^2-m_1^2)(-p^2-m_2^2)}=\left(\frac{1}{-p^2-m_1^2}-\frac{1}{-p^2-m_2^2}\right)$. }
\begin{align}
    \sum_{\sc=\pm}\bigg(
\raisebox{8 pt}{\begin{tikzpicture}[baseline={(current bounding box.center)}, line width=1. pt, scale=3.0]
    \draw[blue3, thick] (0.0,0) -- (0.35,0);
    \draw[red3, thick] (0.35,0) -- (0.7,0);
    \draw[draw=lightgray2, fill=lightgray2] (0, 0) circle (.035cm);
    \draw[draw=lightgray2, fill=lightgray2] (0.35, 0) circle (.035cm);
    \draw[draw=lightgray2, fill=lightgray2] (0.7, 0) circle (.035cm);
   \node[scale=0.3, transform shape] at (0,0.11) {$\sa$};
  \node[scale=0.3, transform shape] at (0.7,0.11) {$\sb$};
   \node[scale=0.3, transform shape] at (0.35,0.11) {$\sc$};
\end{tikzpicture}}
\bigg)=\frac{1}{m_1^2-m_2^2}\,\bigg(\raisebox{8 pt}{\begin{tikzpicture}[baseline={(current bounding box.center)}, line width=1. pt, scale=3.0]
    \draw[red3, thick] (0.0,0) -- (0.5,0);
    \draw[draw=lightgray2, fill=lightgray2] (0, 0) circle (.035cm);
    \draw[draw=lightgray2, fill=lightgray2] (0.5, 0) circle (.035cm);
   \node[scale=0.3, transform shape] at (0,0.11) {$\sa$};
  \node[scale=0.3, transform shape] at (0.5,0.11) {$\sb$};
\end{tikzpicture}}-\raisebox{8 pt}{\begin{tikzpicture}[baseline={(current bounding box.center)}, line width=1. pt, scale=3.0]
    \draw[blue3, thick] (0.0,0) -- (0.5,0);
    \draw[draw=lightgray2, fill=lightgray2] (0, 0) circle (.035cm);
    \draw[draw=lightgray2, fill=lightgray2] (0.5, 0) circle (.035cm);
   \node[scale=0.3, transform shape] at (0,0.11) {$\sa$};
  \node[scale=0.3, transform shape] at (0.5,0.11) {$\sb$};
\end{tikzpicture}}\bigg)\,. \label{eq:mix_factor}
\end{align}
One may be concerned about the boundary terms from the integration by parts. The term at $\tau \to -\infty$ vanishes due to the Bunch-Davies vacuum, with the standard $i\epsilon$-prescription the integrand decays sufficiently fast. The contribution from the IR boundary at $\tau \to \tau_0$ also vanishes, namely,
\begin{align}
    &\sum_{\sc=\pm}i\sc\,{\tau_0^{-2}}\, \big[\mathcal{G}'_{1,\sa\sc}(k,\tau_1,\tau_3)\mathcal{G}_{2,\sc\sb}\,(k,\tau_3,\tau_2)-\mathcal{G}_{1,\sa\sc}(k,\tau_1,\tau_3)\mathcal{G}'_{2,\sc\sb}\,(k,\tau_3,\tau_2)\big]\bigg|_{\tau_0}=0\,,
\end{align}
this comes from the fact that, once one end reaches the boundary, the time-ordering becomes unique, which gives
\begin{align}
    \mathcal{G}'_{n,\sa+}(k,\tau,\tau_0)=\mathcal{G}'_{n,\sa-}(k,\tau,\tau_0)\,,\qquad\mathcal{G}_{n,\sa+}(k,\tau,\tau_0)=\mathcal{G}_{n,\sa-}(k,\tau,\tau_0)\,.
\end{align}
Since the terms inside the bracket are the same for $\sc=\pm$, and can therefore be taken outside the summation. The remaining overall factor of $\sc$ then makes the sum vanish. 

Now we can apply the property \eqref{eq:mix_factor} of the mixed propagator, setting $m_2=0$ for the inflaton field, $m_1=m$ for the $\sigma$ field and taking $\tau_2\to\tau_0$. The bispectrum shown in Figure \ref{figure: tree_example} then can be written as 
\begin{align}
\raisebox{2 pt}{\begin{tikzpicture}[baseline={(current bounding box.center)}, line width=1. pt, scale=2.5]
    \draw[blue3, thick] (0, 0) -- (0.5,0);
    \draw[black] (0, 0) -- (-0.285, 0.305);
    \draw[black] (0, 0) -- (-0.285, -0.305);
    \node[draw, rectangle, minimum width=5.0pt, minimum height=5.0pt, inner sep=0pt, anchor=south east] at (-0.28,0.3) {};
    \node[draw, rectangle, minimum width=5.0pt, minimum height=5.0pt, inner sep=0pt, anchor=north east] at (-0.28,-0.3) {};
    \draw[black] (0.5, 0) -- (0.85, 0);
    \draw[draw=lightgray2, fill=lightgray2] (0, 0) circle (.035cm);
    \draw[draw=lightgray2, fill=lightgray2] (0.5, 0) circle (.035cm);
    \node[draw, rectangle, minimum width=5.0pt, minimum height=5.0pt, inner sep=0pt, anchor=south west] at (0.85,-1.0pt) {};
    \draw[->,color=lightgray2,line width=0.7pt] (0.65,-0.15) to[bend left=15] (0.5,-0.05);
    \node at (0.28,0.08){\textcolor{blue3}{$\sigma$}};
    \node at (0.75,-0.17){$\lambda\varphi\sigma$};
\end{tikzpicture} }
=\frac{\lambda}{m^2}\,\left(~~\raisebox{2 pt}{\begin{tikzpicture}[baseline={(current bounding box.center)}, line width=1. pt, scale=2.5]
    \draw[black] (0, 0) -- (-0.285, 0.305);
    \draw[black] (0, 0) -- (-0.285, -0.305);
    \node[draw, rectangle, minimum width=5.0pt, minimum height=5.0pt, inner sep=0pt, anchor=south east] at (-0.28,0.3) {};
    \node[draw, rectangle, minimum width=5.0pt, minimum height=5.0pt, inner sep=0pt, anchor=north east] at (-0.28,-0.3) {};
    \draw[black] (0.0, 0) -- (0.5, 0);
    \draw[draw=lightgray2, fill=lightgray2] (0, 0) circle (.035cm);
    \node[draw, rectangle, minimum width=5.0pt, minimum height=5.0pt, inner sep=0pt, anchor=south west] at (0.5,-1.0pt) {};
\end{tikzpicture}}~~~-\raisebox{2 pt}{\begin{tikzpicture}[baseline={(current bounding box.center)}, line width=1. pt, scale=2.5]
    \draw[black] (0, 0) -- (-0.285, 0.305);
    \draw[black] (0, 0) -- (-0.285, -0.305);
    \node[draw, rectangle, minimum width=5.0pt, minimum height=5.0pt, inner sep=0pt, anchor=south east] at (-0.28,0.3) {};
    \node[draw, rectangle, minimum width=5.0pt, minimum height=5.0pt, inner sep=0pt, anchor=north east] at (-0.28,-0.3) {};
    \draw[blue3] (0.0, 0) -- (0.5, 0);
    \draw[draw=lightgray2, fill=lightgray2] (0, 0) circle (.035cm);
    \node[draw, rectangle, minimum width=5.0pt, minimum height=5.0pt, inner sep=0pt, anchor=south west] at (0.5,-1.0pt) {};
    \node at (0.28,0.08){\textcolor{blue3}{$\sigma$}};
\end{tikzpicture}}~~\right)\,,\label{eq:mix_diagram}
\end{align}
which reduces to the difference of two contact diagrams. Apparently, the first diagram carries no information about the heavy field and therefore cannot encode any CC signals. 

The second is the three-point function $\langle\varphi\varphi\sigma\rangle$, whose explicit form depends on the couplings. 
Let us start with the IR convergent three-point couplings and take $a^2\varphi'^2\sigma$ as an example, since $\varphi'\sim\tau$ and $\sigma\sim\tau^{3/2}$ at the late time, the contact three-point function then is  
\begin{align}
    \raisebox{2 pt}{\begin{tikzpicture}[baseline={(current bounding box.center)}, line width=1. pt, scale=2.5]
    \draw[black] (0, 0) -- (-0.285, 0.305);
    \draw[black] (0, 0) -- (-0.285, -0.305);
    \node[draw, rectangle, minimum width=5.0pt, minimum height=5.0pt, inner sep=0pt, anchor=south east] at (-0.28,0.3) {};
    \node[draw, rectangle, minimum width=5.0pt, minimum height=5.0pt, inner sep=0pt, anchor=north east] at (-0.28,-0.3) {};
    \draw[blue3] (0.0, 0) -- (0.5, 0);
    \draw[draw=lightgray2, fill=lightgray2] (0, 0) circle (.035cm);
    \node[draw, rectangle, minimum width=5.0pt, minimum height=5.0pt, inner sep=0pt, anchor=south west] at (0.5,-1.0pt) {};
      \draw[->,color=lightgray2,line width=0.7pt] (0.2,-0.15) to[bend left=15] (0.05,-0.05);
    \node at (0.4,-0.17){$a^2 \varphi'^2\sigma$};
\end{tikzpicture}}\sim \left(\sigma_{s}(\tau_0)\int^{\tau_0} d\tau e^{ik_{12}\tau}\sigma^*_{s}(\tau)\right)\sim \mathcal{O}({\tau}_0^{4})=0\,,
\end{align}
that vanishes in the boundary limit $\tau_0 \to 0$. This is true for any IR-convergent coupling, because  external massive propagators contribute a factor of $\sigma_s(\tau_0)$, and these heavy fields decay rapidly outside the horizon. 

What about cubic interactions that are also IR divergent? The only possibility is $a^4 \varphi^2 \sigma$, which was computed in \cite{Cespedes:2025ple} using wavefunction coefficients for light mass $m<3/2H$ and was found to give a constant contribution. Now let us consider the case of heavy fields within the SK formalism. The scalar mode function $\varphi_k$ and the heavy field mode function $\sigma_k$ are given by
    \begin{align}
    \varphi_k(\tau)&=\frac{1}{\sqrt{2k^3}}(1+ik\tau)e^{-ik \tau}\,,\\
    \sigma_k (\tau)&=-i\,e^{-\frac{\pi}{2} \mu+i \frac{\pi}{4}} \frac{\sqrt{\pi}}{2} (-\tau)^{3 / 2} H_{i \mu}^{(1)}(-k \tau)\,,
\end{align}
and this contact three-point diagram with the interaction $g a^4\varphi^2\sigma$ reads
\begin{align}
    \raisebox{2 pt}{\begin{tikzpicture}[baseline={(current bounding box.center)}, line width=1. pt, scale=2.5]
    \draw[black] (0, 0) -- (-0.285, 0.305);
    \draw[black] (0, 0) -- (-0.285, -0.305);
    \node[draw, rectangle, minimum width=5.0pt, minimum height=5.0pt, inner sep=0pt, anchor=south east] at (-0.28,0.3) {};
    \node[draw, rectangle, minimum width=5.0pt, minimum height=5.0pt, inner sep=0pt, anchor=north east] at (-0.28,-0.3) {};
    \draw[blue3] (0.0, 0) -- (0.5, 0);
    \draw[draw=lightgray2, fill=lightgray2] (0, 0) circle (.035cm);
    \node[draw, rectangle, minimum width=5.0pt, minimum height=5.0pt, inner sep=0pt, anchor=south west] at (0.5,-1.0pt) {};
        \node at (-0.07, 0.23) {$k_1$};
    \node at (-0.07, -0.23) {$k_2$};
    \node at (0.3, 0.08) {$s$};
      \draw[->,color=lightgray2,line width=0.7pt] (0.2,-0.15) to[bend left=15] (0.05,-0.05);
    \node at (0.4,-0.17){$a^4 \varphi^2\sigma$};
\end{tikzpicture}}&= \frac{g}{2k_1^3k_2^3}\,\Im\left[\sigma^*_{s}(\tau_0)\,\int^{\tau_0} \frac{d\tau}{\tau^4} (1+ik_1\tau)(1+ik_2\tau)e^{-ik_{12}\tau}\sigma_{s}(\tau)\right]\nonumber\\
&=\frac{g}{2k_1^3k_2^3}\times\frac{1}{2\mu}\Im\left[\int^{\tau_0}\frac{d\tau}{\tau^4}(-\tau)^{3/2+i\mu}(-\tau_0)^{3/2-i\mu}\right]+\mathcal{O}(\tau_0)\nonumber\\
&=\frac{1}{4k_1^3k_2^3}\times\frac{g}{m^2}\,,
\end{align}
where we expanded the massive fields around IR and neglected higher order terms in $\tau_0$ which vanish at the boundary. This is simply a constant and therefore does not contain any oscillatory CC signals. In summary, we have confirmed that, in the general case, the diagram in Figure \ref{figure: tree_example} does not produce any CC signal when the quadratic interaction is $\lambda \varphi\sigma$. 
\paragraph{Field redefinition} We have shown, by explicitly checking the bulk time integrals, that  CC signals vanish for this diagram. This can also be understood more directly through a field redefinition \footnote{We thank David Stefanyszyn for related discussions.}. For example, consider the following action
\begin{align}
    S[\varphi,\sigma]=\int d^4x\sqrt{-g}\left[-\frac{1}{2}(\partial_\mu\varphi)^2-\frac{1}{2}(\partial_\mu\sigma)^2-\frac{1}{2}m^2\sigma^2+\lambda\varphi\sigma+g_3\varphi^2\sigma\right]\,,
\end{align}
the specific form of the cubic interaction is not essential to our discussion, we use $\varphi^2 \sigma$ merely as an illustrative example. We can take the transformation that 
\begin{align}
    \widetilde{\sigma}=\sigma-\frac{\lambda}{m^2}\varphi\,,
\end{align}
under which the action becomes
\begin{align}
    S[\varphi,\widetilde{\sigma}]=&\int d^4x\sqrt{-g}\left[
-\frac{1}{2}(\partial_\mu\varphi)^2
-\frac{1}{2}(\partial_\mu\widetilde{\sigma})^2
-\frac{1}{2}m^2\widetilde{\sigma}^2
+\cancel{\frac{\lambda}{m^2}\widetilde{\sigma}\,\Box\varphi}
+g_3\varphi^2\widetilde{\sigma}
+\frac{g_3\lambda}{m^2}\varphi^3
\right]\nonumber\\
&+O(\lambda^2)\,,
\end{align}
where we have used integration by parts to rewrite the quadratic interaction as $\widetilde{\sigma}\Box\varphi$. At the tree level, one can then use the on-shell equation of motion for the inflaton $\Box\varphi = \mathcal{O}(\lambda)$, from which it follows that no massive field enters at $\mathcal{O}(\lambda)$, and the only relevant coupling after this redefinition is the self-interaction $\varphi^3$.  This is precisely the result we found above from the relation \eqref{eq:mix_diagram}.

Having understood the tree-level example, one immediately sees why the fermionic bubble with Yukawa interactions also gives no CC signal. Using the decomposition relation \eqref{fermionDecom}, the fermionic bubble contribution can be written as a sum of tree-level CC diagrams of exactly the type discussed above. Since each of these vanishes, the total CC signal in \eqref{fermion_bispectrum} also vanishes. 

A natural next question about the phenomenology of fermions in the cosmological collider is that: \textit{can fermion signals still come from the Yukawa interactions?} At least two important possibilities suggest themselves. First, the vanishing of the fermionic CC signal from the bubble diagram is closely tied to the decomposition relation, which relies on the full de Sitter isometry. If one introduces, for example, a fermionic chemical potential through $\bar{\Psi}\gamma^0\gamma^5\Psi$, the decomposition relation \eqref{fermionDecom} no longer holds, and nonvanishing contributions may then arise. Second, one may consider the triangle loop with three Yukawa vertices. Since our decomposition applies only to the bubble diagram, such triangle diagrams could in principle generate a CC signal. We leave a detailed analysis of these two possibilities to future work.

\section{Conclusions}\label{sec: conclusion}
\textit{Fermions matter}: their rich phenomenology in cosmological collider physics has long attracted significant interest. However, since fermionic contributions to cosmological correlators begin at the loop level, exact signals have remained out of reach. In this work, we derive for the first time the complete analytical expressions for cosmological collider signals arising from the fermionic bubble loop with arbitrary couplings. We use two different methods: spectral decomposition expresses the bubble loop signals as sums of tree-level diagrams, while the MB representation reduces the momentum and time integrals to sums over residues from different pole types. Both methods give identical results for both the non-local and local signals. As a first application, we apply our formula to compute the bispectrum with the Yukawa coupling. Surprisingly, we find that all cosmological collider signals vanish, contrary to results reported in previous work based on approximate methods. The vanishing of the signal admits a natural interpretation that through spectral decomposition, the loop contribution can be mapped onto tree-level counterparts whose apparent signals are removed by field redefinitions.

\vskip 5pt
There are several interesting directions for future exploration:
\begin{enumerate}
    \item[$\bullet$] In this work, we devote most of our discussion to the signal contributions. We also present the background pieces in the Appendix \ref{app: MP_background} computed using the MB representation, though these deserve more thorough investigation. Several alternative methods are also available for calculating the background, including using the relation between fermionic and scalar bubbles \eqref{eq: fermion_scalar_relation} to borrow results from the scalar case, or computing directly from the fermionic spectral density. An important distinction from the signal part is that the background contributions are divergent, which requires careful treatment of renormalisation.
    \item[$\bullet$] As we have emphasised throughout, the bispectrum with one Yukawa coupling vanishes due to the spectral decomposition and field redefinitions. There are several ways to avoid this null result. One possibility is to introduce the chemical potential, which breaks the spectral relation we employed and also enhances the signals from fermionic loops. Another is to go beyond the bubble topology and consider other diagrams such as triangle loops. Both directions require new computational tools and open interesting avenues for further study.
    \item[$\bullet$] We have focused on spin $1/2$ fermions in this work. From a phenomenological perspective, spin $3/2$ fields such as the gravitino are also of considerable interest, particularly in supergravity scenarios where they naturally arise as the superpartners of the graviton. Compared to spin $1/2$ fields, gravitino loops involve richer tensor structures and may lead to qualitatively distinct signatures in cosmological correlators. With the existing formulae for higher-spin fermionic fields available \cite{Anninos:2025mje}, it would be interesting to extend our analysis to gravitino loops and study their signatures in cosmological collider physics. 
    \item[$\bullet$] The vanishing signal studied in this work arises from the Yukawa coupling, but the underlying mechanism is not restricted to this example. Similar arguments can also be applied to scalar bubbles with $\sigma^2\varphi$ couplings, as well as to two vertex topologies and other classes of interactions. A natural future direction is to formulate general rules that determine when cosmological collider signals vanish for arbitrary diagrams.
\end{enumerate}

We leave them for future work.
\paragraph{Acknowledgments.}
We would like to thank Sadra Jazayeri, Mang Hei Gordon Lee, Fumiya Sano, David Stefanyszyn, Alessandro Strumia, Xi Tong, Dong-Gang Wang, Xiangwei Wang,  Yi Wang and Zhong-Zhi Xianyu for helpful discussions.
S.A. is supported by the Japan Science and Technology Agency (JST) 
as part of Adopting Sustainable Partnerships for Innovative Research Ecosystem (ASPIRE), 
grant JPMJAP2318, and by JSPS KAKENHI Grant Number 26K07078. 
Z.Q. is supported by the NSFC through Grant No. 12275146, the National Key R\&D Program of China (2021YFC2203100), and the Dushi Program of Tsinghua University.
M.Y. is supported by IBS under the project code, IBS-R018-D3, and by JSPS Grant-in-Aid
for Scientific Research Number JP23K20843. Y.Z. is supported by IBS under the project code, IBS-R018-D3.
\newpage
\begin{appendix}
\setlength{\abovedisplayskip}{11pt}
\setlength{\belowdisplayskip}{11pt}

\section{Mathematical formulae}\label{App:MathFormulae}
In this Appendix, we collect some mathematical formulae used in the main text. Most of them can also be found in the handbook of Mathematical Functions \cite{NIST:DLMF}.
\subsection*{The relation between Whittaker-$W$ and Hankel function}
First, the Hankel functions of the first and second kinds can both be written in terms of the Whittaker-$W$ function with its first parameter set to zero,
\begin{align}
    &H^{(1)}_\mu(z)=e^{-(2\mu+1)\pi i/4}\left(\frac{2}{\pi z}\right)^{\frac{1}{2}}W_{0,\mu}(-2iz)~,\label{eq:H1toW}\\
    &H^{(2)}_\mu(z)=e^{(2\mu+1)\pi i/4}\left(\frac{2}{\pi z}\right)^{\frac{1}{2}}W_{0,\mu}(2iz)~. \label{eq:H2toW}
\end{align}
The fermion mode functions in \eqref{eq:Fermion_Modefunction} are written in terms of the Whittaker function with the first parameter $\pm{1}/{2}$. Using recurrence relations, they can in turn be rewritten in terms of Hankel functions. More explicitly, we can use the following relation:
\begin{align}
    2\mu\,W_{\kappa,\mu}(z)-\sqrt{z}\,W_{\kappa+\frac{1}{2},\mu+\frac{1}{2}}(z)+\sqrt{z}\,W_{\kappa+\frac{1}{2},\mu-\frac{1}{2}}(z)=0~,
\end{align}
take $\kappa=-1/2$, it reads
\begin{align}
    2\mu\,W_{-\frac{1}{2},\mu}(z)=\sqrt{z}\,W_{0,\mu+\frac{1}{2}}(z)-\sqrt{z}\,W_{0,\mu-\frac{1}{2}}(z)~,
\end{align}
using the relation (\ref{eq:H1toW}), we then arrive at
\begin{align}
    W_{-\frac{1}{2},\mu}(-2  i z)=\frac{1}{2\mu}\bigg[e^{(2\mu+1)\pi  i/4}\sqrt{\pi} z\, H^{(1)}_{\mu+\frac{1}{2}}(z)-e^{(2\mu-1)\pi  i/4}\sqrt{\pi}\,z\,H^{(1)}_{\mu-\frac{1}{2}}(z)\bigg]~.
\end{align}    
For the other mode function, with first parameter equal to $1/2$, we can use another identity
\begin{align}
    \left(\kappa-\mu-\frac{1}{2}\right)\sqrt{z}\,W_{\kappa-\frac{1}{2},\mu+\frac{1}{2}}(z)+2\mu W_{\kappa,\mu}(z)-\left(\kappa+\mu-\frac{1}{2}\right)\sqrt{z} W_{\kappa-\frac{1}{2},\mu-\frac{1}{2}}(z)=0~,
\end{align}
then take $\kappa={1}/{2}$, it reads
\begin{align}
    2\,W_{\frac{1}{2},\mu}(z)=\sqrt{z}\,W_{0,\mu+\frac{1}{2}}(z)+\sqrt{z}\,W_{0,\mu-\frac{1}{2}}(z)~,
\end{align}
applying \eqref{eq:H1toW} again yields
 \begin{align}
    W_{\frac{1}{2},\mu}(-2  i z)=\frac{1}{2}\bigg[e^{(2\mu-1)\pi  i/4}\sqrt{\pi}\,z\,H^{(1)}_{\mu-\frac{1}{2}}(z)+e^{(2\mu+1)\pi  i/4}\sqrt{\pi}\,z\,H^{(1)}_{\mu+\frac{1}{2}}(z)\bigg]~.
\end{align}   
These are relations between the Whittaker-$W$ and Hankel function used in \eqref{eq: WhittakerToHankel}.
\subsection*{The relation between Hankel function and other functions}
The Hankel functions are also related to the modified Bessel function of the second kind, $K_{\mu}$ through the following relation: 
\begin{subequations}
\begin{align}
    &K_\mu(z)= \frac{\pi}{2}\,i^{\mu+1}H^{(1)}_{\mu}(i z)\,, ~~~~~~~\qquad -\pi <\arg z \leq \frac{\pi}{2}\,,\\
    &K_\mu (z) =\frac{\pi}{2}\,(-i)^{\mu+1}H^{(2)}_{\mu}(-i z)\,, \qquad -\frac{\pi}{2}< \arg z \leq \pi\,.
\end{align} \label{eq: HankelToBesselK}
\end{subequations}
\subsection*{Integrals of products of Bessel functions}
We will use the following integration formula in the Fourier transform of the two-point functions later
\begin{align}
    \int_{0}^{\infty}\dd x\,x^{\nu+1}K_{\mu}(a x) K_{\mu}(b x) J_{\nu}(c x)=\frac{\sqrt{\pi}c^{\nu}\Gamma(\nu+\mu+1)\Gamma(\nu-\mu+1)}{2^{3/2}(ab)^{\nu+1}(u^2-1)^{\frac{1}{2}\nu+\frac{1}{4}}}\times P^{-\nu-\frac{1}{2}}_{\mu-\frac{1}{2}}(u)\,, \label{intFormula1}
\end{align}
in which $P^\nu_\mu$ denotes the associated Legendre-$P$ function, the variables satisfy the constraint $2abu=a^2+b^2+c^2$. The remaining conditions are 
\begin{align}
    \text{Re}(a+b)>|\text{Im}c|\,,\quad \text{Re}(\nu\pm \mu)>-1\,,\quad \Re \nu>-1\,.
\end{align}
Similarly, another useful integral identity involving the Legendre-$Q$ function is 
\begin{align}
    \int_{0}^{\infty}\dd x\,x^{\nu+1}K_{\mu}(a x) I_{\mu}(b x) J_{\nu}(c x)=\frac{(ab)^{-\nu-1}c^{\nu}e^{-(\nu+\frac{1}{2})\pi i}}{\sqrt{2\pi}(u^2-1)^{\frac{1}{2}\nu+\frac{1}{4}}}\times Q^{\nu+\frac{1}{2}}_{\mu-\frac{1}{2}}(u)\,, \label{intFormula2}
\end{align}
where $2abu=a^2+b^2+c^2$,  and $I_\mu(x)$ denotes the modified Bessel function of the first kind, defined by $I_{\mu}(x)=i^{-\mu} J_{\mu}(i x)$. The conditions for the above formula are
\begin{align}
    \text{Re}\,a>|\text{Re}b|+|\text{Im}c|\,,\quad \text{Re}(\nu+ \mu)>-1\,,\quad \text{Re}\,\nu>-1\,.
\end{align}
\subsection*{Momentum integrals}
The integral frequently used in Section \ref{sec: MBmethod} for evaluating momentum loop integrals is
\begin{align}\label{eq_loopint1}
L(s;a,b) \equiv \int \frac{\mathrm{d}^3 \mathbf{p}_1}{(2 \pi)^3}\,p_1^{-2a} p_2^{-2b}=\frac{s^{3-2a-2b}}{(4\pi)^{3/2}}
 \Gamma\left[
 \begin{matrix}
     a+b-\frac{3}{2},\frac32-a,\frac32-b\\
     3-a-b,a,b
 \end{matrix}
 \right],
\end{align}
and from which we can directly derive another way
\begin{align}\label{eq_loopint2}
\int \frac{\mathrm{d}^3 \mathbf{p}_1}{(2 \pi)^3}\,(\mathbf p_1\cdot\mathbf p_2) p_1^{-2a} p_2^{-2b}
=& \int \frac{\mathrm{d}^3 \mathbf{p}_1}{(2 \pi)^3}\,\frac12 (s^2-p_1^2-p_2^2) \times  p_1^{-2a} p_2^{-2b}\nonumber\\
=&~ \frac12 \left[ s^2L(s;a,b) -L(s;a-1,b) - L(s;a,b-1) \right]\nonumber\\
=&~-\frac{s^{5-2a-2b}}{(4\pi)^{3/2}}
 \Gamma\left[
 \begin{matrix}
     a+b-\frac{5}{2},\frac52-a,\frac52-b\\
     4-a-b,a,b
 \end{matrix}
 \right].
\end{align}
\subsection*{The transformation of hypergeometric functions}
Within the decomposition method, we also make use of the following transformation,
\begin{align}
    \frac{\sin\left(\frac{a-b}{2}\pi\right)}{\sqrt{\pi}}\pregFq{2}{1}{a,b}{\frac{a+b+1}{2}
    }{\frac{1-x}{2}}=\frac{2^{a-1}x^{-b}}{\Gamma(a)}\,\pregFq{2}{1}{\frac{b}{2},\frac{b+1}{2}}{\frac{2-a+b}{2}}{\frac{1}{x^2}}-\frac{2^{b-1}x^{-a}}{\Gamma(b)}\,\pregFq{2}{1}{\frac{a}{2},\frac{a+1}{2}}{\frac{2+a-b}{2}}{\frac{1}{x^2}}\,. \label{eq:hyperTrans}
\end{align}
\section{Detailed derivations}\label{App: derivations}
\subsection{Fourier transforms of propagators} \label{App: Fourier_Trans}
The expressions for the scalar two-point function in both momentum and position space are well known. However, explicit derivations of the corresponding Fourier transforms are rarely presented in the literature. As a warm-up, in this subsection we derive the scalar two-point function in position space by carrying out the Fourier transform explicitly, and then we will proceed in the next section to the transform of the fermionic bubble.
The scalar two-point function in momentum space is given by \eqref{eq:scalar_wightman}. Taking its Fourier transform yields
\begin{align}
    \int \frac{\dd^3 k}{(2\pi)^3}e^{i \mathbf{k}\cdot \mathbf{x}}  H^{(1)}_{i\mu}(-k\tau_1) H^{(2)}_{i\mu}(-k\tau_2)
    &=\int_{0}^{\infty}\dd k\,\frac{2 k^2}{\pi^4}\frac{\sin(k x)}{k x}K_{i\mu}(i k \tau_1) K_{i\mu}(-i k \tau_2)\nonumber\\
    &=\int_{0}^{\infty}\frac{\dd k}{\pi^4} \frac{{\sqrt{2\pi}}\,k^{2}}{\sqrt{kx}} J_{\frac{1}{2}}(k x) K_{i\mu}(i k \tau_1) K_{i\mu}(-i k \tau_2)\,,
\end{align}
where in the first line we have carried out the ($\theta$,\,$\varphi$) integrations and converted the Hankel functions into the modified Bessel function of the second kind using (\ref{eq: HankelToBesselK}). In the second line we have rearranged the expression into the form of (\ref{intFormula1}), then it can be directly calculated as 
\begin{align}
   \int \frac{\dd^3 k}{(2\pi)^3}e^{i \mathbf{k}\cdot \mathbf{x}}  H^{(1)}_{i\mu}(-k\tau_1) H^{(2)}_{i\mu}(-k\tau_2)
   =\frac{\Gamma\left(\frac{3}{2}-i\mu\right)\Gamma\left(\frac{3}{2}+i\mu\right)}{2\pi^3(\tau_1\tau_2)^{3/2}(1-{Z}_{-+})}\,\pFq{2}{1}{\frac{1}{2}-i\mu, \frac{1}{2}+i\mu}{2}{\frac{1+Z_{-+}}{2}}\,,
\end{align}
we have used the relation between Legendre $P$ and the hypergeometric function and $Z_{-+}$ is the  dimensionless embedding distance defined as:
\begin{align}
    Z_{-+}\equiv 1-\frac{|\bfx_1-\bfx_2|^2-(\tau_1-\tau_2-i\epsilon)^2}{2\tau_1\tau_2}\,, \label{eq: embedding_distance}
\end{align}
where we have made the $i\epsilon$-prescription explicit, and one can check it meets all the conditions required for applying (\ref{intFormula1}).  
Finally, we apply the identity of the hypergeometric function:
\begin{align}
    \pregFq{2}{1}{a,b}{c}{z}=(1-z)^{c-a-b}\pregFq{2}{1}{c-a,c-b}{c}{z}\,,
\end{align}
and the Fourier transform is expressed as 
\begin{align}
    \mathcal{G}_{i\mu,-+}(x_1,x_2)&\equiv\int \frac{\dd^3 k}{(2\pi)^3}e^{i \mathbf{k}\cdot \mathbf{x}} \langle 0| \sigma_{\mathbf{k}}(\tau_1)\sigma_{\mathbf{-k}}(\tau_2)| 0\rangle' \nonumber\\
    &= \frac{\Gamma\left(\frac{3}{2}-i\mu\right)\Gamma\left(\frac{3}{2}+i\mu\right)}{16\pi^2}\times\pFq{2}{1}{\frac{3}{2}-i\mu, \frac{3}{2}+i\mu}{2}{\frac{1+Z_{-+}}{2}}\,,\label{eq: scalar_Wightman}
\end{align}
which is exactly the Wightman function in the position space \cite{Spradlin:2001pw}. As for other types of propagators $\mathcal{G}_{i\mu,\sa\sb}$, they take the same form and we only need to replace the argument by the corresponding one $Z_{\sa\sb}$ as
\begin{align}
    &Z_{+-}=\left(Z_{-+}\right)^*=1-\frac{|\bfx_1-\bfx_2|^2-(\tau_1-\tau_2+i\epsilon)^2}{2\tau_1\tau_2}\,,\\
    &Z_{++}=\left(Z_{--}\right)^*=1-\frac{|\bfx_1-\bfx_2|^2-(|\tau_1-\tau_2|-i\epsilon)^2}{2\tau_1\tau_2}\,.
\end{align}
Throughout this paper, we take the $\mathcal{G}_{\nu,-+}$ component as our starting point in the discussion of bubble functions and omit SK indices for brevity. The remaining cases can be obtained straightforwardly, either by taking the appropriate complex conjugate or by following an analogous derivation.
\subsection{The relation between scalar and fermionic bubbles}\label{app: relations}
\subsection*{Relation I}
The fermionic bubble integrand in \eqref{eq:BubbleFunction_MomentumSpace} consists of two terms. The first, proportional to $(\mathbf{p}_1 \cdot \mathbf{p}_2)$, can be obtained straightforwardly by taking spatial derivatives of the scalar propagators, as follows
\begin{align}
    &-\frac{\pi^2}{8}(\tau_1 \tau_2)^4 \int\frac{\mathrm{d}^3\mathbf{p}_1}{(2\pi)^3} (\mathbf{p}_1 \cdot \mathbf{p}_2)\,H^{(1)}_{\nu^*_{1\lambda}}(-p_1\tau_1) H^{(2)}_{\nu^*_{1\lambda}}(-p_1\tau_2) H^{(1)}_{\nu_{2\lambda}}(-p_2\tau_1) H^{(2)}_{\nu_{2\lambda}}(-p_2\tau_2)\nonumber\\
    &={2\tau_1\tau_2}\int \dd^3 x\,e^{-i\mathbf{s}\cdot \bfx}\,\partial_i\mathcal{G}_{\nu_{1\lambda}^*}(x_1,x_2)\times\partial_i\mathcal{G}_{\nu_{2\lambda}}(x_1,x_2)\,.\label{eq:bubble_spatial}
\end{align}
For the first line, we make use of the following identity,
\begin{align}
	k\,e^{i\pi\frac{\nu_\lambda-\nu^*_{\lambda}}{2}}H_{\nu_\lambda}^{(1)}(-k \tau)
	= i\left(\frac{d}{d\tau}+\frac{\nu^*_{\lambda}}{\tau}\right)H_{\nu^*_{\lambda}}^{(1)}(-k\tau)\equiv i\,\mathscr{D}_{\tau,\nu^*_\lambda}H_{\nu^*_{\lambda}}^{(1)}(-k\tau)\,,
\end{align}
The first line of \eqref{eq:BubbleFunction_MomentumSpace} can then be expressed in terms of the time operator $\mathscr{D}_{\tau,\nu}$ acting on the scalar propagators as
\begin{align}
    &\frac{\pi^2}{8}(\tau_1 \tau_2)^4 \int\frac{\mathrm{d}^3\mathbf{p}_1}{(2\pi)^3}   p_1 p_2\,\,e^{(\nu_{1\lambda}-\nu_{2\lambda})i\pi}H^{(1)}_{\nu_{1\lambda}}(-p_1\tau_1) H^{(2)}_{\nu^*_{1\lambda}}(-p_1\tau_2) H^{(1)}_{\nu^*_{2\lambda}}(-p_2\tau_1) H^{(2)}_{\nu_{2\lambda}}(-p_2\tau_2)\nonumber\\
    &=\frac{(\tau_1 \tau_2)^4}{2}\int \dd^3 x\,e^{-i\mathbf{s}\cdot \bfx}\,\mathscr{D}_{\tau_1,\nu^*_{1\lambda}}\left[(\tau_1\tau_2)^{-\frac{3}{2}}\mathcal{G}_{\nu_{1\lambda}^*}(x_1,x_2)\right]\times\mathscr{D}_{\tau_1,\nu_{2\lambda}}\left[(\tau_1\tau_2)^{-\frac{3}{2}}\mathcal{G}_{\nu_{2\lambda}}(x_1,x_2)\right]\,,\label{eq: bubble_time}
\end{align}
combining (\ref{eq: bubble_time}) and (\ref{eq:bubble_spatial}), we obtain the Fourier transform of the fermionic bubble function, expressed as a set of operators acting on the product of scalar propagators: 
\begin{align}
    \int\frac{\dd^3 s}{(2\pi)^3}\,e^{i\mathbf{s}\cdot \bfx}\,\mathcal{Q}_{-+}(s,\tau_1,\tau_2)
    =&\sum_{\lambda=\pm}{2(\tau_1 \tau_2)^4}\Bigg\{(\tau_1\tau_2)^{-3}\partial_i\mathcal{G}_{\nu_{1\lambda}^*}(x_1,x_2)\times\partial_i\mathcal{G}_{\nu_{2\lambda}}(x_1,x_2)\nonumber\\
    -&\mathscr{D}_{\tau_1,\nu^*_{1\lambda}}\left[(\tau_1\tau_2)^{-\frac{3}{2}}\mathcal{G}_{\nu_{1\lambda}^*}(x_1,x_2)\right]\,\mathscr{D}_{\tau_1,\nu_{2\lambda}}\left[(\tau_1\tau_2)^{-\frac{3}{2}}\mathcal{G}_{\nu_{2\lambda}}(x_1,x_2)\right]\Bigg\}\,. \label{eq: Q_functionofScalar}
\end{align}
Plugging into the expression of scalar propagator (\ref{eq: scalar_Wightman}) and sum over the helicities, we finally get the bubble function in position space as 
\begin{align}
    \mathcal{Q}(x_1,x_2)&\equiv\int\frac{\dd^3 s}{(2\pi)^3}\,e^{i\mathbf{s}\cdot \bfx}\,\mathcal{Q}_{-+}(s,\tau_1,\tau_2)=-4\Big[f_{m_1}(Z)f_{m_2}(Z)+g_{m_1}(Z)g_{m_2}(Z)\Big]\,, \label{eq: Q_position}
\end{align}
where $f_m(z)$ and $g_m(z)$ are defined in \eqref{def: fm} and \eqref{def: gm} respectively.
This result agrees exactly with the expressions obtained in previous works \cite{Cui:2021iie, Chen:2018xck}. To derive \eqref{eq: Q_position} from \eqref{eq: Q_functionofScalar}, one needs to use suitable transformations of hypergeometric functions, and the equivalence can be checked numerically.
\subsubsection*{Relation II}
Through the above relation, we can express the fermionic bubble as operators acting on the scalar propagators. However, because the time operators act on each Hankel function individually, they cannot be pulled outside the Fourier transform. As a result, the right-hand side of \eqref{eq: Q_functionofScalar} is not written in the form of scalar bubbles. Instead, we can introduce another class of operators that act on the entire $\tau_i$-dependent integrand, which gives
\begin{align}
    &\left(\partial_{\tau_i}+\frac{1-i\lambda(m_1+m_2)}{\tau_i}\right)H^{(j)}_{1/2-i\lambda m_2}(-p_2\tau_i)H^{(j)}_{1/2-i\lambda m_1}(-p_1\tau_i)\nonumber\\
    &=-p_2 H^{(j)}_{-1/2-i\lambda m_2}(-p_2\tau_i)H^{(j)}_{1/2-i\lambda m_1}(-p_1\tau_i)-p_1 H^{(j)}_{-1/2-i\lambda m_1}(-p_1\tau_i)H^{(j)}_{1/2-i\lambda m_2}(-p_2\tau_i)\,,
\end{align}
which is valid for both types of Hankel functions ($j=1,2$). Applying this above formula, we find
\begin{align}
    &\mathcal{D}_{\lambda,\tau_1}\mathcal{D}_{\lambda,\tau_2}\left[H^{(2)}_{\nu_{2\lambda}}(-p_2\tau_2)H^{(1)}_{\nu_{2\lambda}}(-p_2\tau_1)H^{(2)}_{\nu_{1\lambda}}(-p_1\tau_2)H^{(1)}_{\nu_{1\lambda}}(-p_1\tau_1)\right]\nonumber\\
    &=p^2_1 H^{(2)}_{\nu^*_{1\lambda}}(-p_1\tau_2)H^{(1)}_{\nu^*_{1\lambda}}(-p_1\tau_1)H^{(2)}_{\nu_{2\lambda}}(-p_2\tau_2)H^{(1)}_{\nu_{2\lambda}}(-p_2\tau_1)+(p_1\leftrightarrow p_2,m_1\leftrightarrow m_2)\nonumber\\
     &+ p_1 p_2 e^{(\nu_{1\lambda}-\nu_{2\lambda})i\pi} H^{(2)}_{\nu^*_{1\lambda}}(-p_1\tau_2)H^{(2)}_{\nu_{2\lambda}}(-p_2\tau_2)H^{(1)}_{\nu^*_{2\lambda}}(-p_2\tau_1)H^{(1)}_{\nu_{1\lambda}}(-p_1\tau_1)+(m_1\rightarrow m_2)\,,
\end{align}
where $\mathcal{D}_{\lambda,\tau_i}$ is a shorthand notation for the time operator introduced above and also defined in \eqref{def: Doperator}. Finally, after summing over the two helicities and using the fact that the momentum integral is invariant under the exchange of $p_1$ and $p_2$, we arrive at
\begin{align}
        \mathcal{Q}_{-+}(s,\tau_1,\tau_2)&=\frac{\pi^2(\tau_1\tau_2)^4}{16}\sum_{\lambda=\pm}\int_{\mathbf{p_1}}\Bigg\{\mathcal{D}_{\lambda,\tau_1}\mathcal{D}_{\lambda,\tau_2}\bigg[H^{(2)}_{\nu_{2\lambda}}(-p_2\tau_2)H^{(1)}_{\nu_{2\lambda}}(-p_2\tau_1)H^{(2)}_{\nu_{1\lambda}}(-p_1\tau_2)H^{(1)}_{\nu_{1\lambda}}(-p_1\tau_1)\bigg]\nonumber\\
        &~~~-s^2 H^{(1)}_{\nu^*_{1\lambda}}(-p_1\tau_1) H^{(2)}_{\nu^*_{1\lambda}}(-p_1\tau_2) H^{(1)}_{\nu_{2\lambda}}(-p_2\tau_1) H^{(2)}_{\nu_{2\lambda}}(-p_2\tau_2)\Bigg\}\,,
\end{align}
We have used $\mathbf{p}_1\cdot \mathbf{p}_2=-(p_1^2+p_2^2-s^2)/2$ and after rewriting the product of Hankel functions in terms of the scalar propagator \eqref{eq:scalar_wightman}, we finally obtain the relation given in the main text, with the overall operators acting on the scalar bubble \eqref{eq: fermion_scalar_relation}.

\subsection{Spectral decomposition method}\label{sec: spectral_app}
\subsubsection*{Steps for the mode $M$}
In going from \eqref{eq: fermionStep1} to \eqref{eq: fermionStep2}, we have omitted several intermediate steps, which we provide here. To transform back to momentum space, we require $Q^1_{\nu}$ rather than $Q^2_{\nu}$, and therefore we use the following transformation:
\begin{align}
    Q_\nu^2(Z)=\frac{1}{\sqrt{Z^2-1}}\left[\frac{\nu(\nu-1)}{2\nu+1}Q_{\nu+1}^1(Z)-\frac{(\nu+2)(\nu+1)}{2\nu+1}Q_{\nu-1}^1(Z)\right]\,, \label{eq: Q2toQ1}
\end{align}
then \eqref{eq: fermionStep1} becomes
\begin{align}
    Q_M=-\sum_{k=0}^{\infty}&\sum_{\sa=\pm}\frac{\sech (\pi m_1)\sech (\pi m_2)}{8\pi^{7/2}\sqrt{Z^2-1}}\times\Gamma\left[\bgm 2+i\sa m_1, 2+i\sa m_2\\ 1/2+i\sa m_1, 1/2+i \sa m_2\edm\right]\nonumber\\
       &~~\times\Bigg[\mathcal{A}^{(1)}_k Q_{2+2k+i\sa M}^1(-Z)+\mathcal{A}^{(2)}_kQ_{2k+i\sa M}^1(-Z)\Bigg]\,,
\end{align}
with coefficients defined as 
\begin{align}
&\mathcal{A}^{(1)}_k=\Gamma\left[\bgm3/2+2k+i\sa M,2+2k+i\sa M\\2k+i\sa M,4+2k+i\sa M\edm\right]\times f_{k,2} \,,\nonumber\\
 &\mathcal{A}^{(2)}_k= \frac{\Gamma(3/2+2k+i\sa M)}{\Gamma(2+2k+i \sa M)}\times\left(f_{k,1} -f_{k,2} \right)\,.
\end{align}
One can check that the second term vanishes, $\mathcal{A}^{(2)}_{0}=0$. We can therefore rewrite the above expression as
\begin{align}
    Q_M=-\sum_{k=0}^{\infty}&\sum_{\sa=\pm}\frac{\sech (\pi m_1)\sech (\pi m_2)}{8\pi^{7/2}\sqrt{Z^2-1}}\,\Gamma\left[\bgm 2+i\sa m_1, 2+i\sa m_2\\ 1/2+i\sa m_1, 1/2+i \sa m_2\edm\right]\,\left(\mathcal{A}^{(1)}_{k}+\mathcal{A}^{(2)}_{k+1}\right) Q_{2+2k+i\sa M}^1(-Z)\,.
\end{align}
After simplification, we get the expression given in the main text \eqref{eq: fermionStep2}.
\subsubsection*{Steps for the mode $\delta$}
We now consider the other case $\sa=-\sb$, and define $\delta = m_1 - m_2$, then
\begin{align}
\mathcal{Q}_{\delta}&=\sum_{\sa=\pm}\frac{\left(-2Z\right)^{-3-i\sa \delta}}{2\pi^3\cosh(\pi m_1)\cosh(\pi m_2)}\,\Gamma\left[\bgm2+i\sa m_1,2-i\sa m_2\\\frac{1}{2}+i\sa m_1,\frac{1}{2}-i\sa m_2\edm\right]\nonumber\\
    &\times\Bigg\{\prod_{j=1}^2 Z\,\mathcal{D}\left(1-i(-1)^j\sa m_j,-\frac{1}{2},\frac{1}{Z^2}\right)+\prod_{j=1}^{2}\left(Z^2-1\right)^{\frac{1}{2}}\mathcal{D}\left(2-i(-1)^{j}\sa m_j,-\frac{3}{2},\frac{1}{Z^2}\right)\nonumber\\
    &~~~~~~-\left[\left(Z^2-1\right)\mathcal{D}\Big(2+i\sa m_1,-\frac{3}{2},\frac{1}{Z^2}\Big)\mathcal{D}\left(1-i\sa m_2,-\frac{1}{2},\frac{1}{Z^2}\right)+(m_1\leftrightarrow-m_2)\right]
    \Bigg\}\,.
\end{align}
To apply our formula \eqref{eq: product_2F1}, the second argument of $\mathcal{D}(a_1,a_2,z)$ must be the same in both cases, we therefore need to perform the following transformation,
\begin{align}
  \mathcal D\Big(2+\nu,-\frac{3}{2},\frac{1}{Z^2}\Big) = \frac{Z^2}{Z^2-1} \mathcal D\Big(1+\nu,-\frac12,\frac{1}{Z^2}\Big) + \frac{1-\nu}{(1+2\nu)(Z^2-1)}\mathcal D\Big(2+\nu,-\frac12,\frac{1}{Z^2}\Big),  
\end{align}
then it becomes 
\begin{align}
\mathcal{Q}_{\delta}&=-\sum_{\sa=\pm}\frac{\left(-2Z\right)^{-3-i\sa \delta}}{2\pi^3\cosh(\pi m_1)\cosh(\pi m_2)}\,\Gamma\left[\bgm2+i\sa m_1,2-i\sa m_2\\\frac{1}{2}+i\sa m_1,\frac{1}{2}-i\sa m_2\edm\right]\nonumber\\
    &\times\Bigg\{\prod_{j=1}^2 Z\,\mathcal{D}\left(1-i(-1)^j\sa m_j,-\frac{1}{2},\frac{1}{Z^2}\right)-\prod_{j=1}^{2}\left(Z^2-1\right)^{\frac{1}{2}}\mathcal{D}\left(2-i(-1)^{j}\sa m_j,-\frac{3}{2},\frac{1}{Z^2}\right)\nonumber\\
    &~~~~+\left[\frac{1-i\sa m_1}{1+2i\sa m_1}\mathcal{D}\Big(2+i\sa m_1,-\frac{1}{2},\frac{1}{Z^2}\Big)\mathcal{D}\left(1-i\sa m_2,-\frac{1}{2},\frac{1}{Z^2}\right)+(m_1\leftrightarrow -m_2)\right] 
    \Bigg\}\,,
\end{align}
we can now apply the identity \eqref{eq: product_2F1}, which gives
\begin{align}
    \mathcal{Q}_{\delta}&=-\sum_{\sa=\pm}\sum_{k=0}^\infty\frac{\left(-2Z\right)^{-3-2k-i\sa \delta}}{2\pi^3\cosh(\pi m_1)\cosh(\pi m_2)}\times\Gamma\left[\bgm2+i\sa m_1,2-i\sa m_2\\\frac{1}{2}+i\sa m_1,\frac{1}{2}-i\sa m_2\edm\right]\nonumber\\
    &\times\Bigg\{ \tilde{f}_{k,1}\,Z^2\,\mathcal{D}\Big(2+2k+i\sa \delta,-\frac{1}{2},\frac{1}{Z^2}\Big)-\tilde{f}_{k,2}\,(Z^2-1)\,\mathcal{D}\Big(4+2k+i\sa\delta,-\frac{3}{2},\frac{1}{Z^2}\Big)\nonumber\\
    &~~~~~+\left[\frac{1-i\sa m_1}{1+2i\sa m_1}\tilde{f}_{k,3}+\left(m_1\leftrightarrow-m_2\right)\right]\,\mathcal{D}\Big(3+2k+i\sa\delta,-\frac{1}{2},\frac{1}{Z^2}\Big)\Bigg\}\,,
\end{align}
where we defined the notations as 
\begin{align}
    &\tilde{f}_{k,1}\equiv f_k\Big(1+i \sa m_1,1-i\sa m_2,-\frac{1}{2}\Big)\,,\nonumber\\
    &\tilde{f}_{k,2}\equiv f_k\Big(2+i \sa m_1,2-i\sa m_2,-\frac{3}{2}\Big)\,,\nonumber\\
    &\tilde{f}_{k,3}\equiv f_k\Big(2+i\sa m_1,1-i\sa m_2,-\frac{1}{2}\Big)\,.
\end{align}
Next, we convert this expression into the Legendre-$Q$ form and obtain
\begin{align}
    \mathcal{Q}_{\delta}=&-\sum_{k=0}^{\infty}\sum_{\sa=\pm}\frac{1}{4\pi^{7/2}\cosh(\pi m_1)\cosh(\pi m_2)}\times\Gamma\left[\bgm2+i\sa m_1,2-i\sa m_2,\frac{5}{2}+2k+i\sa\delta\\\frac{1}{2}+i\sa m_1,\frac{1}{2}-i\sa m_2,3+2k+i\sa\delta\edm\right]\nonumber\\
    &\Bigg\{\frac{\tilde{f}_{k,1}\,(2+2k+i\sa\delta)}{3+4k+2i\sa\delta}\frac{Z\,Q_{2k+i\sa\delta}^1(-Z)}{\sqrt{Z^2-1}} +\frac{\tilde{f}_{k,2}}{3+2k+i\sa\delta}\,Z\,Q_{1+2k+i\sa\delta}^2(-Z)\nonumber\\
    &~~~-\left[\frac{1-i\sa m_1}{1+2i\sa m_1}\tilde{f}_{k,3}+\left(m_1\leftrightarrow-m_2\right)\right]\,\frac{\,Q_{1+2k+i\sa\delta}^1(-Z)}{\sqrt{Z^2-1}}\Bigg\}\,,
\end{align}
we need to use an identity for the Legendre-$Q$ function to absorb the additional factor of $z$, as
\begin{align}
zQ^\mu_\nu(z) = \frac{\nu-\mu+1}{2\nu+1}\,Q^\mu_{\nu+1}(z) + \frac{\nu+\mu}{2\nu+1}\,Q^\mu_{\nu-1}(z)\,,    
\end{align}
combining this identity with the relation between $Q_\nu^2$ and $Q_\nu^1$ given in \eqref{eq: Q2toQ1}, we finally obtain three terms:
\begin{align}
    \mathcal{Q}_{\delta}=\sum_{\sa=\pm}\sum_{k=0}^{\infty}\,\left[\mathcal{B}^{(1)}_k Q_{-1+2k+i\sa\delta}^1(-Z)+\mathcal{B}^{(2)}_k Q_{1+2k+i\sa\delta}^1(-Z)+\mathcal{B}^{(3)}_k Q_{3+2k+i\sa\delta}^1(-Z)\right]\,,
\end{align}
noting also that the coefficients satisfy $\mathcal{B}^{(1)}_{0}=0$ and $\mathcal{B}^{(3)}_{-1}=0$, the three terms can be combined into a single term, yielding
\begin{align}
    \mathcal{Q}_{\delta}=\sum_{\sa=\pm}\sum_{k=0}^{\infty}\left[\mathcal{B}^{(1)}_{k+1}+\mathcal{B}^{(2)}_k+\mathcal{B}^{(3)}_{k-1}\right] Q^1_{1+2k+i\sa\delta}(-Z)\,.
\end{align}
After simplifying the coefficients, we finally obtain the result presented in the main text \eqref{eq: density_delta}.
\subsection{Mellin-Barnes method}
\subsubsection{Non-local signals}\label{app: MB_nonlocal}
In the main text, the non-local signals from the MB method are given by \eqref{MB_NLS}, whose coefficients are
{\small
\begin{align}
\nonumber\mathcal{A}_{1\lambda}&=\ \frac{2^{q_{12}+5}}{ \pi^{7 / 2}} \left\{\cos \frac{\pi \bar{q}_{12}}{2}+\cos \left[\frac{\pi}{2}\left(\mathsf{c}_{12}-q_{12}\right)-i \lambda \pi\left(\mathsf{c}_1 {m}_1-\mathsf{c}_2 {m}_2\right)\right]\right\}\frac{(-1)^{n_{1234}}}{n_{1}!n_{2}!n_{3}!n_{4}!}\\
\nonumber &\times \Gamma\left[\begin{array}{c}
-i\lambda (\mathsf{c}_1{m}_1-\mathsf{c}_2{m}_2)-n_{1234}-\frac{5}{2} ,i\lambda\mathsf{c}_1{m}_1+n_{12}+2,-i\lambda\mathsf{c}_2{m}_2+n_{34}+2\\
4+n_{1234}+i\lambda (\mathsf{c}_1{m}_1-\mathsf{c}_2{m}_2),-i\lambda\mathsf{c}_1{m}_1-n_{12}-\frac{1}{2},i\lambda\mathsf{c}_2{m}_2-n_{34}-\frac{1}{2}
\end{array}\right]\\
\nonumber &\times \Gamma\left[5+q_1-\frac{\mathsf{c}_{12}}{2}+i\lambda (\mathsf{c}_1{m}_1-\mathsf{c}_2{m}_2)+2 n_{13}, 5+q_2+\frac{\mathsf{c}_{12}}{2}+i\lambda (\mathsf{c}_1{m}_1-\mathsf{c}_2{m}_2)+2 n_{24}\right]\\
&\times \Gamma\left[\frac{\mathsf{c}_1}{2}-i\lambda\mathsf{c}_1{m}_1-n_1,-\frac{\mathsf{c}_1}{2}-i\lambda\mathsf{c}_1{m}_1-n_2,\frac{\mathsf{c}_2}{2}+i\lambda\mathsf{c}_2{m}_2-n_3,-\frac{\mathsf{c}_2}{2}+i\lambda\mathsf{c}_2{m}_2-n_4\right]\,,\label{eq: coeff_A1}
\end{align}}
and 
{\small
\begin{align}
\nonumber\mathcal{A}_{2\lambda}&=\ \frac{2^{q_{12}+5}}{\pi^{7 / 2}}\left\{\cos \frac{\pi \bar{q}_{12}}{2}+\cos \left[\frac{\pi}{2}(\mathsf{c}_{12}-q_{12})-i\lambda \pi(\mathsf{c}_1{m}_1-\mathsf{c}_2{m}_2)\right]\right\}\frac{(-1)^{n_{1234}}}{n_{1}!n_{2}!n_{3}!n_{4}!}\\
\nonumber &\times \Gamma\left[\begin{array}{c}
\frac{\mathsf{c}_{12}}{2}-i\lambda (\mathsf{c}_1{m}_1-\mathsf{c}_2{m}_2)-n_{1234}-\frac{5}{2},-\frac{\mathsf{c}_1}{2}+i\lambda\mathsf{c}_1{m}_1+n_{12}+\frac{5}{2},-\frac{\mathsf{c}_2}{2}-i\lambda\mathsf{c}_2{m}_2+n_{34}+\frac{5}{2} \\
4+n_{1234}-\frac{\mathsf{c}_{12}}{2}+i\lambda (\mathsf{c}_1{m}_1-\mathsf{c}_2{m}_2),\frac{\mathsf{c}_1}{2}-i\lambda\mathsf{c}_1{m}_1-n_{12},
\frac{\mathsf{c}_2}{2}+i\lambda\mathsf{c}_2{m}_2-n_{34}
\end{array}\right]\\
\nonumber &\times \Gamma\left[5+q_1-\frac{\mathsf{c}_{12}}{2}+i\lambda (\mathsf{c}_1{m}_1-\mathsf{c}_2{m}_2)+2 n_{13}, 5+q_2-\frac{\mathsf{c}_{12}}{2}+i\lambda (\mathsf{c}_1{m}_1-\mathsf{c}_2{m}_2)+2 n_{24}\right]\\
&\times \Gamma\left[\frac{\mathsf{c}_1}{2}-i\lambda\mathsf{c}_1{m}_1-n_1,\frac{\mathsf{c}_1}{2}-i\lambda\mathsf{c}_1{m}_1-n_2,\frac{\mathsf{c}_2}{2}+i\lambda\mathsf{c}_2{m}_2-n_3,\frac{\mathsf{c}_2}{2}+i\lambda\mathsf{c}_2{m}_2-n_4\right].\label{eq: coeff_A2}
\end{align}}

\subsubsection{Local signals}\label{app: local_steps}
\subsubsection*{Calculation of $f_\lambda$}
To calculate the local signals, the key intermediate step is to compute $f_\lambda(s_1,s_3;m)$ in \eqref{def_fm_2}. We now give the detailed derivation leading to \eqref{f_m_2}.
First, we note the following identity
\begin{align}
(x+a)_m\,\Gamma(x)=\sum_{t=0}^m(-1)^t \binom{m}{t}(1-a-t)_t\,\Gamma(x+m-t) , \label{F1}   
\end{align}
for example, by applying this identity, we can rewrite
\begin{align}
&\left(s_{12}-\frac{1}{2}\right)_m \Gamma\left(s_2-\frac{\nu_{1\lambda}^*}{2}\right)=\sum_{t_1=0}^m(-1)^{t_1}\binom{m}{t_1}\left(\frac{3}{2}-s_1-\frac{\nu_{1\lambda}^*}{2}-t_1\right)_{t_1} \Gamma\left(s_2- \frac{\nu_{1\lambda}^*}{2}+m-t_1\right),
\end{align}
and similarly for the other terms. We then obtain
\begin{align}
&\nonumber  f_\lambda(s_1,s_3;m)\nonumber\\
&=\sum_{t_1, t_2=0}^m(-1)^{t_{12}}
\int_{-i \infty}^{+i \infty} \frac{\mathrm{~d} s_2}{2 \pi i}\,\Gamma\left[s_2+\frac{ \nu_{1\lambda}^*}{2}, s_2-\frac{\nu_{1\lambda}^*}{2}+m-t_1,\frac{5}{2}-m-s_{123}+\frac{\nu_{2\lambda}}{2}, \frac{5}{2}-s_{123}-\frac{\nu_{2\lambda}}{2}-t_2\right]
\nonumber\\
&\times\Bigg\{\binom{m}{t_1}\binom{m}{t_2} \times\left[\left(\frac{3}{2}-s_1-\frac{\nu_{1\lambda}^*}{2}-t_1\right)_{t_1}\left(\frac{3}{2}-s_3-\frac{\nu_{2\lambda}}{2}-t_2\right)_{t_2}{\hat\Gamma}_{\lambda}\left(s_1, s_3 \right)\right.\nonumber\\
&\left.\qquad+\left(1-s_1-\frac{\nu_{1\lambda}^*}{2}-t_1\right)_{t_1}\left(1-s_3-\frac{\nu_{2\lambda}}{2}-t_2\right)_{t_2}\hat{\Gamma}_{-\lambda}(s_1, s_3)\right]\Bigg\}\,.
\end{align}
We next apply Barnes’ lemma\cite{BarnesLemma1}:
\begin{align}
\int_{-i\infty}^{+i \infty} \frac{\mathrm{~d} s}{2 \pi i} \,\,\Gamma\Big[a+s, b+s, c-s, d-s\Big]=\Gamma\left[\begin{array}{c}
a+c, a+d, b+c, b+d \\
a+b+c+d
\end{array}\right]\,, \label{BL}  
\end{align}
for the $s_2$-integral. 
Finally, we arrive at the expression \eqref{f_m_2}.
\subsubsection*{Coefficients}
The coefficient of the local signals in \eqref{MB_LS} is
{\small
\begin{align}
\nonumber \mathcal{B}_{m,\lambda}&=\ \frac{1}{ \pi^{7 / 2}} \left\{\cos \frac{\pi \bar{q}_{12}}{2}+\cos \left[i\lambda \pi(\mathrm{c}_1{m}_1-\mathrm{c}_2{m}_2)+\frac{\pi (q_{12}-\mathrm{c}_{12})}{2}\right]\right\}\\
\nonumber &\times  \frac{(-1)^{m+n_{12}}}{2^{2m}\,m!n_{1}!n_{2}!}\times\Gamma\left[\begin{array}{c}
-n_1+\frac{\mathrm{c}_1}{2} -i\lambda\mathrm{c}_1{m}_1,-n_2+\frac{\mathrm{c}_2}{2} +i\lambda\mathrm{c}_2{m}_2 \\
\frac{3}{2}+m
\end{array}\right]\\
\nonumber &\times \Gamma\left[5+q_1+2 n_{12}-\frac{\mathrm{c}_{12}}{2}+i\lambda (\mathrm{c}_1{m}_1-\mathrm{c}_2{m}_2), q_2+2 m-2 n_{12}+\frac{\mathrm{c}_{12}}{2}-i\lambda (\mathrm{c}_1{m}_1-\mathrm{c}_2{m}_2)\right]\\
\nonumber &\times \sum_{t_1, t_2=0}^m\frac{(-1)^{t_{12}}\binom{m}{t_1}\binom{m}{t_2}}{\Gamma\left(5+2 n_{12}-t_{12}-\frac{\mathrm{c}_{12}}{2}+i\lambda (\mathrm{c}_1{m}_1-\mathrm{c}_2{m}_2)\right)}\times  \left\{\left(\frac{3}{2}+n_1-\frac{\mathrm{c}_1}{2}-t_1\right)_{t_1}\left(\frac{3}{2}+n_2-\frac{\mathrm{c}_2}{2}-t_2\right)_{t_2}\right.\\
\nonumber &\times \Gamma\left[\frac{5}{2}-m+n_{12}+i\lambda\left(\mathrm{c}_1 {m}_1-\mathrm{c}_2 {m}_2\right), \frac{5}{2}+m+n_{12}-t_{12}-\frac{\mathrm{c}_{12}}{2}\right]\\
\nonumber &\times \Gamma\left[\frac{5}{2}+n_{12}-t_1-\frac{\mathrm{c}_1}{2}-i\lambda \mathrm{c}_2 {m}_2, \frac{5}{2}+n_{12}-t_2-\frac{\mathrm{c}_2}{2}+i\lambda \mathrm{c}_1 {m}_1\right]\\
\nonumber &+\left(1+n_1-t_1\right)_{t_1}\left(1+n_2-t_2\right)_{t_2}\times \Gamma\left[\frac{5}{2}-m+n_{12}-\frac{\mathrm{c}_{12}}{2}+i\lambda\left(\mathrm{c}_1 {m}_1-\mathrm{c}_2 {m}_2\right), \frac{5}{2}+m+n_{12}-t_{12}\right]\\
&\times \Gamma\left[\frac{5}{2}+n_{12}-t_1-\frac{\mathrm{c}_2}{2}-i\lambda \mathrm{c}_2{m}_2, \frac{5}{2}+n_{12}-t_2-\frac{\mathrm{c}_1}{2}+i\lambda \mathrm{c}_1{m}_1\right]\Bigg\}\,.
\end{align}}
\subsubsection{Background}\label{app: MP_background}
In the main body, we focused on the signal contributions of the bubble loop. In this subsection, we also provide the necessary derivations from the MB integrals for the background contributions. These background terms come from two sources: the factorized integral $\mathcal{I}_{\mathrm{F}}$ or $\mathcal{I}_{\mathrm{TO}}$. We discuss them one by one below.

\subsubsection*{Background from factorised part}
As mentioned in Section \ref{sec: MBmethod}, the factorised parts associated with the poles in \eqref{UV2} contribute to the background. To proceed, we change the integration variable to $S=s_{13}$ so that \eqref{I_FIII_2} becomes
\begin{align}
\nonumber\mathcal{I}_{\mathrm{F,UV}}^{q_1 q_2} =&\ \frac{r_1^{5+q_1} r_2^{q_2}}{ \pi^{7 / 2}} \int_{-i \infty}^{+i \infty}\frac{\mathrm{d} S}{2 \pi i} \frac{\mathrm{~d} s_1}{2 \pi i}\left[\cos \frac{\pi \bar{q}_{12}}{2}+\cos \left(2 \pi S-\frac{\pi q_{12}}{2}\right)\right]\left(\frac{r_1}{r_2}\right)^{-2 S}  \\
&\times \sum_{m=0}^{\infty} \frac{(-1)^m}{m!}\left(\frac{r_2}{2}\right)^{2 m}\,\Gamma\left[\begin{array}{c}
q_1+5-2 S, q_2+2m+2 S \\
\frac{3}{2}+m
\end{array}\right] \sum_{\lambda=\pm}\,f_\lambda\left(s_1, S-s_1;m\right),
\end{align}
where now the poles~\eqref{UV2} are located at
\begin{align}
S=-\frac{q_2+2 m+n}{2}, \quad n=0,1, \cdots . \label{background_poles_factorised}
\end{align}
Let us define the last term as 
\begin{align}
h_m(S) \equiv \sum_{\lambda=\pm} \int_{-i\infty}^{+i\infty} \frac{\mathrm{d} s_1}{2 \pi i} \, f_\lambda\left(s_1, S - s_1;m\right), \label{def_gm_2}
\end{align}
which is given explicitly by
\begin{align}
\nonumber h_m(S)=&\ \sum_{\lambda=\pm}\sum_{t_1, t_2=0}^m\binom{m}{t_1}\binom{m}{t_2} \times \frac{(-1)^{t_{12}}}{\Gamma\left(5-2 S-t_{12}\right)}\times I (S)\\
&\nonumber \times \Gamma\left[\frac{\nu_{1\lambda}^*+\nu_{2\lambda}+5}{2}-m-S, \frac{-\nu_{1\lambda}^*-\nu_{2\lambda}+5}{2}+m-S-t_{12} \right]\\
 &\times \Gamma\left[
\frac{\nu_{1\lambda}^*-\nu_{2\lambda}+5}{2}-S-t_2,\frac{-\nu_{1\lambda}^*+\nu_{2\lambda}+5}{2}-S-t_1\right] \,,
\end{align}
where we used the expression for $f_\lambda$ and grouped together all terms contributing to the $s_1$-integral into
\begin{align}
\nonumber I(S)\equiv & \int_{-i\infty}^{+i \infty} \frac{\mathrm{~d} s_1}{2 \pi i}\left\{\left(\frac{3}{2}-s_1-\frac{\nu_{1\lambda}^*}{2}-t_1\right)_{t_1}\left(\frac{3}{2}-S+s_1-\frac{\nu_{2\lambda}}{2}-t_2\right)_{t_2}\hat{\Gamma}\left(s_1, S-s_1\right)\right.\\
 &\left.+\left(1-s_1-\frac{\nu_{1\lambda}^*}{2}-t_1\right)_{t_1}\left(1-S+s_1-\frac{\nu_{2\lambda}}{2}-t_2\right)_{t_2}\hat{\Gamma}_{-\lambda}(s_1, S-s_1)\right\}.
\end{align}
Applying a similar trick, using \eqref{F1} together with Barnes' lemma \eqref{BL}, the $s_1$ integrals can be evaluated, yielding
\begin{align}
\nonumber h_m(S)=&\ \sum_{\lambda=\pm}\,\Gamma\left[3+ \frac{i \lambda\delta}{2}-m-S,S-\frac{1}{2}-\frac{i \lambda \delta}{2}\right] \sum_{t_1, t_2=0}^m \frac{\binom{m}{t_1}\binom{m}{t_2}\,(2S-1+t_{12})}{\Gamma\left[2 S+t_{12}, 5-2 S-t_{12}\right]}\\
\nonumber &\times \Gamma\left[\frac{5}{2}-S-\frac{i \lambda M}{2}-t_1,\frac{5}{2}-S+ \frac{i \lambda M}{2}-t_2, 2+m-S-t_{12}- \frac{i \lambda\delta}{2}\right]\\
&\times \Gamma\left[S-\frac{1}{2}+t_{12}+ \frac{i \lambda\delta}{2}, S+ \frac{i \lambda M}{2}+t_1, S- \frac{i \lambda M}{2}+t_2\right].\label{gm_2}
\end{align}
Now taking the poles \eqref{background_poles_factorised}, and assuming that $q_2$ is an integer as is usually the case, we obtain the background contribution from the factorised part as
\begin{align}
\nonumber\mathcal{I}_{\mathrm{F,BG}}^{q_1 q_2} =&\ \frac{r_1^{5+q_{12}}}{\pi^{7 / 2}} \cos \left(\frac{\pi \bar{q}_{12}}{2}\right) \sum_{m, n=0}^{\infty} \frac{(-1)^m}{m!} \frac{1+(-1)^n}{2n!}\left(\frac{r_2}{2}\right)^{2 m}\left(\frac{r_1}{r_2}\right)^{2 m+n}\\
&\times \Gamma\left[\begin{array}{c}
5+q_{12}+2 m+n \\
\frac{3}{2}+m
\end{array}\right] \times h_m\left( -\frac{q_2+2 m+n}{2}\right)\,.\label{I_FBG_2}
\end{align}
The factors $1/\Gamma(2S + t_{12})$ and $2S - 1 + t_{12}$ in \eqref{gm_2} make it clear that $h_m(S)$ vanishes at
\begin{align}
S = \frac{1}{2} - m - \frac{k}{2}, \quad (k = 0, 1, 2, \cdots)\,.
\end{align}
Therefore, in \eqref{I_FBG_2}, the only nonzero contribution comes from $(q_2, n) = (-2, 0)$, assuming that $q_2$ is an integer with $q_2 \geq -2$. The expression then simplifies to
\begin{align}
\nonumber \mathcal{I}_{\mathrm{F,BG}}^{q_1, -2} =
& -  \frac{r_1^{3+q_{1}}}{\pi^{7 / 2}} \cos \left(\frac{\pi q_{1}}{2}\right) \sum_{m=0}^{\infty} \frac{(-1)^m}{m!} \left(\frac{r_1}{2}\right)^{2 m}\times \Gamma\left[\begin{array}{c}
3+q_{1}+2 m \\
\frac{3}{2}+m
\end{array}\right]\\
&\times \delta M\left(1+M^2\right) \operatorname{csch} (\pi\delta)\operatorname{csch} \left(\pi M\right)\,.
\end{align}
This part of the background does not contain a total-energy singularity as $k_T \equiv k_{1234}$ approaches zero.
\subsubsection*{Background from Time-ordered part}
The time-ordered term in \eqref{TO} also contributes to the background.
Its evaluation proceeds in much the same way: the Hankel function is first rewritten in terms of the MB representation \eqref{MB}, after which the time and loop integrations are performed.
In the time-ordered case, however, the time integration involves a  nested integral, which may be evaluated using the following formula
\begin{align}
\nonumber &\int_{-\infty}^0 \mathrm{~d} \tau_1\mathrm{~d} \tau_2 \, e^{ \pm i\left( k_{1} \tau_1+ k_{2} \tau_2\right)}\left(-\tau_1\right)^{q_1-1}\left(-\tau_2\right)^{q_2-1}\theta\left(\tau_2-\tau_1\right)=e^{\mp i\frac{\pi}{2}q_{12} } k_1^{-q_{12}}\,\pdressFq{2}{1}{q_2,q_{12}}{1+q_2}{-\frac{k_2}{k_1}}\,,
\end{align}
where ${}_2\mathcal{F}_1$ is the dressed hypergeometric function \eqref{eq: dress_pFq}. 
Then, we obtain
\begin{align}
\nonumber \mathcal{I}_{\text{TO}}^{q_1q_2}\equiv &\ \mathcal{I}_{++,\text{TO}}^{q_1q_2}+\mathcal{I}_{--,\text{TO}}^{q_1q_2}\\
\nonumber=&\ -\frac{r_1^{10+q_{12}} }{16\pi^{7/2}}\int_{-i \infty}^{+i \infty}\prod_{i=1}^4 \frac{\mathrm{~d} s_i}{2 \pi i}\left(\frac{r_1}{2}\right)^{-2 s_{1234}}\sin \left[\pi\left(s_{13}-s_{24}\right)\right] \sin \left(\frac{\pi q_{12}}{2}-\pi s_{1234}\right)\\
&\times\sum_{\lambda= \pm}\left\{  \widetilde{\Gamma}_{1\lambda}\left(\{s_i\}\right) \Gamma\left[\begin{array}{c}
-s_{12}+2,-s_{34}+2 \\
s_{12}-\frac{1}{2}, s_{34}-\frac{1}{2}
\end{array}\right]+\widetilde{\Gamma}_{2\lambda}\left(\{s_i\}\right)\Gamma\left[\begin{array}{c}
-s_{12}+\frac{5}{2},-s_{34}+\frac{5}{2} \\
s_{12}, s_{34}
\end{array}\right]\right\}\nonumber\\
 &\times \Gamma\left[
 \begin{matrix}
     s_{1234}-\frac{5}{2}\\
     -s_{1234}+4
 \end{matrix}
 \right]\,\pdressFq{2}{1}{q_2+5-2s_{24}, q_{12}+10-2s_{1234}}{q_2+6-2s_{24}}{-\frac{r_1}{r_2}}\,.\label{I_TO2}
\end{align}
As in the factorised contributions, \eqref{I_TO2} contains the spectral poles \eqref{s_pole_1} and \eqref{s_pole_2}, as well as the UV poles \eqref{uv_pole}.
The former, however, do not contribute to the integral because of the factors $\sin\left[\pi\left(s_{13}-s_{24}\right)\right]$ and $1/\Gamma[s_{12}-1/2, s_{34}-1/2]$ (or $1/\Gamma[s_{12}, s_{34}]$) appearing in \eqref{I_TO2}.
Taking the UV poles at $s_{1234}=5/2-m$ and then performing the $s_4$ integral, we obtain
\begin{align}
\nonumber \mathcal{I}_{\mathrm{TO}}^{q_1 q_2}=&-\frac{2r_1^{q_{12}+5}}{\pi^{7 / 2}}\cos \left(\frac{\pi q_{12}}{2}\right) \int_{-i \infty}^{+i \infty} \prod_{i=1}^2 \frac{\mathrm{~d} s_i}{2 \pi i}\sum_{m=0}^{\infty} \frac{(-1)^m}{m!}\left(\frac{r_1}{2}\right)^{2 m} \frac{\cos (2 \pi s_{13})}{\Gamma\left(\frac{3}{2}+m\right)}\\
&\times \pdressFq{2}{1}{q_2+2m+2 s_{13}, q_{12}+5+2m}{q_2+1+2m+2s_{13}}{-\frac{r_1}{r_2}}\sum_{\lambda=\pm} f_\lambda(s_1,s_3;m)\,,
\end{align}
where we have used the function $f_\lambda(s_1,s_3;m)$ defined in \eqref{def_fm_2}, together with the result~\eqref{f_m_2}.
Changing the integration variable to $S \equiv s_{13}$ and then performing the $s_1$ integral, we get
\begin{align}
\nonumber \mathcal{I}_{\mathrm{TO}}^{q_1 q_2}=&\ -\frac{2r_1^{q_{12}+5}}{\pi^{7 / 2}}\cos \left(\frac{\pi q_{12}}{2}\right) \int_{-i \infty}^{+i \infty} \frac{\mathrm{~d} S}{2 \pi i} \sum_{m=0}^{\infty} \frac{(-1)^m}{m!}\left(\frac{r_1}{2}\right)^{2 m} \frac{\cos (2 \pi S)}{\Gamma\left(\frac{3}{2}+m\right)}\\
&\times \pdressFq{2}{1}{q_2+2m+2 S, q_{12}+5+2m }{q_2+1+2m+2S}{-\frac{r_1}{r_2}}\times h_m(S),
\end{align}
where $h_m(S)$ is defined in \eqref{def_gm_2} and explicitly evaluated in \eqref{gm_2}. The $S$-integral is divergent. To see this, we collect all terms involving $S$ into
\begin{align}
\widetilde{h}_m\left(S ; \frac{r_1}{r_2}\right) \equiv \frac{\cos (2 \pi S)}{\Gamma\left(\frac{3}{2}+m\right)} \times\pdressFq{2}{1}{q_2+2m+2 S, q_{12}+5+2m }{q_2+1+2m+2S}{-\frac{r_1}{r_2}}\times h_m(S)\,,    
\end{align}
and use asymptotic behavior of special functions, especially   
\begin{align}
\lim _{x \rightarrow \infty}\pFq{2}{1}{a,b+x}{c+x}{z}=\lim _{x \rightarrow \infty} \sum_{n=0}^{\infty} \frac{(a)_n(b+x)_n}{(c+x)_n} \frac{z^n}{n!}=\sum_{n=0}^{\infty} \frac{(a)_n z^n}{n!}=(1-z)^{-a} ,    
\end{align}
and
\begin{align}
\lim _{z \rightarrow \pm i \infty} \Gamma(a+z)\Gamma (b-z) \sim 2 \pi\,e^{ \pm i \pi(a-b) / 2} e^{-\pi|z|}|z|^{a+b-1}\,.     \end{align}
Then, we find that the integrand behaves at $|S|\rightarrow\infty$ as 
\begin{align}
\nonumber \lim _{S \rightarrow \pm i \infty} \widetilde{h}_m\left(S ; \frac{r_1}{r_2}\right)&=  \frac{e^{2 \pi|S|}}{4S } \times \frac{\Gamma\left(5+q_{12}+2 m\right)}{\Gamma\left(\frac{3}{2}+m\right)}\left(\frac{r_2}{r_{12}}\right)^{5+q_{12}+m} \times \lim _{S \rightarrow \pm i \infty} h_m(S)\\
&= \begin{cases}-\pi^{5 / 2} S\left(\frac{r_2}{r_{12}}\right)^{5+q_{12}} \Gamma\left(5+q_{12}\right), & m=0 \\ \mathcal{O}\left(1\right), & m=1,2, \cdots\end{cases}
\end{align}
It is now clear that the leading divergence in this type of fermionic bubble diagram is \textit{quadratic}.

\end{appendix}
\bibliographystyle{JHEP}
\bibliography{ref}

\end{document}